\documentclass[pra, aps, amssymb, twocolumn, superscriptaddress, notitlepage, longbibliography]{revtex4-2}

\usepackage[utf8]{inputenc}
\usepackage[T1]{fontenc}
\usepackage{bm,amssymb,amsmath,mathrsfs,amstext,latexsym,physics}
\usepackage{cancel,hhline}
\usepackage[usenames,dvipsnames]{xcolor}
\usepackage[caption=false,subrefformat=parens,labelformat=empty]{subfig}
\usepackage{graphicx,tikz}
\usepackage[colorlinks=true,citecolor=RoyalBlue,urlcolor=RoyalBlue,linkcolor=RoyalBlue]{hyperref}

\usepackage[all]{hypcap}
\usepackage{orcidlink}
\hypersetup{
  pdftitle={Universal dynamics of quantum phase transitions with fractional long-range interactions},
  pdfauthor={A. Grabarits, L. D. Carr, A. del Campo}}

\let\Im\relax

\DeclareMathOperator{\Im}{Im}

\begin{document}

\author{Andr\'as Grabarits\orcidlink{0000-0002-0633-7195}}
\email{andras.grabarits@uni.lu}
\affiliation{Department  of  Physics  and  Materials  Science,  University  of  Luxembourg,  L-1511  Luxembourg, Luxembourg}
\author{Lincoln D.  Carr\orcidlink{0000-0002-4848-7941}}
\affiliation{Quantum Engineering Program, Colorado School of Mines, 1523 Illinois St., Golden, CO 80401, USA}
\affiliation{Department of Physics, Colorado School of Mines, 1523 Illinois St., Golden, CO 80401, USA}
\affiliation{Department of Applied Mathematics and Statistics, Colorado School of Mines, 1523
Illinois St., Golden, CO 80401, USA}
\affiliation{Institute of Science and Technology Austria (ISTA), Am Campus 1, 3400 Klosterneuburg, Austria}

\author{Adolfo del Campo\orcidlink{0000-0003-2219-2851}}\email{adolfo.delcampo@uni.lu}
\affiliation{Department of Physics and Materials Science, University of Luxembourg, L-1511 Luxembourg, Luxembourg}
\affiliation{Donostia International Physics Center, E-20018 San Sebastian, Spain}

\date{\today}

\title{Universal dynamics of quantum phase transitions with fractional long-range interactions}
\begin{abstract}
The complete statistical characterization of defects generated in a quantum
phase transition has mostly been studied in exactly solvable models, and
remains little explored in nonintegrable systems such as long-range
interacting spin chains. We characterize the full
defect counting statistics in a nonintegrable long-range spin chain, the
fractional Ising model, by time-dependent variational-principle simulations
within the matrix-product-state and matrix-product-operator formalism.
Fractional interactions capture emergent multiscale correlations and long-range
effects in nature, governed by the order of the underlying fractional
derivative. Such
couplings can be approximated on existing quantum simulation platforms,
where the full defect statistics can be measured. The defect number cumulants follow power laws in the driving time, with exponents that vary continuously with the fractional order. At low fractional
order, deep in the long-range regime, the mean defect number decays more slowly
than predicted by the Kibble-Zurek mechanism (KZM), while at intermediate order the KZM holds with the numerically extracted many-body critical exponents. At large fractional order, the exactly solvable free-fermion limit of the model becomes progressively more accurate towards the short-range limit. We derive exact bounds on the kink-number cumulants and on the kink-kink correlations, implying an exact Gaussian limiting distribution for large system sizes. In the many-body simulations, however, kinks are more correlated, consistent with the confinement of domain walls, and these bounds are violated in the long-range regime, also in the bulk of the chain. Nevertheless, the kink distribution still approaches a Gaussian for large defect numbers, with proportionality partially broken between the cumulants.

\end{abstract}

\maketitle

\section{Introduction}\label{sec:intro}

In the Kibble-Zurek (KZ) mechanism, equilibrium properties determine a nonequilibrium phenomenon. The critical exponents of a continuous phase transition fix how the density of topological defects scales with the rate at which the transition is crossed~\cite{kibble1976topology,kibble1980some,zurek1985cosmological,zurek1996cosmological,delcampo2014universality}.
The validity of the KZ mechanism in the quantum domain was theoretically established in a series of
seminal works, from minimal two-level settings to genuinely many-body ones
\cite{damski2005simplest,dziarmaga2005dynamics,zurek2005dynamics,polkovnikov2005universal}.
Subsequent efforts have involved experimental tests in quantum
simulators \cite{cui2016experimental,keesling2019quantum,Cui2020,bando2020probing,ebadi2021quantum,king2022coherent,king2024computational,ZhangWang2025,bayocboc2026}, which have also revealed its breakdown in frustrated lattices~\cite{Ali2024}, and ultracold atomic systems \cite{Weiler08,Lamporesi13,Corman2014,Navon2015,Ko2019,Goo21}, as well as numerical simulations~\cite{Schmitt2022,Zhang2025}, including classical benchmarks of claimed quantum advantage~\cite{Tindall2026}.
The universality unveiled by the KZ scaling extends to the full counting statistics of topological defects, whereby every defect number cumulant follows the same power law as the average \cite{delcampo2018,gomez-ruiz2020full,Mayo21,Balducci2023,grabarits2024nonadiabatic,Kiss25}. 
 In quantum Ising chains, the defects are identical to kinks arising from spontaneous $\mathbb Z_2$ symmetry breaking and separating domains of opposite magnetization.
These results assume that kinks form independently of each other. In a spin chain with fractional long-range interactions, we find that the critical exponents still fix the defect density at intermediate range, while at long range, where interactions bind defects into pairs, the variance of the kink number exceeds twice its mean, the upper limit for any free-fermion state.

Long-range interactions arise naturally on a variety of quantum simulation
platforms~\cite{Defenu23}, and power-law interacting systems have been
 experimentally investigated with trapped-ion crystals
\cite{Ejtemaee13,Pyka13,Ulm13}, in Rydberg quantum simulators
\cite{keesling2019quantum}, and in condensed matter \cite{Du2023}.
In this setting, nonequilibrium scaling laws were found in a pioneering study in the  Lipkin-Meshkov-Glick (LMG) model \cite{Caneva08}. 
The KZ scaling holds in the exactly solvable
long-range Kitaev chain, with anomalous deviations in parts of parameter space \cite{DuttaDutta17,Defenu19}, while in the strong long-range limit the kink density becomes independent of the driving rate~\cite{Defenu18,Gherardini24}. Matrix-product-state
simulations of the long-range Ising chain found power-law scaling with exponents that change by about 25\% with the interaction range \cite{Jaschke17}. The KZ scaling also
persists in classical regular networks of low connectivity
\cite{GrabaritsSammartino25} and under truncated counterdiabatic
driving \cite{delcampo2012assisted,Visuri25,Grabarits25}. Experimentally, trapped-ion quantum simulators have observed the KZ mechanism in the long-range transverse-field Ising model \cite{LiWu23}. 
Beyond the mean
number of defects, their universal statistics has been characterized in
free-fermion models, including long-range Kitaev chains \cite{Balducci2023,Grabarits25,Dhara2026} in agreement with previous beyond-KZM studies \cite{delcampo2018,gomez-ruiz2020full,Mayo21}, also in the presence of noise \cite{Dutta2016,Cui2020,bando2020probing,Dhara2026}. In Rydberg chains, domain-wall statistics were measured after a single sweep~\cite{Bernien17}, and the defect number distribution after slow ramps departs from that of independently formed defects~\cite{Balewski25}.
In the LMG model the defect number follows a negative binomial distribution, which is super-Poissonian \cite{Gherardini24}.
Despite these advances, even the average number of kinks remains poorly
understood in nonintegrable long-range spin systems, while their full counting
statistics has so far been obtained only for free-fermion chains and in the fully connected limit, and has not been explored in nonintegrable spin chains, especially in the most exciting long-range interacting regime. Most pressingly, it remains open whether kink-kink interactions induce a domain-wall confinement strong enough to break the independent kink formation assumed in the free-fermion description, leading to completely different counting statistics.

A prominent route to long-range quantum critical dynamics is provided by
fractional calculus, which extends differentiation and integration to
noninteger order and is widely used to model anomalous transport \cite{MetzlerKlafter00,Zaburdaev15,Berkowitz06} and wave propagation in optics \cite{Longhi15PRL,Zhang15PRL,Liu23NatComm}.  In quantum mechanics, it leads to the fractional
Schr\"odinger equation, originally derived from L\'evy-flight path integrals in which the Brownian paths of the Feynman construction are replaced
by paths with heavy-tailed step-length statistics
\cite{Laskin00PLA,Laskin00PRE}.  The fractional order acts as
a L\'evy index and controls the diffusive character of the quantum states.
A standard realization is the Riesz fractional derivative, defined in momentum space by $\partial_q f(x)\to-|k|^q\tilde f(k)$, with $\tilde f(k)$ the Fourier transform of $f(x)$. For noninteger $q$, the Riesz fractional derivative is no longer local in position space \cite{Ortigueira06,Stickler13}.  Yet, L\'evy flights provide a natural origin for it.
For $0<q<2$, their stable jump-length distribution, with the heavy tail $|x|^{-(1+q)}$, has
the characteristic function $e^{-|k|^{q}}$, so that the composition of many microscopic
steps accumulates a factor $-|k|^{q}$ in the exponent~\cite{Laskin00PRE,LewisCarr25JPhysA}. The nonlocality is
therefore inherited from the stability of the underlying jump statistics rather
than imposed by hand as a prescribed algebraic decay. By the generalized central limit theorem, sums of many heavy-tailed steps approach L\'evy-stable laws, just as sums of steps with finite variance approach the Gaussian, their $q=2$ member~\cite{MetzlerKlafter00,Zaburdaev15}, so heavy-tailed transport is generic in disordered and multiscale media. 

On a lattice, the coupling between two sites at distance $r$ is the corresponding coefficient of the centered discretized fractional derivative, $J_{r,q}$, decaying asymptotically as $r^{-(1+q)}$ with subleading power-law corrections. The interaction is
strongly nonlocal and ferromagnetic for $q<2$, reduces to the nearest-neighbor bond at $q=2$,
and acquires a weak antiferromagnetic tail for $q>2$, so that long- and short-range regimes arise as limits of a single tunable model~\cite{LewisCarr25FractionalIsing,LewisGongCarr25LevyLightCones}. Quantum Ising chains with pure power-law couplings $r^{-(1+q)}$ are Gaussian for $q<2/3$~\cite{DuttaBhattacharjee01,DefenuTrombettoniRuffo17,Koziol21} and, according to functional renormalization-group and quantum Monte Carlo studies, in the short-range class for $q>7/4$~\cite{DefenuTrombettoniRuffo17,ShirataniTodo24}. The fractional couplings share this tail, but whether the fractional chain follows the same thresholds has not been established; classical simulations of the fractional model are consistent with the corresponding classical thresholds~\cite{LewisCarr25FractionalIsing}. For $q\leq1$ the fractional couplings confine domain walls into bound pairs~\cite{Liu19confined,Tan21confinement}. The critical exponents vary continuously with the fractional
order~\cite{LewisCarr25FractionalIsing,LewisGongCarr25LevyLightCones}, and the spreading of the entanglement follows the L\'evy light cones \cite{LewisGongCarr25LevyLightCones}.  These features make
fractional models a natural setting in which to confront the KZM with a continuous family of exponents, which is also relevant for quantum simulators of realistic materials beyond idealized lattices.

In this work, we study the quantum critical dynamics of the fractional Ising chain with its interaction generated by a fractional derivative of continuously tunable
order, by means of time-dependent variational principle (TDVP)
simulations~\cite{Haegeman2016} with a matrix-product-operator (MPO) and
matrix-product-state (MPS)~\cite{Schollwock2011} representation of
the long-range interacting Hamiltonian and its eigenstates, respectively.
Driving the transverse field across the critical point, we obtain the
residual kink density as a function of the driving time and find a power-law exponent that continuously varies with the fractional order.  We furthermore extract the critical exponents from the equilibrium quantities alone, from the finite-size scaling of the gap, and from the entanglement light cone arising from a local perturbation at criticality.  With these exponents,
the Kibble-Zurek prediction accounts for the measured scaling at intermediate- and short-range order, while the measured decay is slower at small fractional order. Going beyond the
mean kink number, we resolve the higher cumulants and reconstruct the full
kink number distribution of a nonintegrable long-range spin
chain. These results are benchmarked against the exactly solvable free-fermion approximation of the model, a fractional $p$-wave superconductor. The description deviates appreciably at small fractional order, improves steadily
above unit order, and becomes quantitatively accurate in the short-range limit.

Within the free-fermion description, we derive the full statistics of both the excitations and the kinks. We show that the kink number relative to the free-fermion ground state coincides with the number of excitations, up to a fractional-order dependent prefactor. We further prove that in any free-fermion state the kink number variance cannot exceed twice its mean and the total correlation of a bond cannot exceed its local variance, implying a Gaussian limiting distribution for the full kink statistics.

In the TDVP simulations, both bounds are violated deep in the long-range regime, also at the center of the chain, where kinks are created close to each other, consistent with the domain-wall confinement induced by the kink-kink interactions. Nevertheless, the kink distribution still approaches a Gaussian for large kink numbers.

The paper is organized as follows. Section~\ref{sec:Ising} introduces
the fractional Ising chain, the numerical scheme used
throughout, and the numerically obtained kink-number
cumulants and full counting statistics. Section~\ref{sec:diagnostics}
then measures $z$ and $\nu$ independently of the driving protocol, through the
finite-size gap scaling and the entanglement light cone, and compares the corresponding KZ exponent with the numerical results of the decay of the mean kink number.
Section~\ref{sec:pwave} closes the argument by developing the free-fermion description, the fractional $p$-wave superconductor, from which
the critical exponents, the exact Landau-Zener kink statistics, the general free-fermion bounds on the kink cumulants and correlations and the
confrontation with the KZM follow analytically, together with the kink-kink correlations, which we compare with the TDVP simulations. All technical details are collected in the appendices.

\section{Fractional Ising model}\label{sec:Ising}

\subsection{Model and numerical method}\label{sec:Ising_model}
We consider the fractional transverse-field Ising chain
\begin{equation}\label{eq:H_Ising}
\begin{split}
    \hat H_\mathrm{Ising}(t)=&-g(t)\sum_{i=1}^L \hat\sigma_i^z\\
    &-\bigl(1-g(t)\bigr)\sum_{i=1}^L\sum_{r=1}^{L}
    J_{r,q}\,\hat\sigma_i^x\hat\sigma_{i+r}^x,
\end{split}
\end{equation}
driven by the linear protocol $g(t)=1-t/\tau_Q$, $t\in[0,\tau_Q]$, with the driving time $\tau_Q$ and open
boundary conditions, so that the second sum runs over the bonds with
$i+r\leq L$. We set $\hbar=1$, so that times are measured in units of the inverse energy scale of Eq.~\eqref{eq:H_Ising}.
The couplings are given by the coefficients of the centered discretized Riesz fractional derivative of order $q$~\cite{Ortigueira06}, expressed as binomial coefficients of
the noninteger argument,
\begin{equation}\label{eq:Jrq}
    J_{r,q}=(-1)^{r+1}\binom{q}{q/2+r}.
\end{equation}
For large separations the couplings decay as
\begin{equation}\label{eq:Jrq_tail}
    J_{r,q}\simeq\frac{\Gamma(q+1)}{\pi}\sin\!\left(\frac{\pi q}{2}\right)
    r^{-(1+q)},\qquad r\to\infty.
\end{equation}
The next subleading correction scales as $r^{-(3+q)}$, so that the profile approaches a
power law only asymptotically and retains additional structure at short and
intermediate range \cite{LewisGongCarr25LevyLightCones}.
For $0<q<2$ all couplings are ferromagnetic and heavy-tailed, describing
L\'evy-type long-range interactions.  At $q=2$ the binomial terminates after
the first bond, $J_{1,2}=1$, and the
standard nearest-neighbor transverse-field Ising model (TFIM) is recovered.  For $q>2$ the tail changes sign while decaying faster than
$r^{-3}$, so that a dominant nearest-neighbor ferromagnetic bond competes with a weak long-range antiferromagnetic background. The equilibrium critical exponents vary smoothly with the fractional order~\cite{LewisCarr25FractionalIsing}.

At $g=1$ the ground state is the fully polarized paramagnet, while below the critical point the $\mathbb Z_2$ symmetry $\hat\sigma^x_i\to-\hat\sigma^x_i$ is spontaneously broken, so that the ramp creates kinks between domains of opposite magnetization along $x$.
At small $q$ the long-range couplings bind kinks into pairs. At $g=0$, reversing $m$ consecutive spins in the bulk of the ordered state creates a kink-antikink pair at the energy cost
\begin{equation}\label{eq:Vm}
\begin{split}
    V(m)&=4\sum_{r\geq1}J_{r,q}\min(r,m)\\
    &=V_0+\frac{4\mathcal A_q}{q(1-q)}\,m^{1-q}+\mathcal O\!\left(m^{-1-q}\right),
\end{split}
\end{equation}
for $q\neq1$, with $\mathcal A_q=\Gamma(q+1)\sin(\pi q/2)/\pi$ the tail amplitude of Eq.~\eqref{eq:Jrq_tail} and $V_0=\frac{q}{q-1}\binom{q}{q/2}$, which equals $4\sum_r rJ_{r,q}$ for $q>1$. For $q<1$ the potential grows as $m^{1-q}$, and logarithmically at $q=1$, confining kinks into pairs~\cite{Liu19confined,Tan21confinement}. For $1<q<2$ it saturates at $V_0$, i.e., the energy cost of two free walls, and for $q>2$ it decreases with $m$, as the antiferromagnetic tail repels the walls at short distance. The free-fermion approximation of Sec.~\ref{sec:JW} does not contain this kink-kink interaction~\cite{DefenuEnssHalimeh19}.

Away from $q=2$ the model is not integrable, and we study it numerically
within the MPS formalism~\cite{Schollwock2011}
using the open-source package ITensors.jl~\cite{ITensor}, for $L=150$ spins. The Hamiltonian is represented as an MPO, built with the OpSum construction of ITensors.jl directly from all terms $J_{r,q}\hat\sigma^x_i\hat\sigma^x_{i+r}$, $1\leq r\leq L-1$, with the exact couplings of Eq.~\eqref{eq:Jrq} rather than from a fit by a sum of exponentials~\cite{Pirvu10}. The conversion to an MPO compresses, at each bond $l$, the coupling matrix $J_{j-i}$ with $i\leq l<j$ between the two halves of the chain by a singular value decomposition with a cutoff of $10^{-15}$~\cite{Chan16}, so that the MPO bond dimension at bond $l$ equals the rank of this matrix plus two. As the couplings form a single smoothly decaying sequence, these matrices have at most $18$ singular values above $10^{-15}$ of the largest one for every $q$. The MPO bond dimension is therefore at most $20$, compared with $L/2+2=77$ for the uncompressed MPO and $12$--$16$ for fits by $10$--$14$ exponentials~\cite{LewisGongCarr25LevyLightCones}, and the couplings are reproduced to a relative accuracy of at least $10^{-8}$. The transverse-field and interaction parts are stored as separate MPOs and combined at each time step with $g(t)$ evaluated at the midpoint of the step. The ground states are obtained by the density-matrix renormalization group (DMRG), and the driven dynamics by the two-site TDVP~\cite{Haegeman2016} implemented in TimeEvoMPS.jl~\cite{TimeEvoMPS}, starting from the paramagnetic product state, with a maximal bond dimension $\chi_{\rm max}=200$, time steps $0.01\leq\delta t\leq0.4$ with $\delta t\leq\tau_Q/10$, and truncation cutoffs between $10^{-7}$ and $10^{-10}$, which bound the discarded weight of each truncation whenever the bond dimension stays below $\chi_{\rm max}$.

To control the truncation, each of the $356$ driving times was simulated with typically four to six combinations of $\delta t$ and cutoff, $1469$ runs in total over $17$ fractional orders. With cutoffs of $10^{-8}$ and below, the cumulants, defined in Sec.~\ref{sec:cumulants_num}, deviate from those of the most accurate run at the same $q$ and $\tau_Q$ by a median of $0.04\%$ for $\kappa_1$ and $0.2\%$ for $\kappa_2$, and by less than $0.3\%$ and $2\%$ in $90\%$ of the cases, while the deviation of $\kappa_3$ stays below $6\%$ of $\kappa_2$ in $90\%$ of the cases. A cutoff of $10^{-7}$ changes $\kappa_2$ by about $4\%$, and the figures show the most accurate run at each driving time. The bond dimension reaches $\chi_{\rm max}$ in $12\%$ of the runs with cutoffs of $10^{-8}$ and below, all at $q\leq0.9$ apart from a few at $q=1.25$ and for $\tau_Q\gtrsim20$. Where saturated and unsaturated runs exist at the same $\tau_Q$, for example at $q=0.2$ and $\tau_Q=30$, their cumulants agree within the deviations given above. For $q\lesssim0.3$ the ratio $\kappa_2/\kappa_1$ exceeds $2$ at slow ramps, with $\kappa_2/\kappa_1\simeq3.1$--$5.1$ at $q=0.1$, $2.2$--$3.0$ at $q=0.2$ and $2.0$--$2.6$ at $q=0.3$ for $\tau_Q\geq50$, so that the violation of the free-fermion bound of Sec.~\ref{sec:pwave} is not an effect of the truncation.

On the open chain each spin feels, on average over the sites, $59\%$ of the total coupling $\sum_{r\neq0}J_{r,q}=J_q$ of the infinite chain at $q=0.2$, with $62\%$ at the center and $34\%$ at the edges, while the average rises to $88\%$ and $99\%$ at $q=0.5$ and $q=1.5$. The missing tail decays only as $L^{-q}$, so that finite-size effects are strongest deep in the long-range regime. The finite-size scaling of the gap up to $L=200$ in Sec.~\ref{sec:gapscaling} shows that $L=150$ already lies in the scaling regime. With the leading correction of Eq.~\eqref{eq:gap_L}, the gaps of all system sizes collapse onto a single curve near the minimum for every $q$, and for $q\geq1.5$ over the whole range of $g$, while at small $q$ the curves of the largest sizes, $L\gtrsim140$, nearly coincide also away from the minimum (Fig.~\ref{fig:gap_collapse}).   In addition, as the bound of App.~\ref{app:bound} holds for every $L$, its violation cannot be produced by any free-fermion state of any size, and an exact time evolution of small chains with all couplings, $L=12$--$20$, already shows a departure from the free-fermion bounds (Sec.~\ref{sec:defects}).

\begin{figure*}[t]
    \centering
    \begin{tabular}{c  c  c c}\includegraphics[width=0.53\columnwidth,trim={0 0 0 0cm},clip]{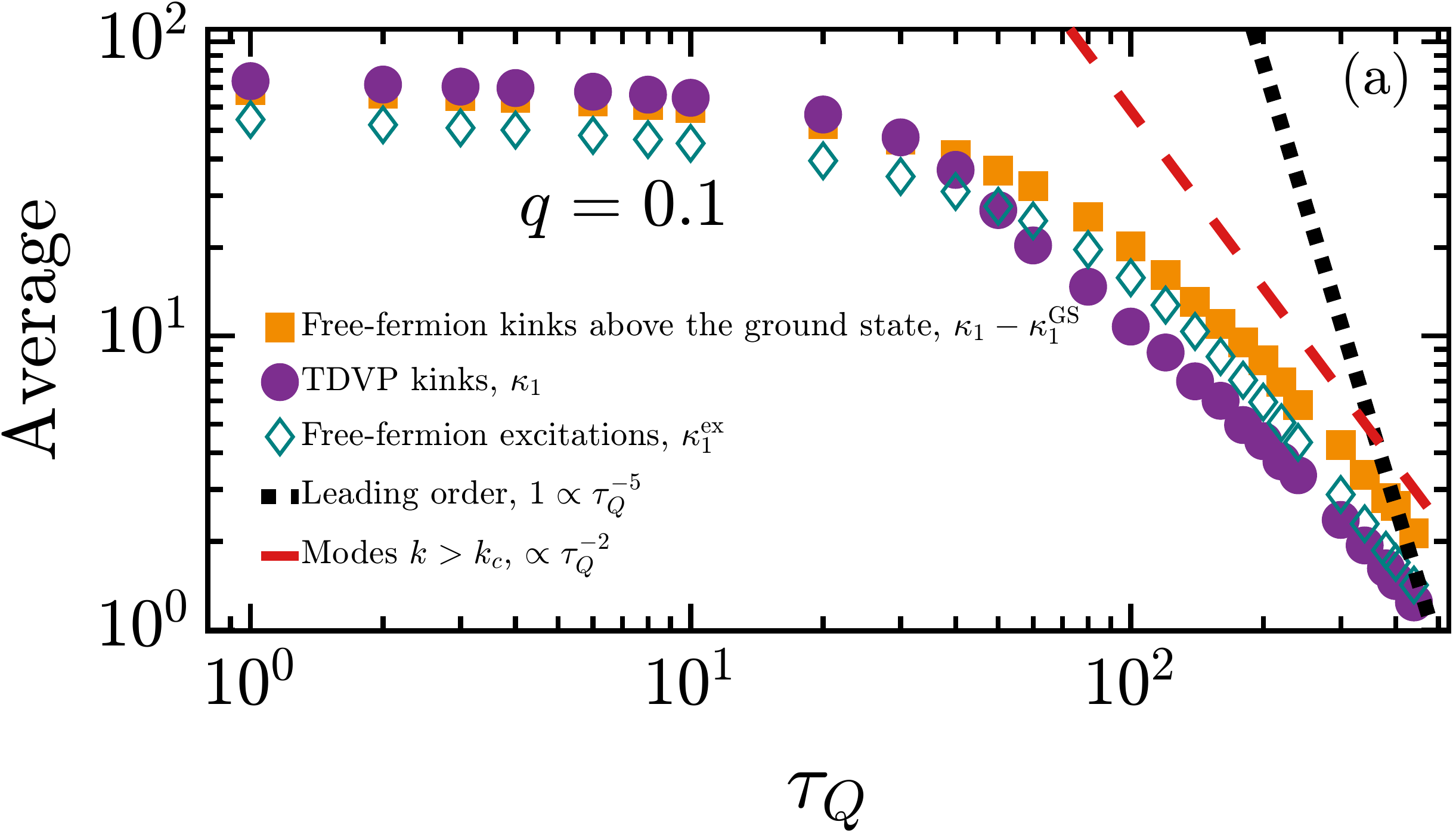} &         
    \includegraphics[width=0.49\columnwidth,trim={2.75cm 0 0 0},clip]{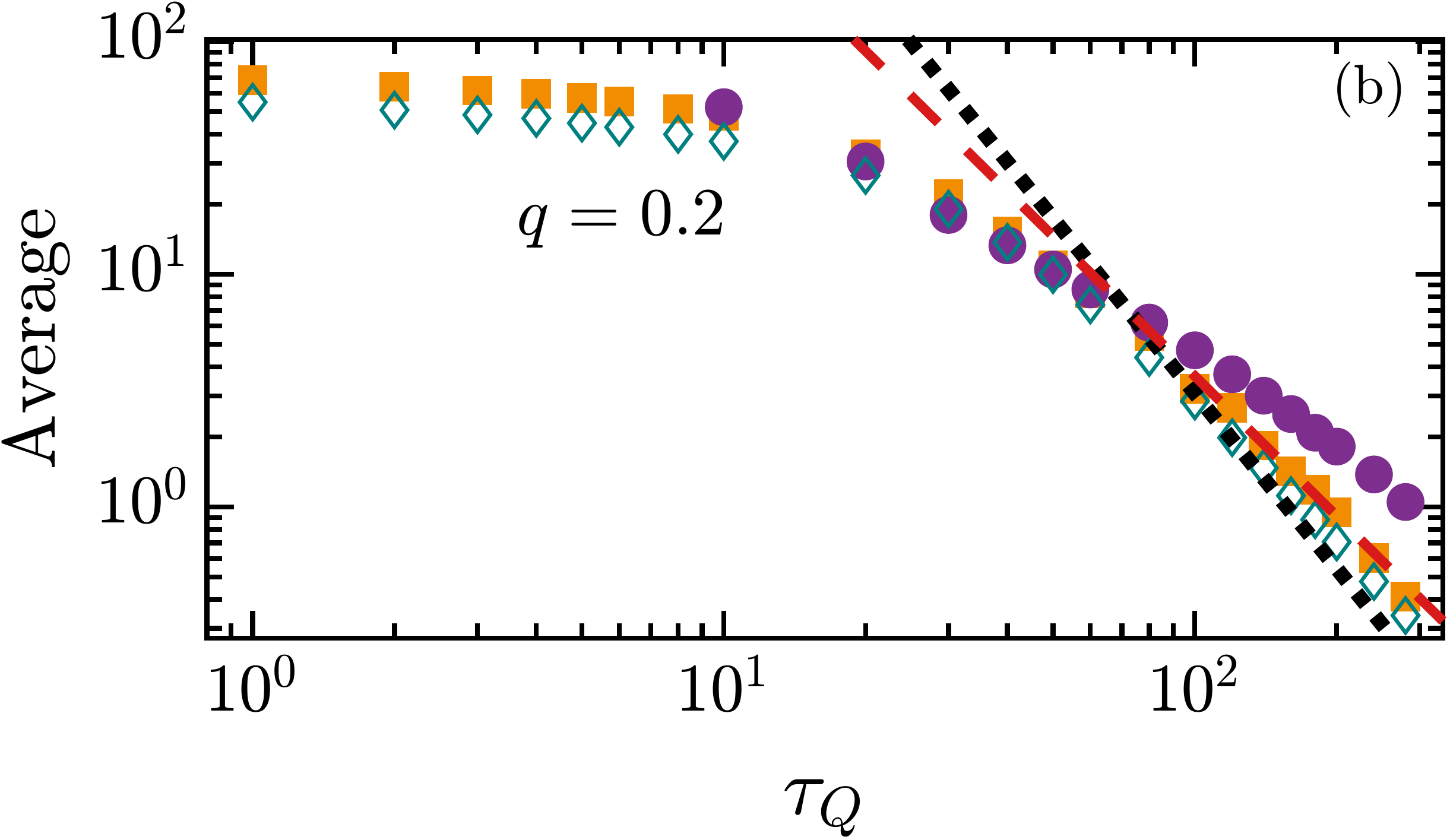} &   
    \includegraphics[width=0.49\columnwidth,trim={2.75cm 0 0cm 0},clip]{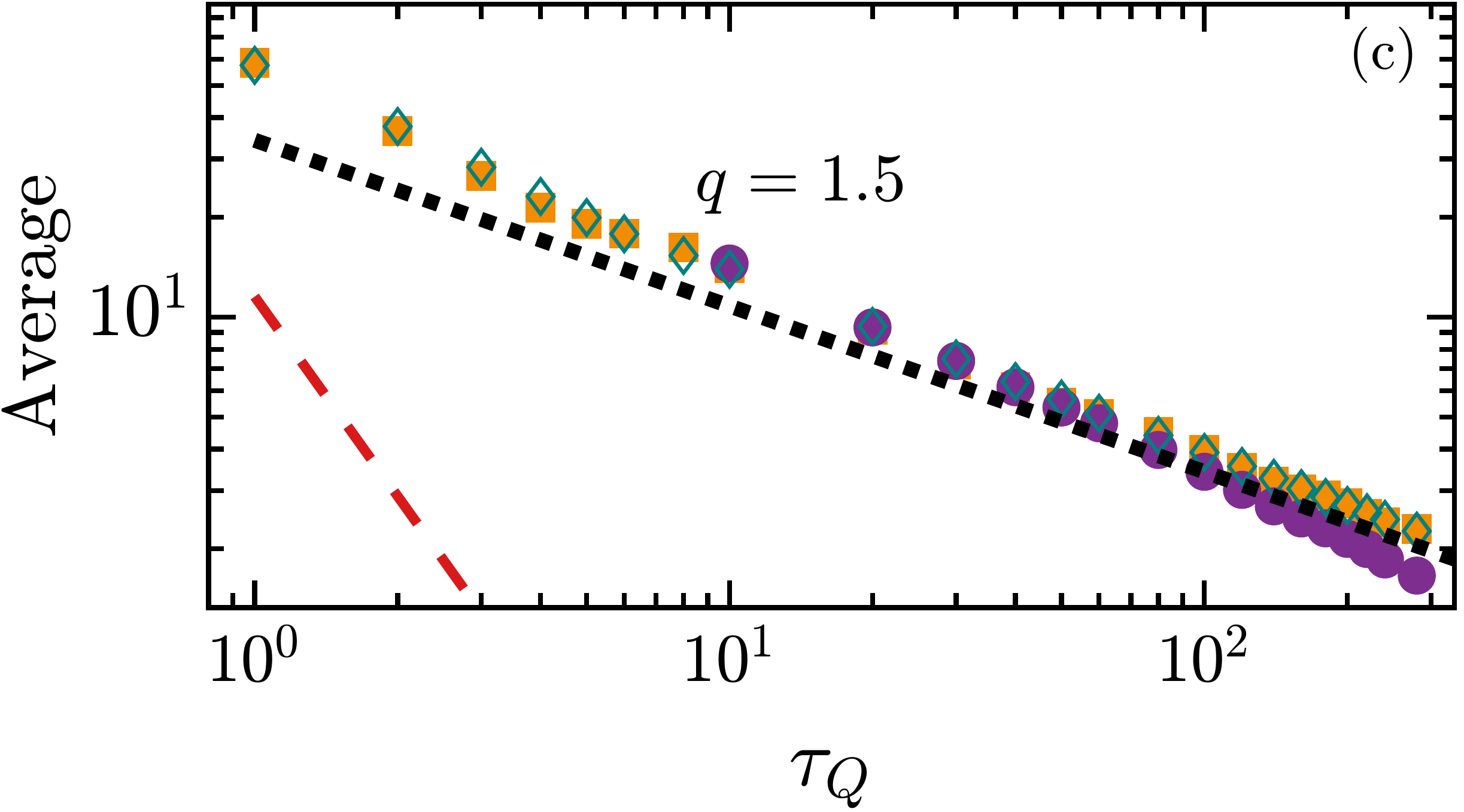} &
      \includegraphics[width=0.49\columnwidth,trim={2.75cm 0 0cm 0},clip]{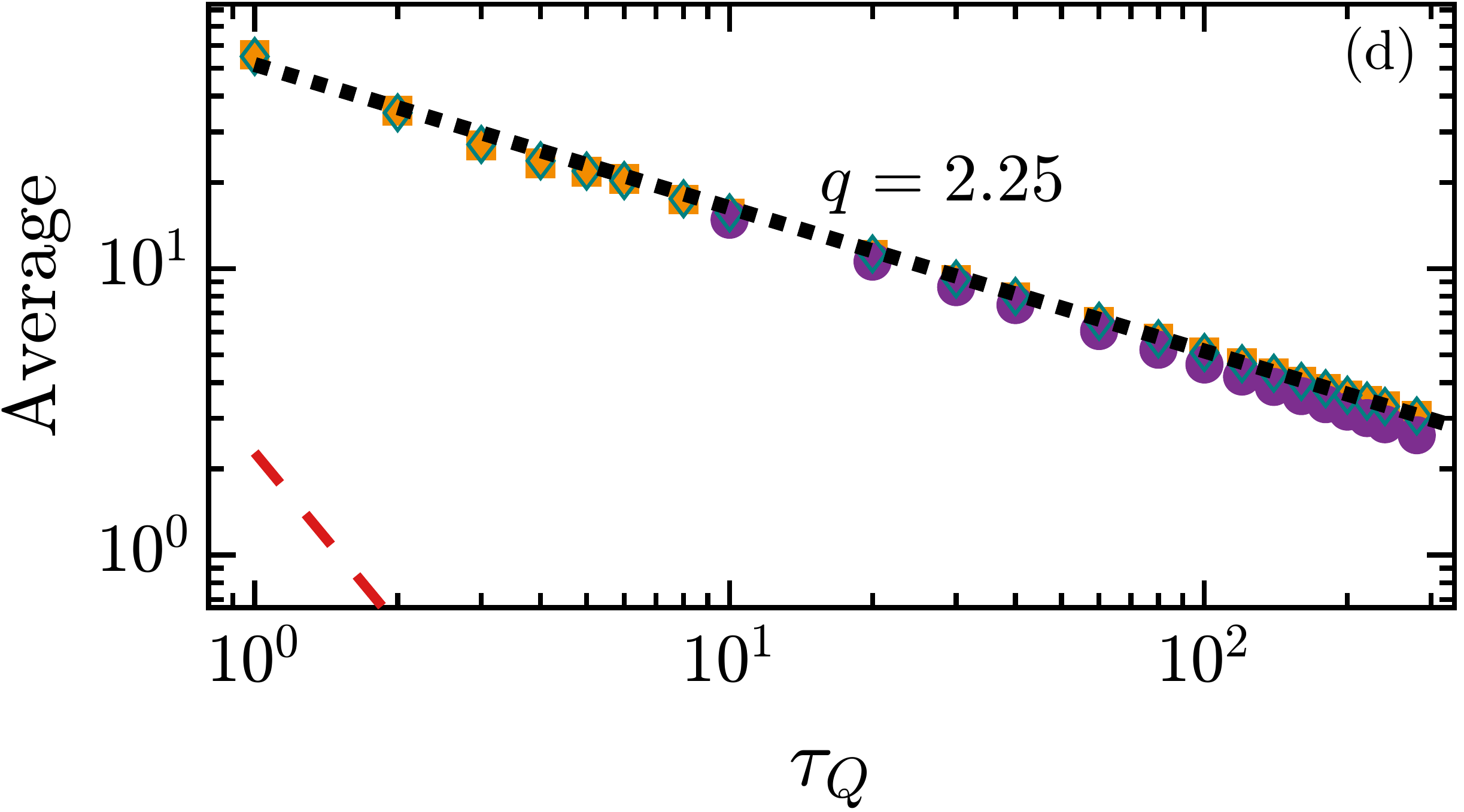} \\
    \end{tabular}
    \caption{Mean kink number, $\kappa_1$, in the long-range regime, (a) $q=0.1$ and (b) $q=0.2$, (c) the intermediate regime, $q=1.5$, and (d) the short-range regime, $q=2.25$. Purple circles show the TDVP results, orange squares the free-fermion kinks relative to the ground state at $g=0$ with open boundaries, and open diamonds the free-fermion excitations with periodic boundary conditions. The black dotted lines give the leading-order prediction $\propto\tau_Q^{-1/(2q_*)}$ of Eq.~\eqref{eq:kappa_scaling}, with $q_*=\min(q,1)$, and the red dashed lines the contribution $\propto\tau_Q^{-2}$ of the modes with $k>k_c$, Eq.~\eqref{eq:nex_nc}, which dominates slow ramps for $q<1/4$. The agreement between the TDVP and free-fermion results improves as $q$ approaches the short-range regime. For $q=0.1$ and $q=0.2$ the red dashed line captures the slow-ramp decay of the free-fermion data, while for $q=1.5$ and $q=2.25$ it lies orders of magnitude below both the numerical results and the leading-order power law.}
    \label{fig:kappa_1}
\end{figure*}

\begin{figure}[t]
    \centering
    \begin{tabular}{c}\includegraphics[width=0.99\columnwidth]{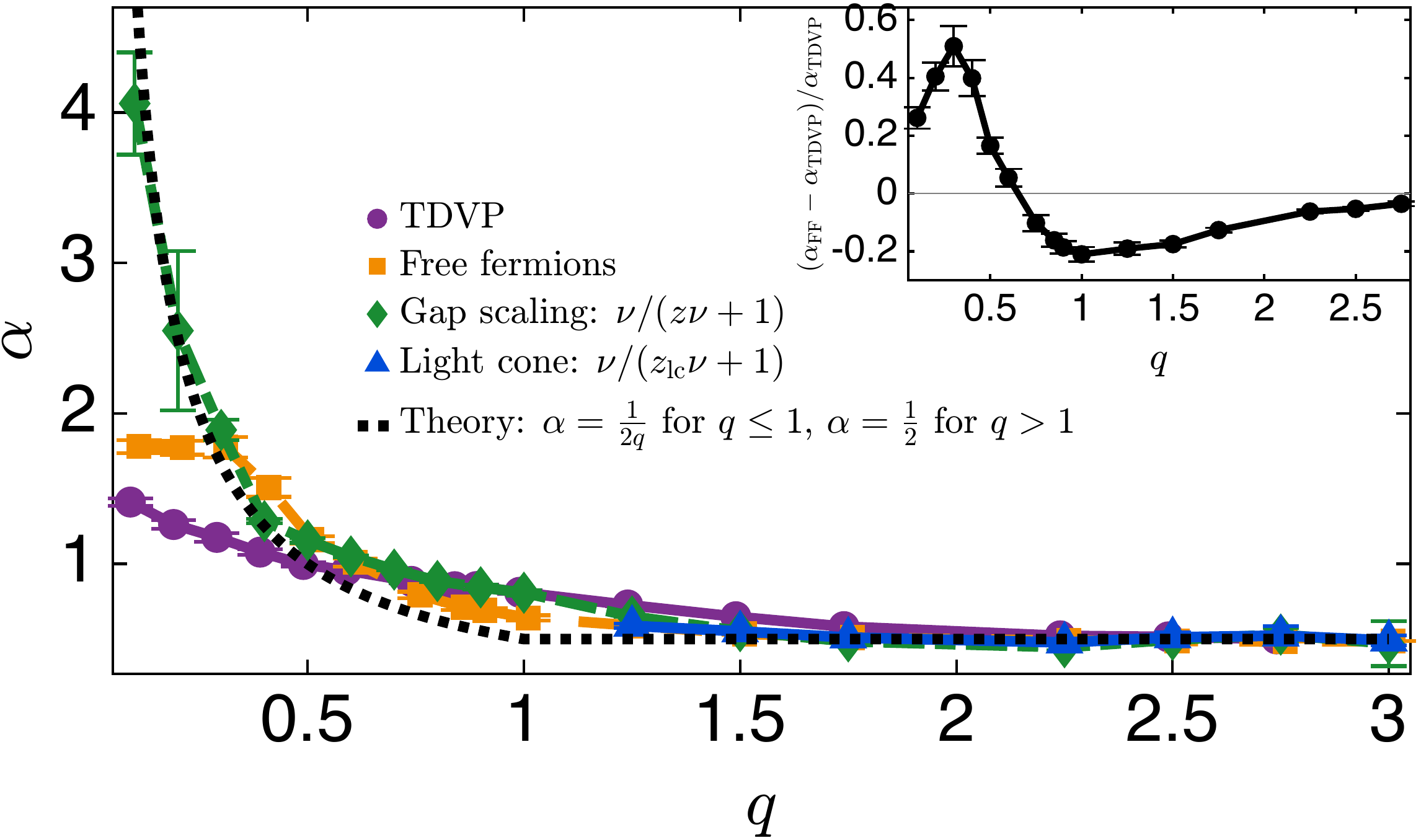}
    \end{tabular}
    \caption{Power-law exponents $\alpha$ of the mean kink number for the TDVP (purple), the numerically extracted KZ exponents from the gap analysis (green) and, for $q>1$, the light-cone (blue) results, and free-fermion numerics on a ring (orange). The free-fermion results deviate clearly for $q<1$, improve gradually for $q>1$, and agree with the TDVP simulations for $q>2$. The KZ exponents obtained from the gap and light-cone analyses match the TDVP results over a wider range of $q$, within the systematic uncertainty of about 10--20\% of the gap fits (Sec.~\ref{sec:gapscaling}).}
    \label{fig:alpha}
\end{figure}

\subsection{Kink number cumulants and full counting
statistics}\label{sec:cumulants_num}

To access universal signatures beyond the KZM, we now turn to
the full counting statistics of kinks \cite{delcampo2018,Cui2020,bando2020probing}. 
Specifically, we focus on the
probability distribution of finding $n$ kinks in the final
state $\ket{\Psi(\tau_Q)}$,
\begin{equation}
P(n)=\langle\Psi(\tau_Q)\lvert\delta_{\hat N,n}\rvert\Psi(\tau_Q)\rangle,
\end{equation}
with the kink-number operator
\begin{equation}\label{eq:Nhat}
    \hat N=\frac{1}{2}\sum_{i=1}^{L-1}
    \left(1-\hat\sigma^x_i\hat\sigma^x_{i+1}\right).
\end{equation}
A complementary characterization is provided by the Fourier transform of $P(n)$, $\tilde P(\theta)=\sum_n e^{i\theta n}P(n)$, and the corresponding cumulants, captured by the cumulant generating function,
\begin{align}
        \ln\tilde P(\theta)=\sum_{\ell=1}^{\infty}\kappa_{\ell}\frac{(i\theta)^{\ell}}{\ell!},
    \qquad\kappa_{\ell}=\partial^{\ell}_{i\theta}\ln\tilde P(\theta)\big|_{\theta=0},
\end{align}
evaluated with the final time-evolved MPS for $\ket{\Psi(\tau_Q)}$. Since the bond projectors in $\hat
N$ commute and are diagonal in the $\hat\sigma^x$ basis, $\tilde P(\theta)$
factorizes into a product of local two-site gates and is evaluated directly on
the MPS at each point of a discrete grid in $\theta$.  As the spectrum of
$\hat N$ is integer-valued and bounded, the inverse discrete Fourier transform
of $\tilde P(\theta)$ returns $P(n)$ without any truncation in the number of
moments.
Further details of the MPS and MPO
construction of the characteristic function are given in
App.~\ref{app:char_function}.

   Besides the kink number distribution, we investigate its first three cumulants, $\kappa_1=\ev{\hat N}$, $\kappa_2=\bigl\langle(\hat N-\kappa_1)^2\bigr\rangle$ and $\kappa_3=\bigl\langle(\hat N-\kappa_1)^3\bigr\rangle$ with $\ev{\cdot}=\bra{\Psi(\tau_Q)}\cdot\ket{\Psi(\tau_Q)}$, which are obtained from the moments $\langle\hat N^m\rangle$, $m=1,2,3$, by contracting powers of the MPO of $\hat N$ with the time-evolved MPS.

Figure~\ref{fig:kappa_1} shows the mean defect number in the long-range regime, for $q=0.1$ and $q=0.2$, the intermediate regime, $q=1.5$, and the short-range regime, $q=2.25$. The TDVP and free-fermion results (Sec.~\ref{sec:JW}) differ strongly for $q=0.2$, come surprisingly close to each other for $q=0.1$, approach each other as $q$ increases, and agree for $q>2$ up to finite-size effects.

Figure~\ref{fig:alpha} shows the exponents $\alpha$ from fits of the TDVP mean to $\tau_Q^{-\alpha}$ over the simulated driving times, together with the KZ exponents from the measured critical exponents (Sec.~\ref{sec:diagnostics}) and the free-fermion results (Sec.~\ref{sec:JW}). For $q\lesssim1$, the TDVP results show substantially smaller exponents, and thus a slower decay, than predicted by the KZ scaling with the measured critical exponents. The free-fermion numerics also do not reach their asymptotic decay in the simulated range of $\tau_Q$ for $q\lesssim0.3$, which is $\tau_Q^{-1/(2q)}$ for $q>1/4$ and $\tau_Q^{-2}$ for $q<1/4$ (Sec.~\ref{sec:defects}). For $1<q<2$, the TDVP results agree with the KZ exponents from the numerically extracted critical exponents to within 11--17\%, at the upper end of the 10--20\% systematic uncertainty of the gap fits, and approach the free-fermion results progressively. The free-fermion exponents of Fig.~\ref{fig:alpha} are obtained on a ring of $L=150$ sites with the couplings truncated at $r\leq L/2$. On the open chain, both the TDVP and free-fermion simulations show fewer kinks near the edges for $q\gtrsim1.5$ (App.~\ref{app:profiles}), which steepens the decay of the mean at $L=150$ by an amount that decreases with system size. In this geometry, free fermions give $\alpha\simeq0.59$ and $0.53$ at $q=1.5$ and $2.25$, against $0.65$ and $0.52$ for TDVP, so that the open boundary accounts for about half of the difference between the TDVP and ring exponents at $q=1.5$ and for all of it at $q=2.25$. For $q>2$, the TDVP and free-fermion exponents are close to $1/2$, while the gap estimate lies below it at $q=2.25$.

\begin{figure*}[t]
    \centering
    \begin{tabular}{c  c  c }\includegraphics[width=0.704\columnwidth,trim={0 0 0cm 0cm},clip]{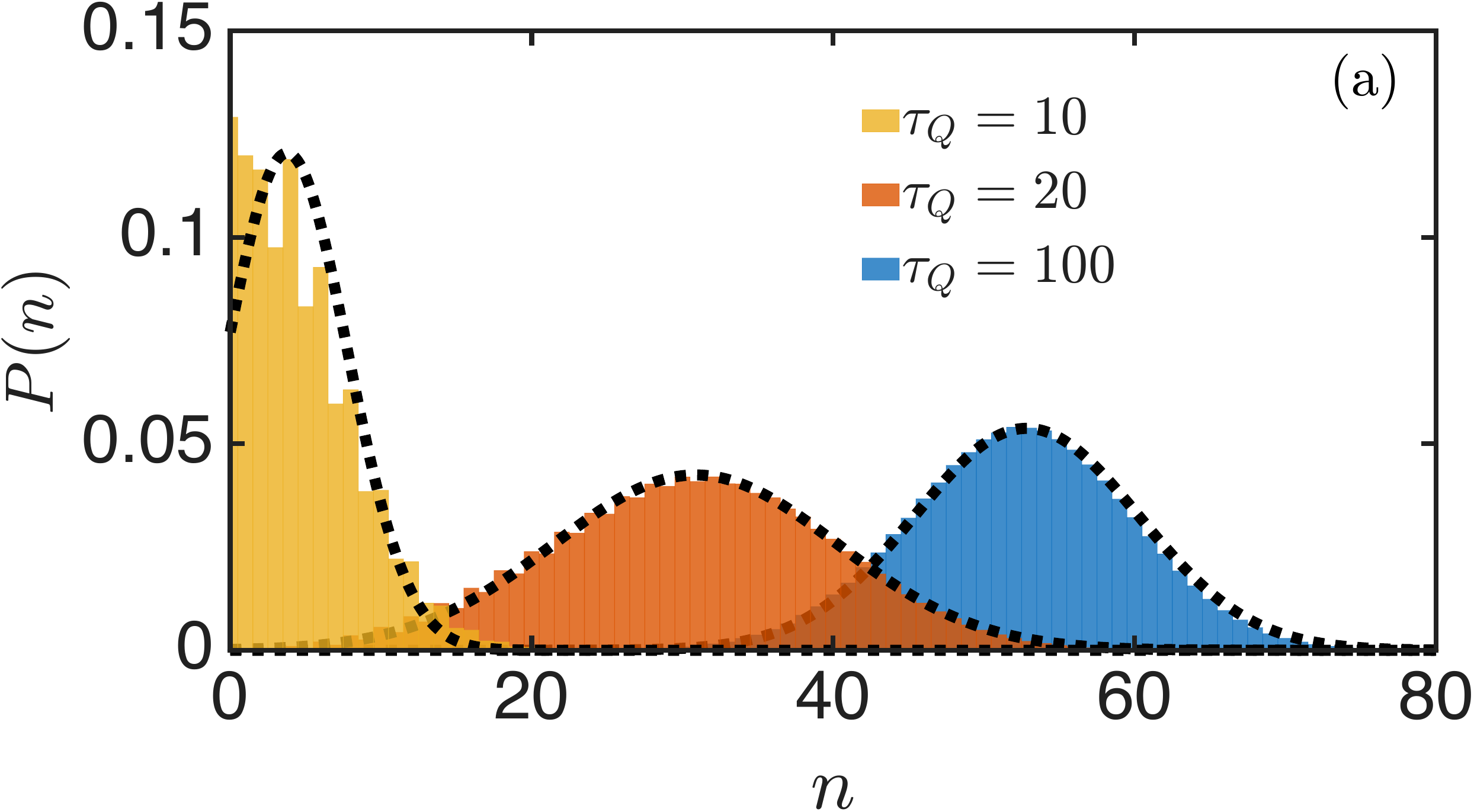} &         \includegraphics[width=0.648\columnwidth,trim={2.25cm 0 0cm 0},clip]{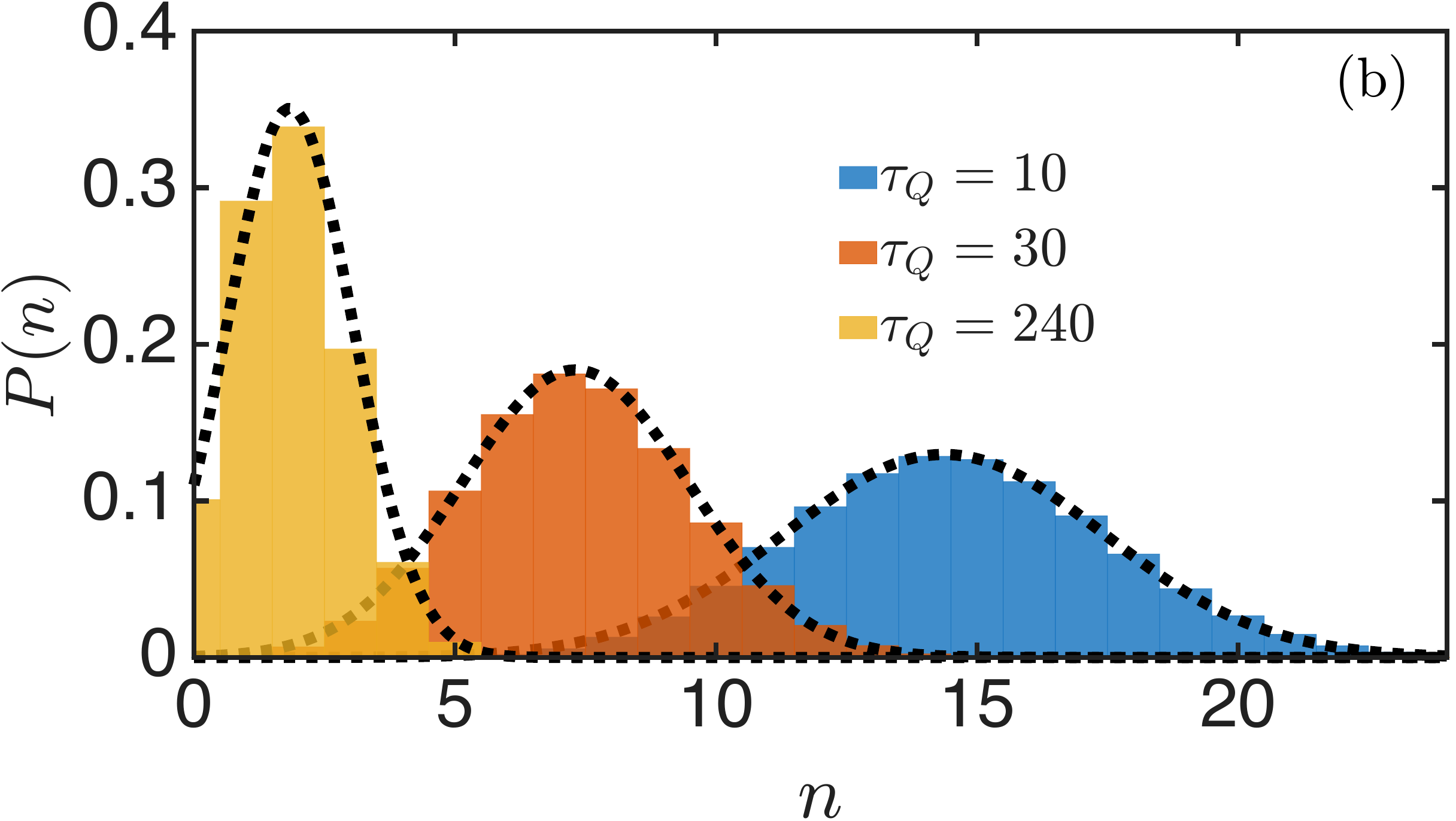} &
      \includegraphics[width=0.648\columnwidth,trim={2.25cm 0 0cm 0},clip]{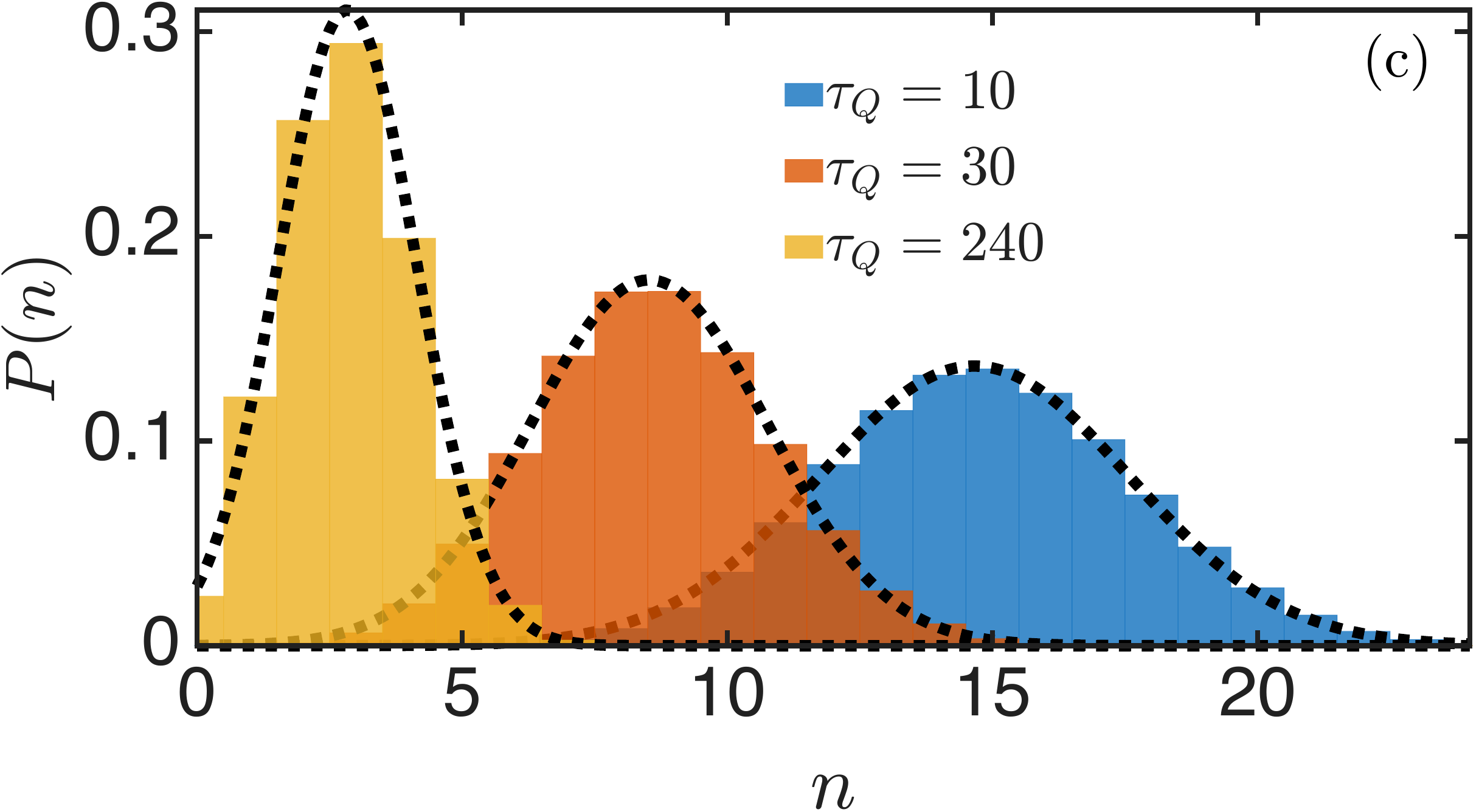} \\
    \end{tabular}
   \caption{Kink number histograms from the characteristic function for (a) $q=0.2$, (b) $q=1.5$, and (c) $q=2.25$, showing close agreement with a Gaussian distribution with the numerically obtained mean and variance. Deviations appear for slow driving, where the mean is small, most visibly at $q=0.2$.}
    \label{fig:Stat_qsmall}
\end{figure*}

\begin{figure*}[t]
    \centering
    \begin{tabular}{c  c  c c}\includegraphics[width=0.53\columnwidth,trim={0 0 0 0cm},clip]{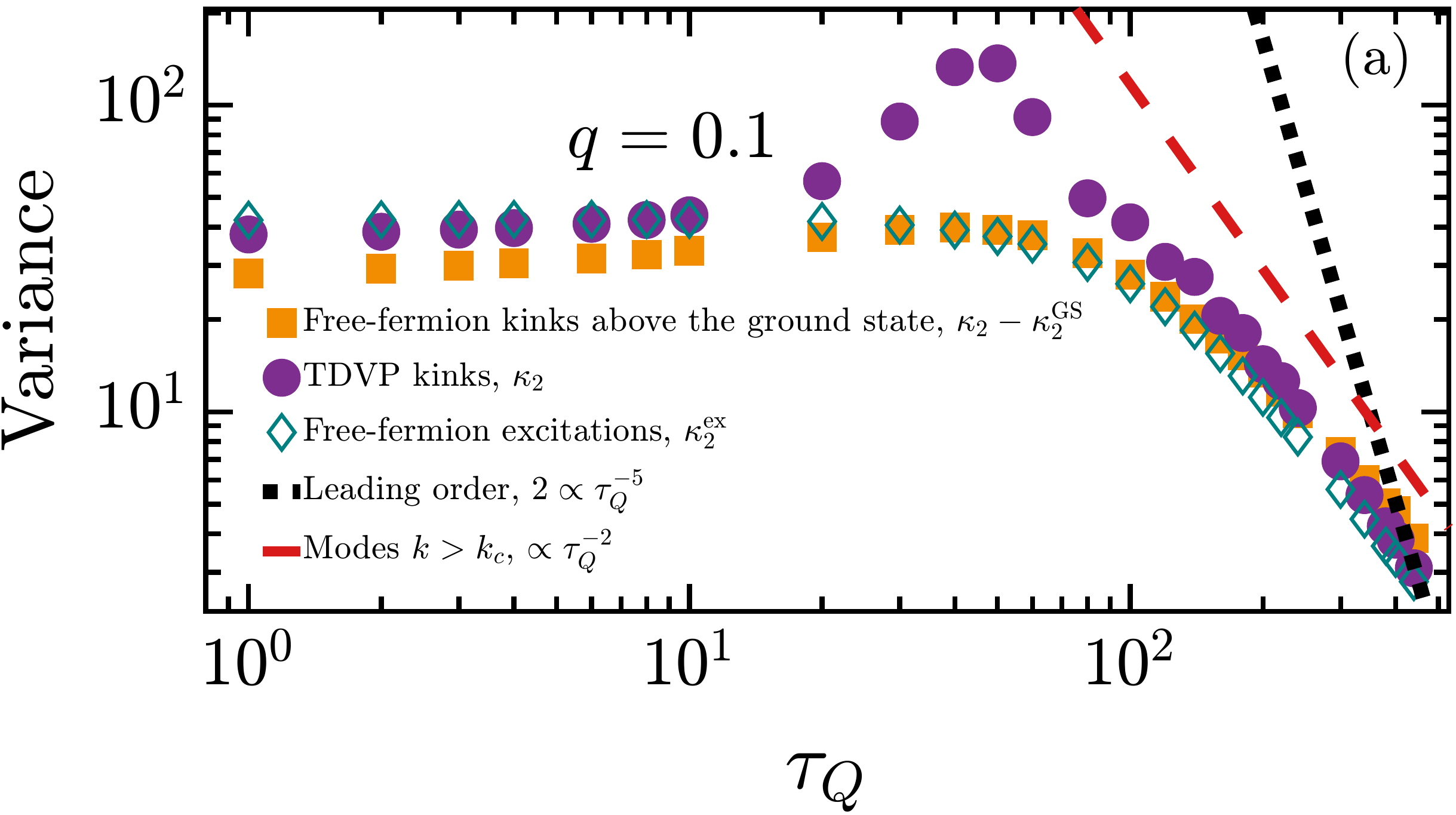} &         
    \includegraphics[width=0.49\columnwidth,trim={2.75cm 0 0 0},clip]{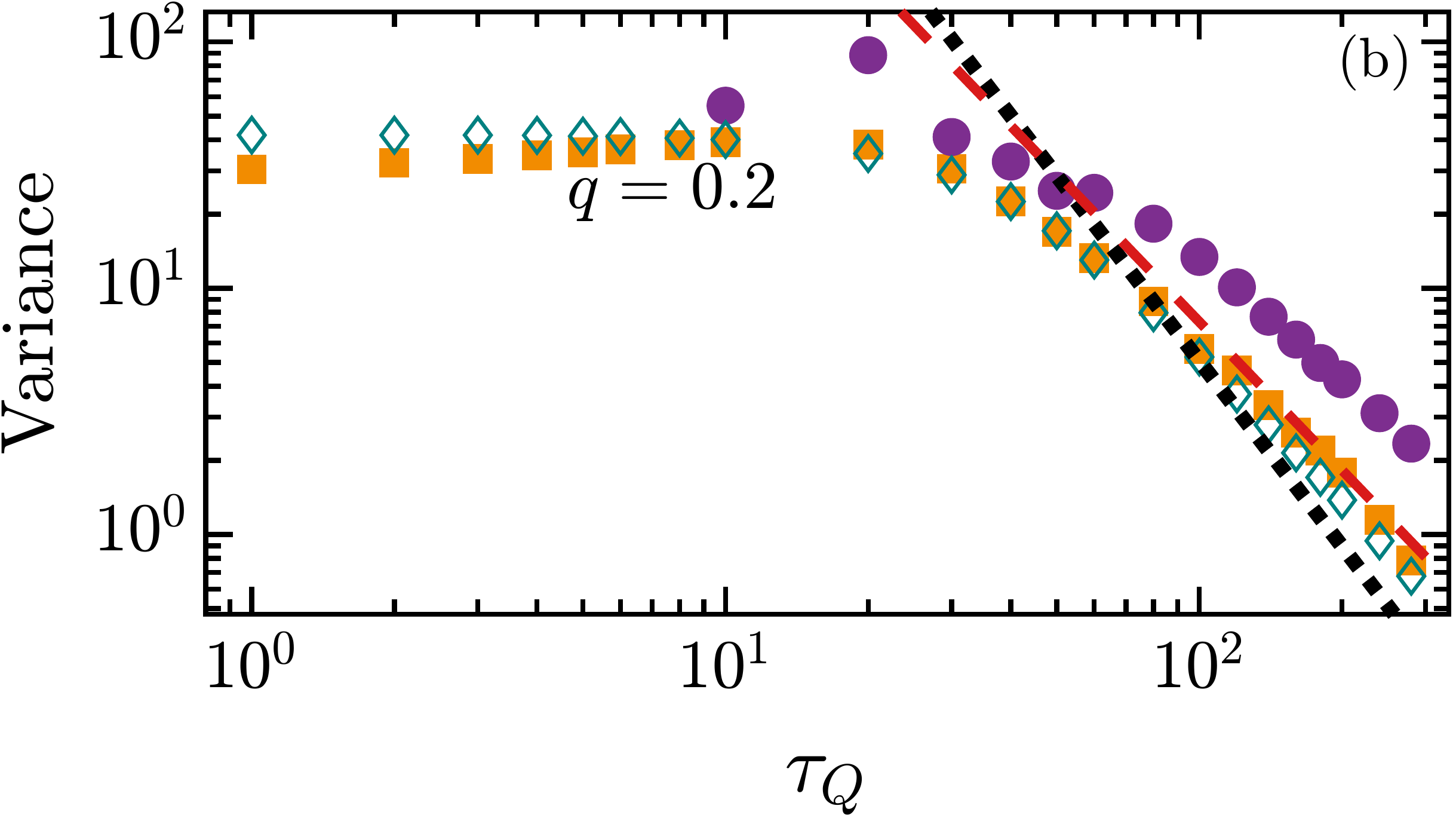} &   
    \includegraphics[width=0.49\columnwidth,trim={2.75cm 0 0cm 0},clip]{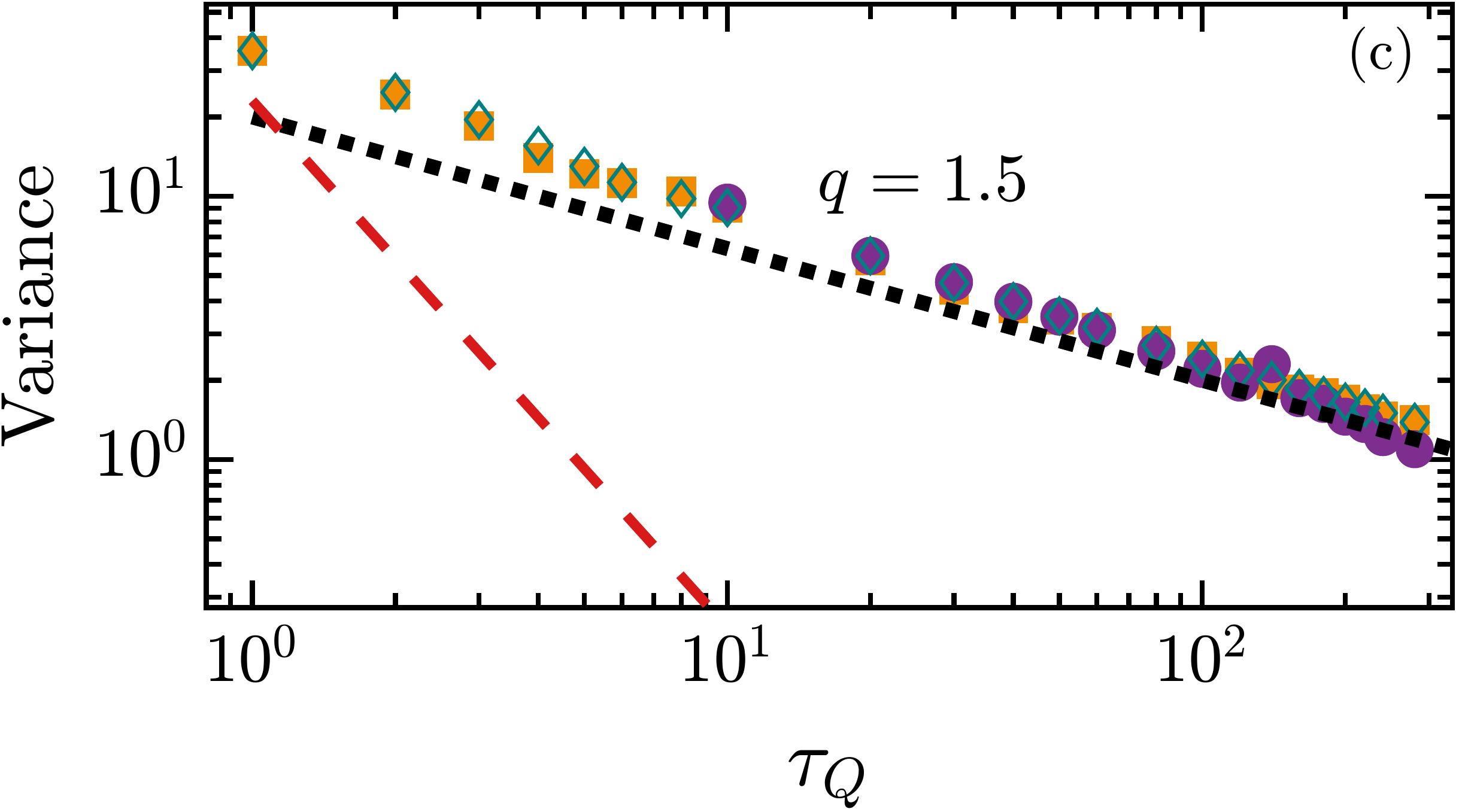} &
      \includegraphics[width=0.49\columnwidth,trim={2.75cm 0 0cm 0},clip]{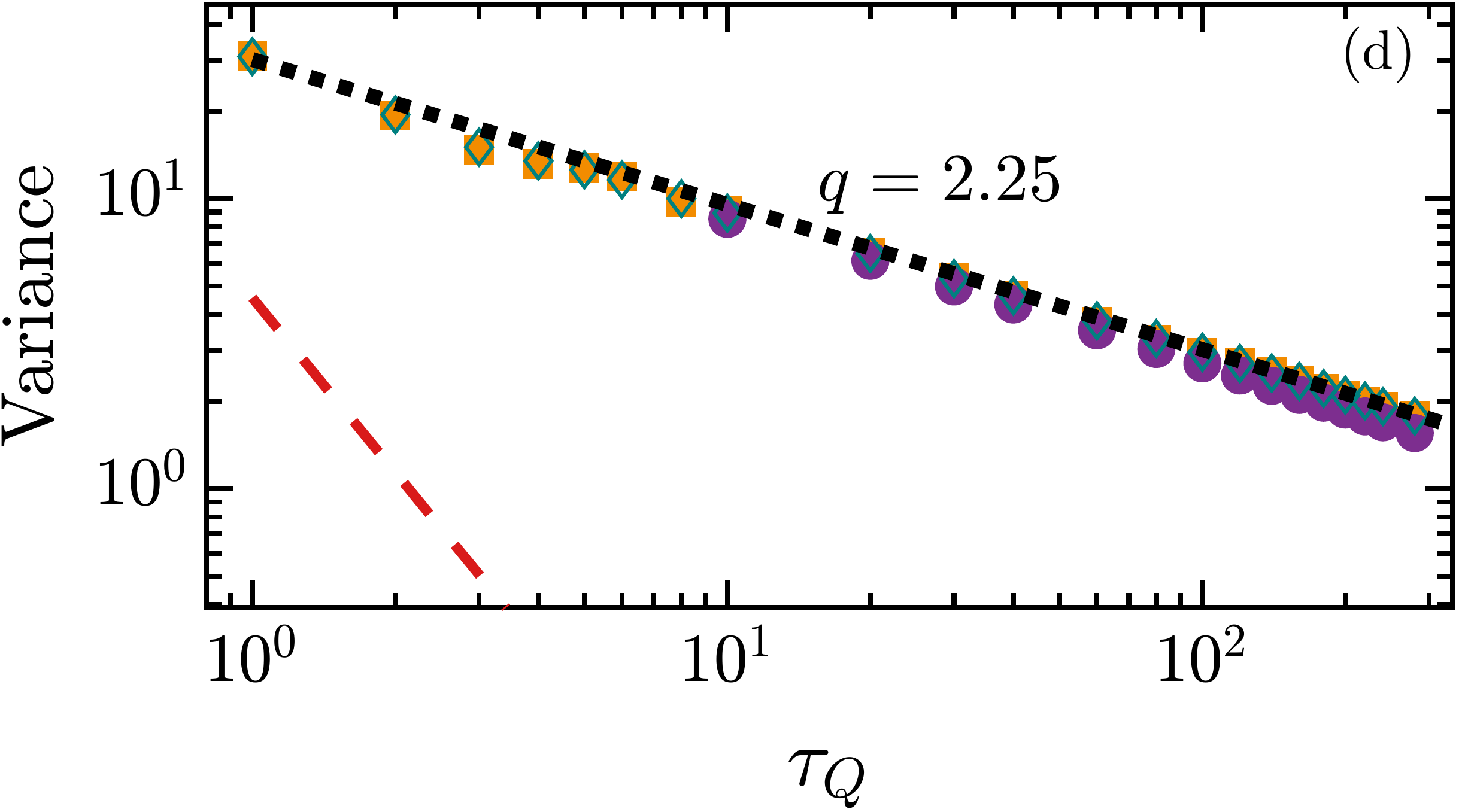} \\
    \end{tabular}
    \caption{Variance $\kappa_2$ of the kink number for (a) $q=0.1$, (b) $q=0.2$, (c) $q=1.5$, and (d) $q=2.25$, with the same conventions as in Fig.~\ref{fig:kappa_1}. For $q=0.1$ and $q=0.2$ the red dashed line agrees well with the power-law decay of the free-fermion data, while for $q=1.5$ and $q=2.25$ it lies far below the numerical results, in the lower left corner of the panels.}
    \label{fig:kappa_2}
\end{figure*}

The corresponding kink number histograms are shown in Fig.~\ref{fig:Stat_qsmall}, in the same
three regimes. When the mean is large, the kink number distribution approaches a Gaussian with mean $\kappa_1$ and variance $\kappa_2$.
    The variance (Fig.~\ref{fig:kappa_2}) and the third cumulant (Fig.~\ref{fig:kappa_3} in App.~\ref{app:cumulants}) approach the free-fermion results at slightly smaller $q$ than the mean (Sec.~\ref{sec:defects}). The higher cumulants thus remain proportional to the mean for $q\gtrsim1.25$, with ratios close to the free-fermion values (Fig.~\ref{fig:ratios} in App.~\ref{app:cumulants}). Deep in the long-range regime, however, the range of $\tau_Q$ values of this proportionality shrinks substantially, and the variance exceeds the free-fermion bound (App.~\ref{app:bound}), while kink-kink correlations also decay slightly slower, consistent with the domain-wall confinement described by Eq.~\eqref{eq:Vm}. At slow ramps, where the mean is small, the histograms depart from the Gaussian form.

\section{Gap scaling and entanglement light cone}\label{sec:diagnostics}
For a linear ramp through the critical point $g_c$, the correlation length $\xi\sim|g-g_c|^{-\nu}$ and the relaxation time $\tau_{\rm eq}\sim\xi^z\sim|g-g_c|^{-z\nu}$ diverge. Adiabaticity fails when $\tau_{\rm eq}$ becomes comparable with the time left before the transition, $\tau_Q|g-g_c|$, which fixes the freeze-out distance $|g-g_c|\sim\tau_Q^{-1/(1+z\nu)}$ and the frozen correlation length $\xi_{\rm KZ}\sim\tau_Q^{\nu/(1+z\nu)}$. The kink density in one dimension then scales as $\kappa_1/L\sim\xi_{\rm KZ}^{-1}\sim\tau_Q^{-\nu/(1+z\nu)}$~\cite{delcampo2014universality}.
In this section, we test whether this KZ scaling, with $\nu$ and $z$ extracted from the finite-size scaling of the gap and, for $z$, also from the entanglement light cone, reproduces the mean kink number of Sec.~\ref{sec:cumulants_num}.

\subsection{Finite-size scaling of the critical energy gap}\label{sec:gapscaling}
We extract both exponents from the universal scaling form of the critical gap,
\begin{equation}\label{eq:gap_scaling}
    \Delta(g,L)=L^{-z}\,\mathcal F\left[(g-g_c)L^{1/\nu}\right],
\end{equation}
with $\mathcal F(x)\sim x^{z\nu}$ for $x\gg1$ and $\mathcal F(x)\sim 1$ for $x\ll 1$ capturing both the low-energy and the close-to-criticality scaling forms, respectively. The leading order at criticality yields $\Delta(g_c,L)\sim L^{-z}$. However, for the accessible sizes we also need to include the next-to-leading power-law correction,
\begin{equation}\label{eq:gap_L}
    \Delta(g_c,L)=a\,L^{-z}\left(1+b\,L^{-\omega}\right),
\end{equation}
where the subleading term with $\omega>0$ absorbs the non-universal
finite-size corrections that would otherwise bias $z$ at finite $L$.
In the thermodynamic limit, the gap closes as    $\Delta(g,L\to\infty)\sim|g-g_c|^{z\nu}$.  To obtain $\nu$, we also study the finite-size critical point $g_c(L)$, defined by the minimum of the gap, which satisfies $d\Delta/dg=L^{1/\nu-z}\,\mathcal F'(x)=0$. As the prefactor never vanishes, $\mathcal F'(x)=0$ fixes $x=(g_c(L)-g_c)L^{1/\nu}$ at an $L$-independent value $a'$. Including the leading correction gives
\begin{equation}\label{eq:gc_L}
    g_c(L)=g_c+a'\,L^{-1/\nu}\left(1+b'\,L^{-\omega'}\right).
\end{equation}
This fixes $\nu$ independently of $z$ and
provides the finite-size critical point $g_c(L)$ used in the light-cone protocol of Sec.~\ref{sec:lightcone}.
Further details on the fits are provided in Fig.~\ref{fig:gapfits} and Tables~\ref{tab:gapfit} and~\ref{tab:gcfit} of App.~\ref{app:gapfits}.

\begin{figure*}[t]
    \centering
    \begin{tabular}{c  c  c c}\includegraphics[width=0.515\columnwidth,trim={0 0 0 0cm},clip]{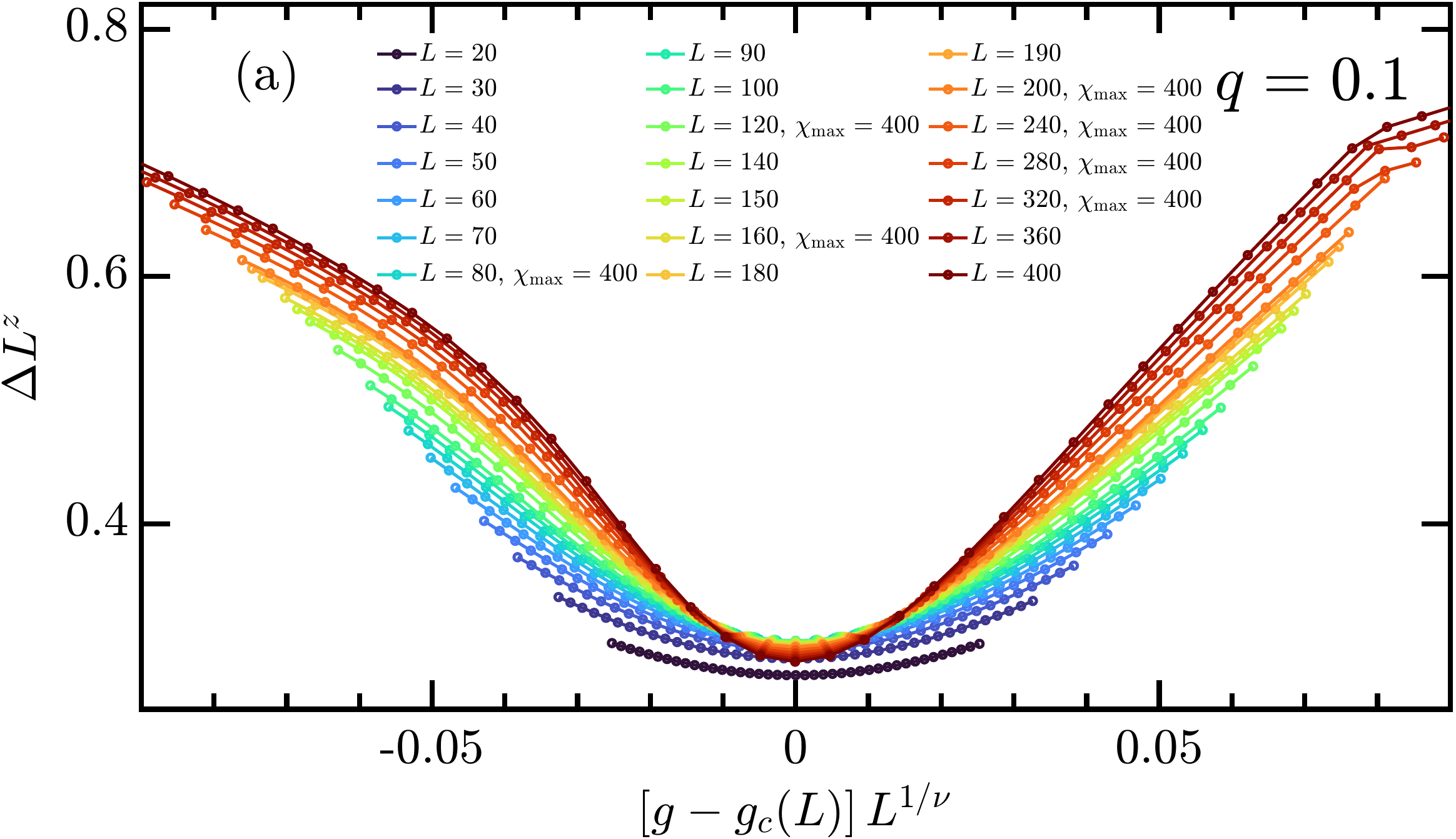} &         
    \includegraphics[width=0.493\columnwidth,trim={2.75cm 0 0 0},clip]{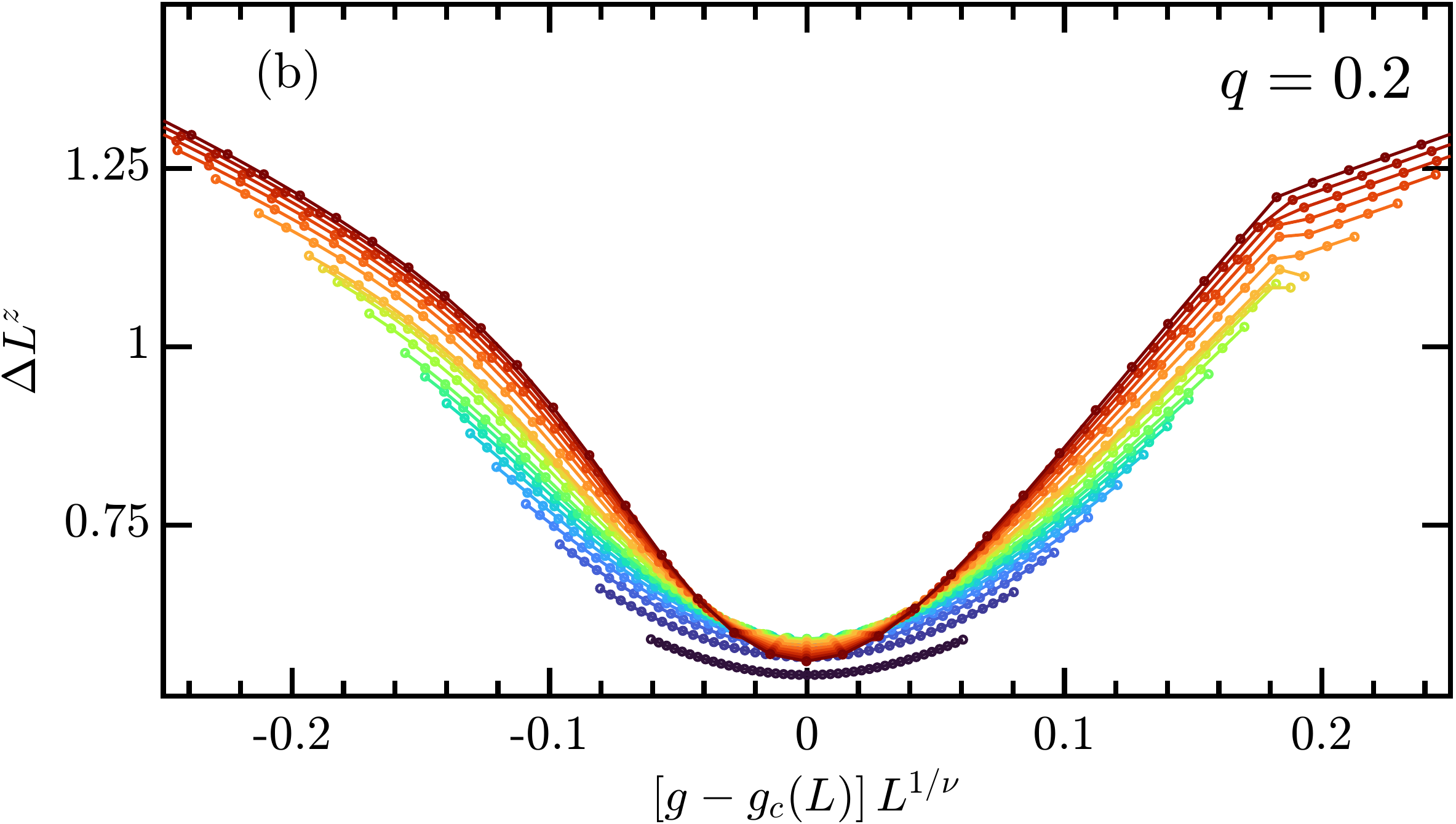} &   
    \includegraphics[width=0.493\columnwidth,trim={1.75cm 0 0cm 0},clip]{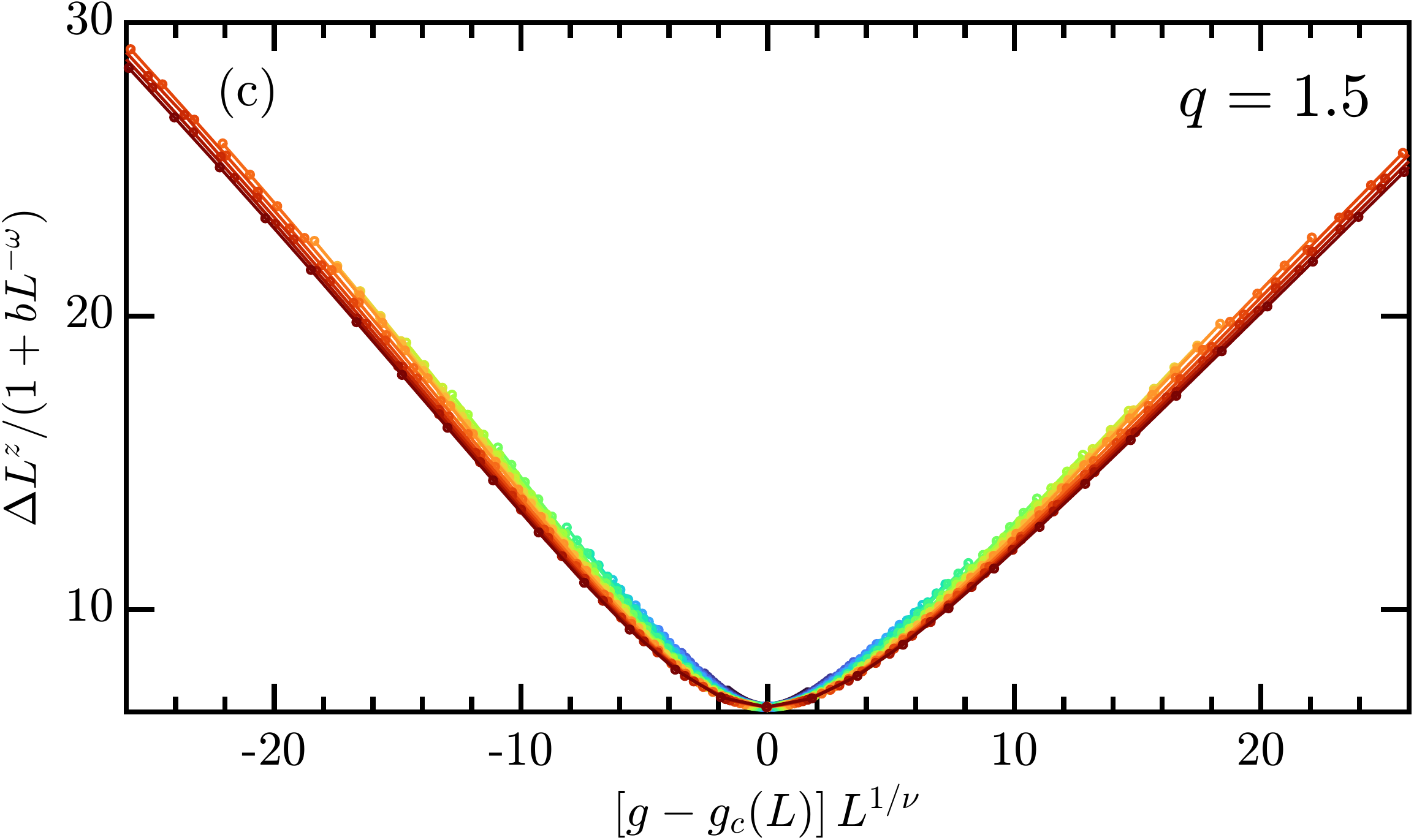} &
      \includegraphics[width=0.493\columnwidth,trim={1.75cm 0 0cm 0},clip]{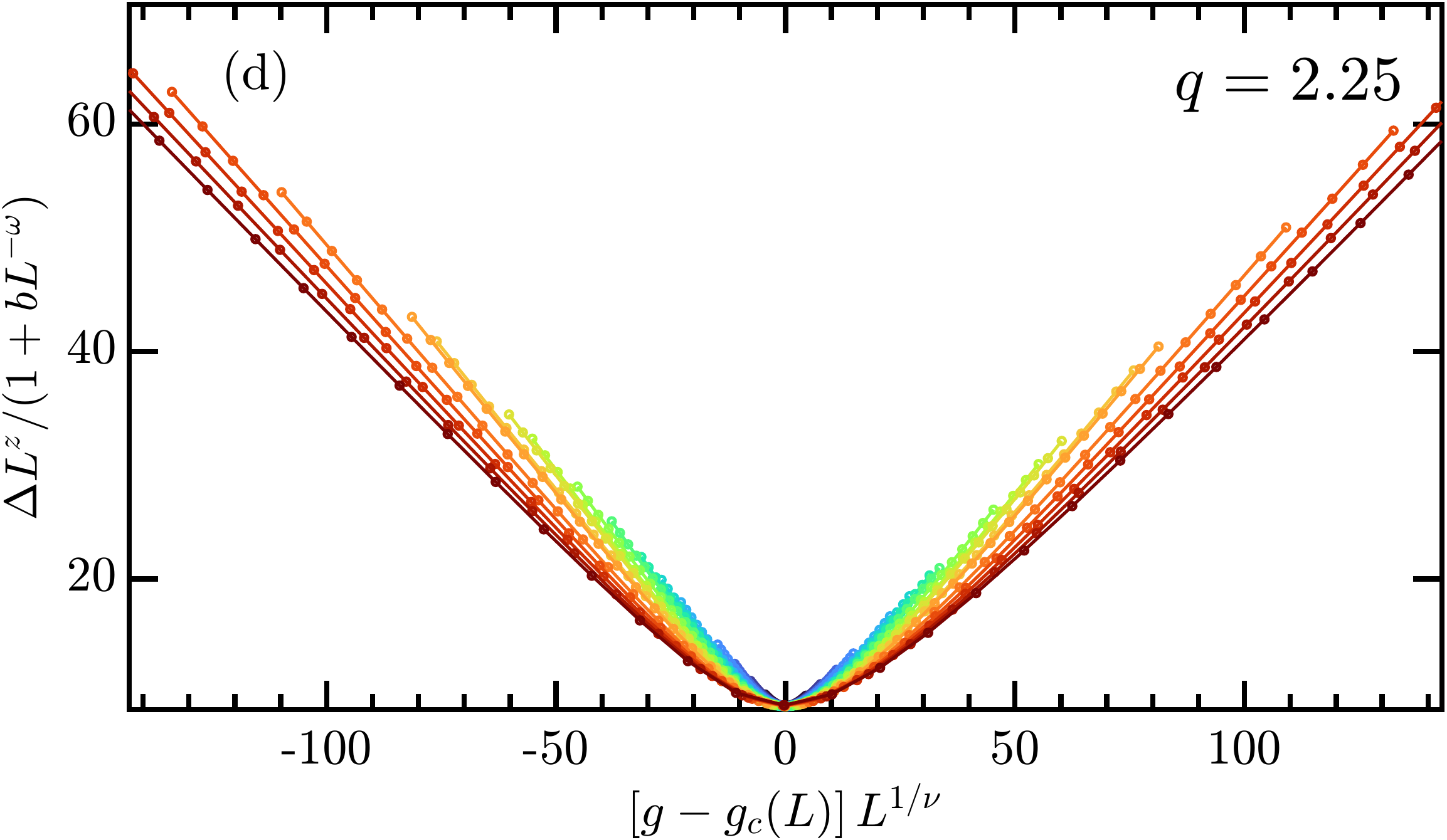} \\
    \end{tabular}
  \caption{Scaling collapse of the low-energy gap for
    (a) $q=0.1$, (b) $q=0.2$, (c) $q=1.5$ and (d) $q=2.25$.
    For each $q$ the data of all system sizes approximately collapse near the gap minimum when
    $\Delta L^{z}/(1+bL^{-\omega})$ is plotted against
    $[g-g_c(L)]L^{1/\nu}$, with $z$ and $b$ from Eq.~\eqref{eq:gap_L} and $\nu$, $g_c$ and $b'$ from Eq.~\eqref{eq:gc_L}. Away from the minimum the curves of small sizes separate, most visibly at small $q$, while the spacing between successive sizes shrinks as $L$ grows.}
  \label{fig:gap_collapse}
\end{figure*}

 In the numerical realization, we used DMRG simulations~\cite{Schollwock2011,ITensor} to search
for the minimum gap in a sufficiently large window around the free-fermion critical point $g_c^{\rm FF}=J_q/(2+J_q),\,J_q=\binom{q}{q/2}$. In the DMRG simulations we use the Hamiltonian of Eq.~\eqref{eq:H_Ising} divided by $1-g$,
\begin{equation}\label{eq:H_DMRG}
    \hat H_{\rm DMRG}=-h\sum_{i=1}^L\hat\sigma_i^z-\sum_{i=1}^{L-1}\sum_{r=1}^{L-i}J_{r,q}\,\hat\sigma_i^x\hat\sigma_{i+r}^x,\quad h=\frac{g}{1-g},
\end{equation}
with a single field parameter $h$ and the MPO of Sec.~\ref{sec:Ising_model}. Here $\Delta=E_2-E_0$ is the gap to the second state of the odd parity sector. The first excited state, which is the lowest odd state, becomes degenerate with the ground state in the ordered phase and has no minimum in $g$. We obtain the excited states by DMRG with penalty terms for the lower states, for system sizes up to $L=200$. We use $12$ sweeps, a bond dimension increased up to $\chi_{\rm max}=200$, a truncation cutoff of $10^{-10}$ and a penalty weight $\max(20,0.1|E_0|)$, and accept a field value only when the ground-state energy variance stays below $10^{-6}$ and the mutual overlaps of the states below $10^{-3}$. As an additional check of convergence, we repeated the calculations with $\chi_{\rm max}=400$, which changes the gap by less than $1\%$, even at $q=0.1$.
 At $q=1.5$ and $L=20$, exact diagonalization gives the minimum $\Delta=0.282$ at $g=0.450$ in the units of Eq.~\eqref{eq:H_Ising}, in agreement with our DMRG value $0.283$.  For each $L$, the minimum of the gap is located by a golden-section search~\cite{Kiefer53,NumericalRecipes} to a relative accuracy of $2\times10^{-4}$ in $h$. The minima of all sizes are then fitted according to Eqs.~\eqref{eq:gap_L} and~\eqref{eq:gc_L} by least squares~\cite{NumericalRecipes}, with next-to-leading order correction terms kept only where the data resolve them. The exponents are listed in Table~\ref{tab:exponents}. As the free-fermion model in Sec.~\ref{sec:pwave} predicts $z\nu=1$ for all $q$, the product of the two independently fitted exponents measures the deviations from this approximation. For $0.4\leq q\leq2.25$ we find $z\nu\simeq0.6$--$0.85$, and $z\nu\simeq0.7$--$1.15$ for $q\geq2.5$, while for $q\leq0.3$ $z$ agrees within errors with the Gaussian value $z=q/2$ of Ising chains with power-law couplings below $q=2/3$~\cite{DuttaBhattacharjee01,DefenuTrombettoniRuffo17,Koziol21}, and $z\nu\simeq0.50$--$0.60$ with the Gaussian $z\nu=1/2$, while $\nu$ lies below $1/q$, so that the KZ exponents fall below the Gaussian value $2/(3q)$. At these orders the gap closes only slowly over the accessible sizes, so that $z$ depends on the correction term of Eq.~\eqref{eq:gap_L}. At $q=1.5$ the fitted $z\simeq0.82$ is close to the functional renormalization-group and Monte Carlo value $z\approx0.83$ for Ising chains with power-law couplings of the same decay~\cite{Defenu23}, and lies between the earlier estimate $z\approx0.75$~\cite{LewisGongCarr25LevyLightCones} and the free-fermion value $z=1$, and the local exponent from exact diagonalization at $L=12$--$20$ drifts upward with $L$, which indicates a slow approach to the asymptotic value probed by the light cone of Sec.~\ref{sec:lightcone}.
The scaling collapse obtained with the resulting $\nu$ and $z$ is shown in
Fig.~\ref{fig:gap_collapse}. There, the gap is rescaled at next-to-leading
order, dividing out the subleading factor of Eq.~\eqref{eq:gap_L} and
subtracting the critical point $g_c(L)$ of Eq.~\eqref{eq:gc_L} at the given finite system size $L$, rather than the extrapolated
$g_c$ alone.
The resulting KZ exponents $\nu/(z\nu+1)$ are plotted in Fig.~\ref{fig:alpha}. Tested on the exactly solvable free-fermion chain of Sec.~\ref{sec:JW}, these fits have a systematic uncertainty of 10--20\%. These KZ exponents deviate appreciably from the TDVP results for $q\lesssim1$, while for $q>1$ they lie at the upper end of this uncertainty, as also discussed in Sec.~\ref{sec:cumulants_num}.

\begin{table}[t]
\caption{Critical exponents $z$ and $\nu$ from the finite-size scaling of the gap, Eqs.~\eqref{eq:gap_L} and~\eqref{eq:gc_L}, the extrapolated critical point $g_c$ in the convention of Eq.~\eqref{eq:H_Ising}, and the resulting KZ exponent $\nu/(z\nu+1)$. The fitted correction terms and the system sizes of each fractional order are given in App.~\ref{app:gapfits}.}
\label{tab:exponents}
\begin{ruledtabular}
\begin{tabular}{ccccc}
$q$ & $z$ & $\nu$ & $g_c$ & $\nu/(z\nu+1)$ \\
\hline
$0.1$ & $0.101(24)$ & $6.87(21)$ & $0.4604$ & $4.06(34)$ \\
$0.2$ & $0.141(25)$ & $4.0(16)$ & $0.4887$ & $2.55(53)$ \\
$0.3$ & $0.177(22)$ & $2.843(27)$ & $0.4948$ & $1.890(68)$ \\
$0.4$ & $0.3392(65)$ & $2.240(11)$ & $0.4955$ & $1.273(14)$ \\
$0.5$ & $0.3458(48)$ & $1.895(27)$ & $0.4942$ & $1.145(16)$ \\
$0.6$ & $0.3618(34)$ & $1.6754(25)$ & $0.4924$ & $1.0431(47)$ \\
$0.7$ & $0.3867(22)$ & $1.5244(40)$ & $0.4903$ & $0.9591(36)$ \\
$0.8$ & $0.4237(19)$ & $1.415(16)$ & $0.4884$ & $0.8845(48)$ \\
$0.9$ & $0.471(18)$ & $1.402(25)$ & $0.4878$ & $0.844(16)$ \\
$1$ & $0.517(23)$ & $1.389(20)$ & $0.4874$ & $0.808(18)$ \\
$1.25$ & $0.6624(27)$ & $1.1424(71)$ & $0.4833$ & $0.6503(23)$ \\
$1.5$ & $0.8157(8)$ & $1.000(20)$ & $0.4844$ & $0.5508(59)$ \\
$1.75$ & $0.9395(27)$ & $0.893(25)$ & $0.4897$ & $0.4856(65)$ \\
$2.25$ & $1.0179(4)$ & $0.787(75)$ & $0.5142$ & $0.437(22)$ \\
$2.5$ & $1.034(12)$ & $1.02(9)$ & $0.5320$ & $0.496(22)$ \\
$2.75$ & $1.009(10)$ & $1.14(21)$ & $0.5562$ & $0.530(45)$ \\
$3$ & $0.9300(51)$ & $0.835(34)$ & $0.5782$ & $0.470(15)$ \\
\end{tabular}
\end{ruledtabular}
\end{table}

\subsection{Entanglement light cone}\label{sec:lightcone}
As a complementary route to the dynamical exponent, $z$, we study the entanglement entropy spreading generated by a local perturbation at criticality. In short-range systems, the propagation is bounded by a strict Lieb-Robinson velocity~\cite{LiebRobinson72}, whereas algebraically decaying couplings admit nonlinear and, for sufficiently slow decay, unbounded light cones~\cite{Hastings2006,FossFeig2015,HaukeTagliacozzo13}, as observed directly in trapped-ion crystals~\cite{Richerme2014,Jurcevic2014}.  We
prepare the ground state of $\hat H_\mathrm{Ising}$ at the critical value $g_c(L)$, given by the position of the gap minimum of Sec.~\ref{sec:gapscaling}, at finite size $L=150$, using the same DMRG numerics as in
Sec.~\ref{sec:gapscaling}, and evolve it with TDVP~\cite{Haegeman2016} under the locally perturbed
Hamiltonian
\begin{equation}\label{eq:pulse}
    \hat H(t)=\hat H_\mathrm{Ising}(g_c(L))
    +h_0\,W_{\rm BH}(t)\,\hat\sigma^z_{j^*},
\end{equation}
with $j^*\equiv\lfloor L/2\rfloor$ the central site.  From a quasiparticle perspective, this local
injection of energy at criticality launches low-energy excitations whose
velocity is governed by the dynamical exponent $z$.  The amplitude $h_0=2$
is modulated by a Blackman-Harris window of duration $T_w=1.2$,
\begin{equation}\label{eq:BH}
    W_{\rm BH}(t)=a_0-a_1\cos\frac{2\pi t}{T_w}
    +a_2\cos\frac{4\pi t}{T_w}-a_3\cos\frac{6\pi t}{T_w},
\end{equation}
for $0\leq t\leq T_w$ and otherwise $W_{\rm BH}(t)=0$, with the 
standard coefficients
$(a_0,a_1,a_2,a_3)=(0.35875,0.48829,0.14128,0.01168)$~\cite{Harris78}.  The window suppresses the high-frequency content of the pulse, so that mainly low-energy quasiparticles are excited. The time evolution is carried out with two-site TDVP up to $t_{\max}=30$, with a time step $\delta t=0.02$, $\chi_{\max}=200$ and a truncation cutoff of $10^{-10}$, and the bond entropy is recorded at $100$ equally spaced time instances.

\begin{figure}[t]
    \centering
    \begin{tabular}{c}\includegraphics[width=0.99\columnwidth]{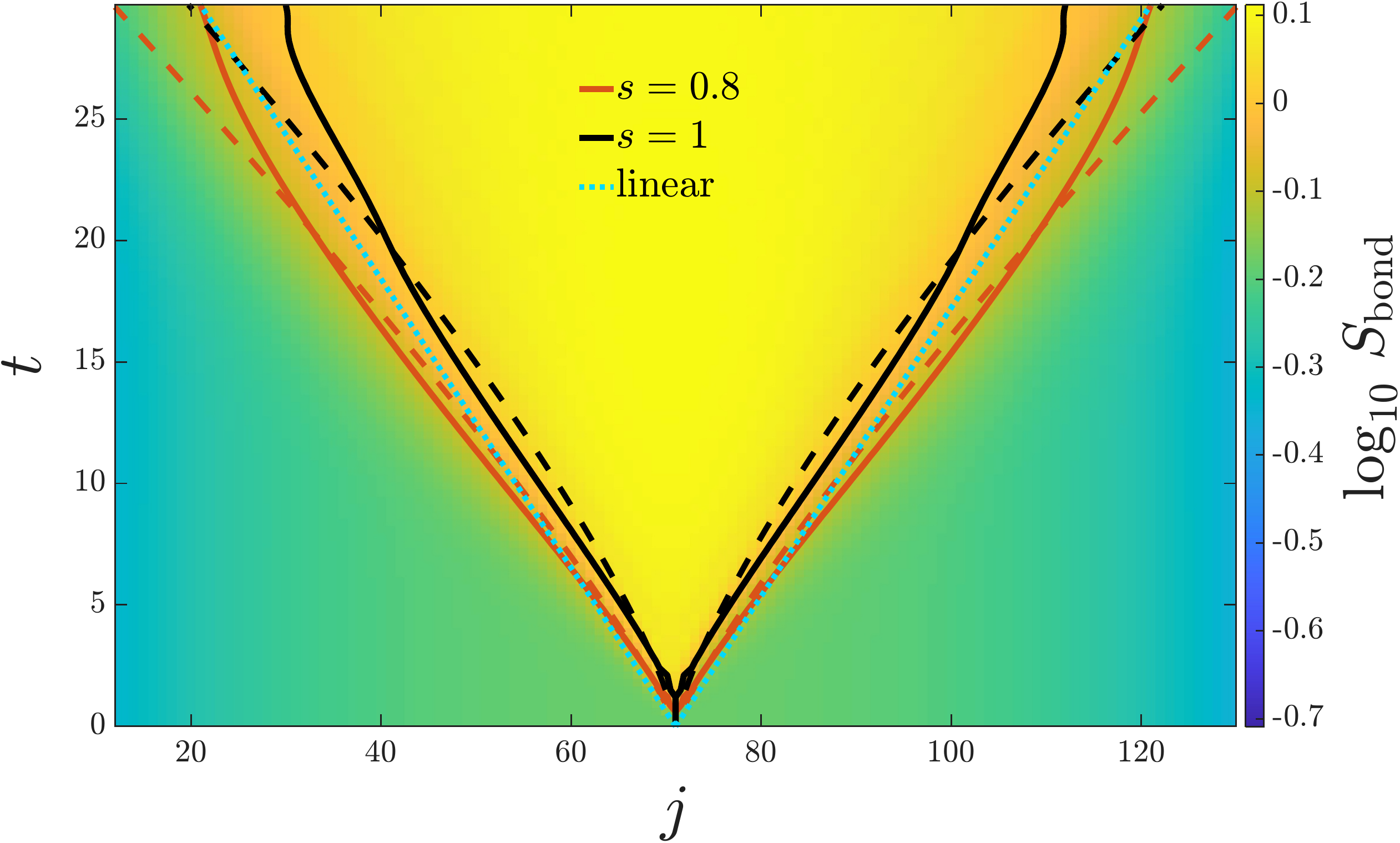}
    \end{tabular}
    \caption{Entanglement light cone at criticality for fractional order
    $q=1.5$, with the local pulse applied at the central site of a chain of
    $L=150$ spins.  The color scale gives the base-ten logarithm of the bond
    entropy $S_{\rm bond}(j,t)$: blue marks the bonds the disturbance has not
    reached, where the entropy retains its critical ground-state value, yellow
    the interior of the cone where it has saturated, and the sharp change
    between them defines the front.  Solid lines are the measured fronts,
    given by the contours $S_{\rm bond}(j,t)=s$, shown for $s=0.8$ (orange) and $s=1$ (black); the dashed lines
    of the same color are the corresponding fits to $|j-j^*|\propto t^{1/z}$ over the whole simulated time window.  The
    cyan dotted line is a strictly ballistic reference, $|j-j^*|\propto t$,
    anchored to the outermost point reached by the front.}
    \label{fig:lightcone}
\end{figure}

The propagation is tracked through the bond entanglement entropy across the cut
between sites $j$ and $j+1$,
\begin{equation}\label{eq:Sbond}
    S_{\rm bond}(j,t)=-\Tr\left[\hat\rho_{\leq j}(t)\ln\hat\rho_{\leq j}(t)\right],
\end{equation}
with $\hat\rho_{\leq j}(t)$ the reduced density matrix of the sites $1,\dots,j$.
Unlike other local perturbation measures, such as magnetization or
higher-order correlation functions, the bond entropy provides a monotonic
signal of quasiparticle passage: as each quasiparticle crosses a bond, it
entangles the two halves of the chain and raises $S_{\rm bond}(j,t)$ above its initial value. Inside the cone, the entropy then saturates into a plateau, while outside it retains its initial value.
The front is defined by the contour of constant bond entropy, $S_{\rm bond}(j,t)=s$~\cite{LewisGongCarr25LevyLightCones}. At each time, it lies at the first bond on either side of $j^*$ at which $S_{\rm bond}$ drops below the threshold $s$, and the two fronts form a cone around the perturbed site.     We repeated the analysis for thresholds from $s=0.75$ to $s=1$ with uniform steps $\Delta s=0.025$, all above the bond entropy of the critical ground state, $S_{\rm bond}\lesssim0.68$, and within the range of the bond entropy reached inside the cone, $S_{\rm bond}\lesssim1.3$ in Fig.~\ref{fig:lightcone}, and fitted the fronts over the whole simulated time window, $t\leq t_{\max}$, once they have left the central bond. In the leading order, the resulting boundaries of this L\'evy light cone obey the universal power-law behavior
\begin{equation}\label{eq:front}
    |j-j^*|\sim t^{1/z}.
\end{equation}
Larger $z$ corresponds to slower quasiparticles and a narrower cone~\cite{LewisGongCarr25LevyLightCones}.
At $q=1.5$, averaging over the thresholds gives $z=0.82(4)$, as demonstrated in Fig.~\ref{fig:lightcone} for $s=0.8$ and $s=1$, in agreement with the gap value $z\simeq0.82$ of Sec.~\ref{sec:gapscaling}.  For comparison, Fig.~\ref{fig:lightcone} also shows a strictly ballistic cone, $|j-j^*|=v\,t$, which is not fitted but passes through the perturbed site at $t=0$ and through the front point farthest from $j^*$ within the fit window, $v=|j-j^*|_{\max}/t_\mathrm{max}$. The measured fronts deviate only weakly from it at $q=1.5$. For $q<1$ the couplings decay more slowly than $r^{-2}$, and the group velocity of the quasiparticles diverges at small momenta. No light cone is then defined, and the front does not determine $z$~\cite{HaukeTagliacozzo13,Defenu23}. Accordingly, Fig.~\ref{fig:alpha} shows the KZ exponents obtained with this $z$ and the $\nu$ of the gap analysis only for $q>1$, where they lie within about $10\%$ of those of the gap analysis.

\section{\texorpdfstring{Free-fermion}{Free-fermion} description}\label{sec:pwave}

In this section, we investigate to what extent the full many-body dynamics can be captured by the simplest free-fermion approximation.  For this, we employ the Jordan-Wigner transformation for the spin operators and truncate the arising fermionic multibody interaction terms at quadratic order, which yields a free-fermion description. The resulting model is a fractional $p$-wave superconductor that is exactly solvable at every $q$. As we show below, the critical exponents, the excitation density, and the complete kink statistics can be expressed in closed forms and provide the analytical benchmark for the simulations of Sec.~\ref{sec:cumulants_num}. We further derive bounds for the variance and the kink-kink correlations that hold for every free-fermion state and compare them with the numerical results of the TDVP simulations.

\subsection{Truncated Jordan-Wigner mapping and the fractional p-wave superconductor}\label{sec:JW}

The Jordan-Wigner transformation $\hat\sigma_i^z=1-2\hat c_i^\dagger \hat c_i$,
$\hat\sigma_i^x=\hat{\mathcal K}_i(\hat c_i^\dagger+\hat c_i)$ with
$\hat{\mathcal K}_i=\prod_{j<i}(1-2\hat c_j^\dagger \hat c_j)$ maps the transverse field to
$-\sum_i\hat\sigma_i^z=2\sum_i \hat c_i^\dagger \hat c_i+\mathrm{const}$ and the interaction
to
\begin{equation}\label{eq:xxJW}
\begin{split}
    \hat\sigma_i^x\hat\sigma_{i+r}^x
    =(\hat c_i^\dagger-\hat c_i)\!\!\prod_{j=i+1}^{i+r-1}\!\!(1-2\hat c_j^\dagger \hat c_j)(\hat c_{i+r}^\dagger+\hat c_{i+r}),
\end{split}
\end{equation}
with the $\hat c_i$ and $\hat c^\dagger_i$ annihilation and creation operators satisfying the usual fermionic anticommutation relations.
The strings $\hat{\mathcal K}_i$ enforce these anticommutation relations at the expense of generating multibody interaction terms. We therefore truncate all such terms to quadratic level, i.e., $\hat{\mathcal K}_i=1$ and so
$\hat\sigma_i^x\hat\sigma_{i+r}^x\approx \hat c_i^\dagger \hat c_{i+r}+\hat c_i^\dagger \hat c_{i+r}^\dagger
+\mathrm{h.c.}$, which becomes exact at fractional order $q=2$. This treats the strings at mean-field level, replacing each by its value in the polarized state, as in Ref.~\cite{Jaschke17}. Away from $q=2$ the truncation is uncontrolled. It removes the kink-kink interactions~\cite{DefenuEnssHalimeh19}, and the resulting long-range Kitaev chain reaches short-range exponents already for $q>1$~\cite{Vodola16}, whereas Ising chains with power-law couplings $r^{-(1+q)}$ do so only for $q>7/4$~\cite{DefenuTrombettoniRuffo17,Defenu23}.
After the truncation, the Hamiltonian takes the form of a $p$-wave superconductor,
\begin{equation}\label{eq:H_MF}
\begin{split}
    \hat H_\mathrm{FF}(t)=\ &2g(t)\sum_i \hat c_i^\dagger \hat c_i
    -\bigl(1-g(t)\bigr)\\
    &\times\sum_i\sum_r J_{r,q}
    \left(\hat c_i^\dagger \hat c_{i+r}+\hat c_i^\dagger \hat c_{i+r}^\dagger+\mathrm{h.c.}\right),
\end{split}
\end{equation}
with hopping
and pairing amplitudes given by the same fractional couplings $J_{r,q}$ as in Eq.~\eqref{eq:Jrq}.
The system is driven by the same linear protocol as in Sec.~\ref{sec:Ising_model}, from the trivial phase at $g=1$ to the fractional $p$-wave phase at $g=0$. Neglecting the open boundaries and imposing antiperiodic boundary conditions in the even-parity sector allows for
momentum-space decomposition via $\hat c_k=L^{-1/2}e^{-i\pi/4}\sum_{j} e^{ijk}\hat c_{j}$, and since
the pairing couples $k$ to $-k$ only, the Hamiltonian decomposes into
independent two-level systems (TLSs) labeled by $k>0$, yielding
\begin{equation}\label{eq:Hk}
    \hat H_\mathrm{FF}(t)=\sum_{k>0}\hat\psi_k^\dagger H_k(t)\,\hat\psi_k,
    \quad
    H_k(t)=\xi_k(t)\,\tau^z+\Delta_k(t)\,\tau^x,
\end{equation}
with the Nambu spinor $\hat\psi_k:=(\hat c_k,\hat c_{-k}^\dagger)^T$ and
$\tau^{x,y,z}$ the Pauli matrices acting inside the given TLSs, and
\begin{align}\label{eq:xiDelta}
    \xi_k(t)&=2g(t)-\bigl(1-g(t)\bigr)\,j_q(k),
    \nonumber\\
    \Delta_k(t)&=\bigl(1-g(t)\bigr)\,d_q(k).
\end{align}
The hopping and pairing dispersions admit the closed forms
\begin{align}\label{eq:jqk}
    j_q(k)&=2\sum_{r\geq1}J_{r,q}\cos(kr)
    =J_q-\left\lvert2\sin\frac{k}{2}\right\rvert^q,\\
    d_q(k)&=2\sum_{r\geq1}J_{r,q}\sin(kr)\\
    &=-2\binom{q}{q/2}\,
    \Im\!\left[{}_2F_1\!\left(-\frac{q}{2},1;\frac{q}{2}+1;e^{ik}\right)\right],\nonumber
\end{align}
with $J_q\equiv\binom{q}{q/2}$, where $j_q$ is even and $d_q$ odd in $k$,
and the latter does not alter the minimum gap as $d_q(0)=0$.

Each TLS~\eqref{eq:Hk} is identical to a spin coupled to the effective
field $\vec h_k=(\Delta_k,0,\xi_k)$, which lies in the $xz$ plane and is
specified by its modulus and by the angle it makes with the $z$ axis,
\begin{equation}\label{eq:theta}
    \epsilon_k=\lvert\vec h_k\rvert=\sqrt{\xi_k^2+\Delta_k^2},
    \qquad
    \tan 2\vartheta_k=\frac{\Delta_k}{\xi_k}.
\end{equation}
As $\vec h_k$ lies in the $xz$ plane, diagonalizing each TLS is identical to a
rotation about the $y$ axis, i.e., to a Bogoliubov rotation by $\vartheta_k$,
$\hat\gamma_k=\cos\vartheta_k\,\hat c_k+\sin\vartheta_k\,\hat c^\dagger_{-k}$ and
$\hat\gamma^\dagger_{-k}=-\sin\vartheta_k\,\hat c_k
+\cos\vartheta_k\,\hat c^\dagger_{-k}$, which preserves the anticommutation
relations and diagonalizes the Hamiltonian,
\begin{equation}\label{eq:Hdiag}
    \hat H=\sum_{k>0}\epsilon_k
    \left(\hat\gamma^\dagger_k\hat\gamma_k
    +\hat\gamma^\dagger_{-k}\hat\gamma_{-k}-1\right).
\end{equation}
The instantaneous ground state is the quasiparticle vacuum,
$\hat\gamma_{\pm k}\ket{\Omega}=0$, which is the paired state
$\ket{\Omega}=\prod_{k>0}\left(\cos\vartheta_k-\sin\vartheta_k\,
\hat c^\dagger_k\hat c^\dagger_{-k}\right)\ket{0}$ with $\ket{0}$ denoting the fermionic
vacuum. Starting from a paramagnet, each block thus stays within the two-dimensional subspace spanned
by $\ket{0}_k$ and $\hat c^\dagger_k\hat c^\dagger_{-k}\ket{0}_k$, the singly
occupied states being decoupled and empty throughout.

As $d_q(0)=0$, the field $\vec h_0$ is purely longitudinal and its
modulus vanishes when $\xi_0=0$, so that the gap closes at $k=0$ for
\begin{equation}\label{eq:gc}
    g_c^{\rm FF}=\frac{J_q}{2+J_q}.
\end{equation}
As shown in App.~\ref{app:lowk}, for low momenta the hopping term scales as $j_q(k)\simeq J_q-k^q$, while the
connection formula of ${}_2F_1$ near $\zeta=e^{ik}=1$ separates the pairing term into an analytic and a non-analytic part,
$d_q(k)=a_q k+c_q k^{q}+\mathcal O\!\left(k^{\min(3,q+2)}\right)$, with
\begin{align}\label{eq:aq_cq}
    a_q&=\frac{q}{2(q-1)}\binom{q}{q/2},\nonumber\\
    c_q&=2\binom{q}{q/2}
    \frac{\Gamma\!\left(\frac{q}{2}+1\right)\Gamma(-q)}{\Gamma\!\left(-\frac q2\right)}
    \sin\!\left(\frac{\pi q}{2}\right),
\end{align}
and the smaller of the two exponents controls the limit,
\begin{equation}\label{eq:dp_effective}
    d_q(k)\simeq c_*\,k^{q_*},\qquad q_*=\min(q,1),
\end{equation}
where
\begin{equation}\label{eq:cstar}
    c_*=
    \begin{cases}
        c_q, & 0<q<1,\\[2mm]
        a_q=\dfrac{q}{2(q-1)}\dbinom{q}{q/2}, & q>1 .
    \end{cases}
\end{equation}
The non-analytic term therefore dominates for $0<q<1$ and the analytic
linear term for $q>1$; the two amplitudes exchange roles at $q=1$, where $a_q$
diverges, and the leading term is dressed by a logarithm,
$d_1(k)=\frac{4}{\pi}\sin\frac k2\ln\cot\frac k4\simeq\frac{2}{\pi}k\ln\frac4k$.
For $q>1$, $a_q=2\sum_r rJ_{r,q}=V_0/2$ is half the energy of two free kinks, Eq.~\eqref{eq:Vm}.
At even integers $q=2,4,\dots$ the hypergeometric series terminates and
$c_q=0$, so the expansion is purely analytic. At
criticality, the kinetic term reduces to $\xi_k(g_c^{\rm FF})\propto k^{q}$ and the
pairing to $\Delta_k(g_c^{\rm FF})\propto k^{q_*}$, both entering $\epsilon_k$ squared,
so that one obtains
\begin{equation}\label{eq:ek_gc}
    \epsilon_k(g_c^{\rm FF})\sim
    \begin{cases}
        k^{q_*}, & q\neq1,\\[1mm]
        k\ln(1/k), & q=1 ,
    \end{cases}
\end{equation}
so that $z=\min(q,1)$, up to the marginal logarithmic correction at $q=1$.
Since $\epsilon_0(g)\sim|g-g_c^{\rm FF}|$, one has $z\nu=1$ and therefore
$\nu=1/z$.

\subsection{Excitation and kink statistics}\label{sec:defects}

Due to the truncation of the Jordan-Wigner strings away from $q=2$, the free-fermion ground state at $g=0$ differs from the ferromagnetic ground state of the full fractional Ising model, Eq.~\eqref{eq:H_Ising}, and already contains an extensive number of kinks. Accordingly, the kink number operator of Eq.~\eqref{eq:Nhat} differs from the excitation number operator at $g=0$, which is given by $\hat N_{\rm ex}=\sum_{k>0}\bigl(\hat\gamma^\dagger_k\hat\gamma_k+\hat\gamma^\dagger_{-k}\hat\gamma_{-k}\bigr)$ in terms of the Bogoliubov operators $\hat\gamma_{\pm k}$ at $g=0$. As we show below, these ground-state kinks are separated from the ones generated during the ramp, which are directly related to the non-adiabatic excitations (App.~\ref{app:kinks}).

Ramping through the critical point $g_c^{\rm FF}$ in Eq.~\eqref{eq:gc}, the time-evolved state becomes $\ket{\Psi_k(t)}=u_k(t)\ket{0}_k+v_k(t)\,
\hat c^\dagger_k\hat c^\dagger_{-k}\ket{0}_k$ with $|u_k|^2+|v_k|^2=1$ and $u_k(0)=1,\,v_k(0)=0$ governed by the Bogoliubov-de Gennes equations,
\begin{equation}\label{eq:EOM}
    i\frac{d}{dt}
    \begin{pmatrix}u_k\\ v_k\end{pmatrix}
    =\begin{pmatrix}
        -\xi_k(t) & \Delta_k(t)\\
        \Delta_k(t) & \xi_k(t)
    \end{pmatrix}
    \begin{pmatrix}u_k\\ v_k\end{pmatrix}.
\end{equation}
As this differential equation takes the Landau-Zener form in the leading order of $k$, the resulting excitation probability $\langle\hat\gamma^\dagger_k\hat\gamma_k\rangle$ in the given pair of $(k,-k)$ modes is given by
\begin{equation}\label{eq:pk}
        p_k=\exp\left[-\frac{4\pi\tau_Q\,d_q^2(k)}{\bigl(2+j_q(k)\bigr)^3}\right]
    \simeq\exp\left(-\lambda_q\tau_Q k^{2q_*}\right),
\end{equation}
with $\lambda_q=4\pi c_*^2/(2+J_q)^3$ valid for small $k$ and $q\neq1$. However,  Eq.~\eqref{eq:pk} only applies to the modes for which $\xi_k$ changes sign during the ramp, i.e., $j_q(k)>0$, restricted to the region $k<k_c$ with $2\sin(k_c/2)=J_q^{1/q}$. The remaining modes, $k_c<k\leq\pi$, do not form avoided level crossings and excitations in them are no longer described within the Landau-Zener approximation. At $g=0$ their quasiparticle energy $\epsilon_k(g=0)=\sqrt{j_q^2(k)+d_q^2(k)}$ is smallest at $k=\pi$, where it equals $2^q-J_q\simeq q\ln2$ for small $q$. As shown in App.~\ref{app:noncrossing} by means of adiabatic perturbation theory, the abrupt end of the ramp at this small gap results in excitation probabilities,
\begin{equation}\label{eq:pk_nc}
    p_k\simeq\frac{d_q^2(k)}{4\tau_Q^2\,\epsilon_k^{6}(g=0)}.
\end{equation}
Approximating the sum over all momentum modes with an integral in the continuum limit, one finds
\begin{equation}\label{eq:nex}
    n_{\rm ex}=\frac{2}{L}\sum_{k>0}p_k
    \approx\mathcal N_q\,\tau_Q^{-1/(2q_*)},\qquad \mathcal N_q=\frac{\Gamma\bigl(1+\frac{1}{2q_*}\bigr)}{\pi\lambda_q^{1/(2q_*)}},
\end{equation}
for $q>1/4$, while for $q<1/4$ and slow ramps, $\tau_Q\gg(q\ln2)^{-2}$, the high-momentum excitations dominate with
\begin{eqnarray}\label{eq:nex_nc}
    n_{\rm ex}&\approx&B_q\,\tau_Q^{-2},\\
    B_q&=&\frac1\pi\int_{k_c}^{\pi}\frac{d_q^2(k)}{4\,\epsilon_k^{6}(g=0)}\,dk=\frac{I_0}{4\pi}q^{-4}(1+1.46\, q)+\mathcal O(q^{-2}),\nonumber
\end{eqnarray}
with $I_0\approx2.10$. Regarding finite-size effects in the continuum approximation, we show in App.~\ref{app:finite_size} that for $q>1/4$ the mean kink and excitation numbers drop below unity before the discreteness of the momenta is resolved, independently of $L$, and Eq.~\eqref{eq:nex} therefore holds whenever these numbers are extensive. For $q<1/4$ the slowest ramps resolve the discreteness first, which leaves the contribution of the modes near $k=\pi$ unaffected.
At the marginal point $q=1$ the logarithmic pairing term modifies the
prefactor but not the power, $n_{\rm ex}\propto\tau_Q^{-1/2}/\ln\tau_Q$, with
the next-to-leading term given in App.~\ref{app:lowk}. The logarithm makes the effective finite-time exponent $1/2+1/\ln\tau_Q+\dots$, which approaches $1/2$ from above.

Similarly to Sec.~\ref{sec:cumulants_num}, for the full counting statistics of $\hat N_{\rm ex}$ we employ the logarithm of the characteristic function, i.e., the cumulant generating function. As the excitations appear as independent Bernoulli random variables with probability $p_k$ in the $k$-th momentum mode, the cumulant generating function decomposes into the sum over these independent modes
\cite{delcampo2018,Cui2020,Balducci2023}
\begin{align}\label{eq:charf}
    \ln\langle e^{i\theta\hat N_{\rm ex}}\rangle
    &=\sum_{k>0}\ln\left[1+(e^{2i\theta}-1)p_k\right]\\
    &=-\frac{L}{2}n_{\rm ex}\,
    {\rm Li}_{1+\frac{1}{2q_*}}\!\left(1-e^{2i\theta}\right),\nonumber
\end{align}
with $\mathrm{Li}_s(x)=\sum_{k=1}^\infty\frac{x^k}{k^s}$ denoting the polylogarithm function. Here, $\langle\dots\rangle$ denotes the average over the full many-body time-evolved state, i.e., the tensor product of the two-level wave functions of Eq.~\eqref{eq:EOM}.
The last equality holds in the continuum limit in the leading order in $\tau_Q$ for $q>1/4$ and $q\neq1$, and reduces to
the ${\rm Li}_{3/2}$ form familiar from the transverse-field Ising chain when
$q_*=1$~\cite{delcampo2018}.  As a result, all excitation cumulants scale with the same exponent,
\begin{equation}\label{eq:kappa_scaling}
    \kappa^{\rm ex}_\ell=C_\ell(q)\,\tau_Q^{-1/(2q_*)},
    \qquad
    \frac{C_\ell}{C_1}=2^{\ell-1}S_\ell\!\left(\frac{1}{2q_*}\right),
\end{equation}
with $C_1=L\mathcal N_q$, $S_2(w)=1-2^{-w}$ and $S_3(w)=1-3\cdot2^{-w}+2\cdot3^{-w}$, so that the
cumulant ratios $R_\ell\equiv\kappa^{\rm ex}_\ell/\kappa^{\rm ex}_1$ are
independent of the quench rate and depend only on $q_*$; for
$q_*=1$, one obtains $R_2\simeq0.586$ and $R_3\simeq0.134$. A Poisson number of independently formed excitations has $\kappa^\mathrm{ex}_2=\kappa^\mathrm{ex}_1$. On the ring the excitations are created in pairs, so that $R_2=2S_2\bigl(\tfrac{1}{2q_*}\bigr)$ is sub-Poissonian for $q>1/2$ and super-Poissonian for $q<1/2$. Pair creation limits this bunching: every free-fermion state satisfies $\kappa^\mathrm{ex}_2<2\kappa^\mathrm{ex}_1$ and $|\kappa^\mathrm{ex}_3|\leq2\kappa^\mathrm{ex}_2$ (see App.~\ref{app:bound} for the derivation), so that $R_2<2$ and $|R_3|\leq2R_2$ also beyond the scaling limit.

For $q<1/4$ and slow driving, only those terms linear in $p_k\propto\tau^{-2}_Q$ contribute to the excitation number cumulants in the leading order, resulting in
\begin{equation}\label{eq:kappa_scaling_nc}
    \kappa^{\rm ex}_\ell=2^{\ell-1}LB_q\,\tau_Q^{-2}+\mathcal O(\tau^{-4}_Q).
\end{equation}
Accordingly, one finds in the leading order that $\kappa^{\rm ex}_2/\kappa^{\rm ex}_1=2$ and $\kappa^{\rm ex}_3/\kappa^{\rm ex}_1=4$, the values of independently created pairs. For a more precise characterization of the cumulant ratios, which also respects the exact bound of App.~\ref{app:bound}, one thus needs the next-to-leading order correction, $\kappa^{\rm ex}_2/\kappa^{\rm ex}_1=2[1-J_0/(4I_0q^4\tau_Q^2)]$ and $\kappa^{\rm ex}_3/\kappa^{\rm ex}_1=4[1-3J_0/(4I_0q^4\tau_Q^2)]$, with $I_0\simeq2.10$ of Eq.~\eqref{eq:nex_nc} and $J_0=\int_{\pi/3}^{\pi}\bigl(\frac{\pi-k}{2}\bigr)^4\bigl[\ln^2\bigl(2\sin\frac k2\bigr)+\bigl(\frac{\pi-k}{2}\bigr)^2\bigr]^{-6}dk\simeq2.72$, keeping both ratios stay below these values.

The number of excited $(k,-k)$ mode pairs, each contributing two excitations,
has a Poisson-binomial distribution, by which $\hat N_{\rm ex}$ approaches in the large-$L$ limit the Gaussian form
\begin{equation}\label{eq:gaussian_limit}
    P_\mathrm{ex}(n)\to\frac{1}{\sqrt{2\pi\kappa^{\rm ex}_2}}
    \exp\left[-\frac{(n-\kappa^{\rm ex}_1)^2}{2\kappa^{\rm ex}_2}\right],
\end{equation}
which also describes the kink statistics of Sec.~\ref{sec:cumulants_num} for large enough mean values.

Regarding the kink number statistics within the free-fermion setting, with periodic boundary conditions, the kink number operator of Eq.~\eqref{eq:Nhat} is exactly quadratic (App.~\ref{app:kinks}),
\begin{equation}\label{eq:Nk_main}
    \hat N=\sum_{k>0}\hat n_k,\qquad
    \hat n_k=1+\hat\psi_k^\dagger\bigl(-\cos k\,\tau^z+\sin k\,\tau^x\bigr)\hat\psi_k,
\end{equation}
where $\hat n_k$ has the eigenvalues $0$ and $2$, corresponding to no kink and a kink pair in the $k$-th mode, respectively. The kink probability in the pair $(k,-k)$ is thus given by
\begin{equation}\label{eq:ptilde_main}
    \tilde p_k=\frac12\langle\Psi_k|\hat n_k|\Psi_k\rangle=\Bigl|\cos\tfrac k2\,u_k+\sin\tfrac k2\,v_k\Bigr|^2,
\end{equation}
and it is related to $p_k$ through the angle $\delta_k$ between the excited quasiparticle state and the kink-pair state at $g=0$, defined as
\begin{equation}\label{eq:sin_delta}
    \sin^2\delta_k=\frac12\left[1-\frac{j_q(k)\cos k+d_q(k)\sin k}{\epsilon_k(g=0)}\right],
\end{equation}
which vanishes at $q=2$, where $j_2(k)=2\cos k$ and $d_2(k)=2\sin k$. In terms of $p_k$ and the relative phase $\Phi_k$ of the amplitudes of the ground and excited states at $g=0$, the kink probability reads
\begin{equation}\label{eq:ptilde_split_main}
    \tilde p_k=\sin^2\delta_k+p_k\cos2\delta_k-\sqrt{p_k(1-p_k)}\,\sin2\delta_k\cos\Phi_k .
\end{equation}
Here the first term gives the ground-state kinks of the free-fermion model at $g=0$, with mean number $\kappa^{\rm GS}_1=2\sum_{k>0}\sin^2\delta_k\propto L$, while the second term stems exclusively from the non-adiabatic excitations, and the third, interference term averages out in the momentum sum, as its phase $\Phi_k$ varies rapidly with $k$ (App.~\ref{app:kinks}). For $q>1/4$ kink production is governed by the small-momentum excitations, where $\delta_k\to0$, so that the kink probability becomes identical to the Landau-Zener excitation probability, $\tilde p_k\simeq p_k$. For $q<1/4$ the modes close to $k=\pi$ dominate instead, and their kink probability above the ground state acquires a momentum-dependent prefactor compared to the corresponding excitation probability, $\tilde p_k-\sin^2\delta_k\simeq w_kp_k$, with $w_k$ reaching $w_\pi\simeq3.32$ at $k=\pi$ for $q\to0$ (for details see App.~\ref{app:kinks}). As $p_k\sin^2\delta_k$ is negligible at leading order, in both cases the cumulant generating function of the kink number separates into the contributions of the ground-state kinks and of the excitations,
\begin{eqnarray}\label{eq:charf_kinks}
    \ln\tilde P(\theta)=\ln\langle e^{i\theta\hat N}\rangle&\simeq&\sum_{k>0}\ln\left[1+\bigl(e^{2i\theta}-1\bigr)\sin^2\delta_k\right]\nonumber\\
    &&+\sum_{k>0}\ln\left[1+\bigl(e^{2i\theta}-1\bigr)\tilde p_k\right].
\end{eqnarray}
 The first sum gives the $\tau_Q$-independent cumulants $\kappa^{\rm GS}_\ell$ of the kink number in the ground state at $g=0$, and the second one the kinks generated during the ramp. For $q>1/4$ the second sum reduces to Eq.~\eqref{eq:charf}, i.e., the kink cumulants measured from the ground state coincide at leading order with those of the excitations, $\kappa_\ell-\kappa^{\rm GS}_\ell\simeq\kappa^{\rm ex}_\ell$. For $q<1/4$, the mean follows as $\kappa_1-\kappa^{\rm GS}_1\simeq(\tilde B_q/B_q)\kappa^{\rm ex}_1$, with $\tilde B_q=\frac1\pi\int_{k_c}^{\pi}w_k\,d_q^2(k)/[4\epsilon_k^6(g=0)]\,dk$, while the higher cumulants are reduced by the overlap of the excited modes with the ground-state kinks (App.~\ref{app:kinks}). As a result, the ratios $\tilde R_\ell\equiv(\kappa_\ell-\kappa^{\rm GS}_\ell)/(\kappa_1-\kappa^{\rm GS}_1)$ stay below those of the excitations, increasing monotonically from $\tilde R_2\simeq1.89$ and $\tilde R_3\simeq3.36$ for $q\to0$ to $\tilde R_2\simeq1.93$ and $\tilde R_3\simeq3.57$ at $q=1/4$ and thus satisfying the general free-fermion bound of App.~\ref{app:bound}.

The kink number cumulant ratios thus exhibit the same behavior as the excitations for $q>1/4$, while for $q<1/4$ the plateau values $\tilde R_2$ and $\tilde R_3$ replace $2$ and $4$, and the next-to-leading order correction carries $q$-dependent prefactors, $(\kappa_2-\kappa^\mathrm{GS}_2)/(\kappa_1-\kappa^\mathrm{GS}_1)=\tilde R_2-\tilde J_0/(2\tilde I_0q^4\tau_Q^2)$ and $(\kappa_3-\kappa^\mathrm{GS}_3)/(\kappa_1-\kappa^\mathrm{GS}_1)\simeq\tilde R_3-3\tilde J_0/(\tilde I_0q^4\tau_Q^2)$, with $\tilde I_0\simeq4.89$ of Eq.~\eqref{eq:app_pi_Bq_smallq} and $\tilde J_0=\int_{\pi/3}^{\pi}w_k^2\bigl(\frac{\pi-k}{2}\bigr)^4\bigl[\ln^2\bigl(2\sin\frac k2\bigr)+\bigl(\frac{\pi-k}{2}\bigr)^2\bigr]^{-6}dk\simeq15.1$, so that both ratios stay below their plateau values.

The mean kink numbers with the free-fermion numerics (exact evolution of the TLSs with $L=150$) and the excitation number from the exact momentum-space analysis are plotted in Fig.~\ref{fig:kappa_1}, showing a progressively improving agreement with the TDVP results as $q$ is increased towards $q=2$ and close agreement for $q\gtrsim2$.

Figure~\ref{fig:kappa_2} compares the second cumulant of the TDVP simulations with the same free-fermion results on the open chain and on the ring and with the analytical predictions. Somewhat differently from the case of the mean, the free-fermion results start matching the TDVP ones slightly earlier, for $q\gtrsim 1.25$.
 Similar observations hold for the third cumulant in App.~\ref{app:cumulants}.

Finally, we show the comparison between the cumulant ratios averaged over the middle half of the considered driving times as a function of the fractional order $q$ in Fig.~\ref{fig:ratios_q}. As for the variance, the free-fermion kink ratios measured from the ground state on the open chain and the analytical ratios $R_\ell$ of Eq.~\eqref{eq:kappa_scaling} on the ring converge to the TDVP results for $q\gtrsim1.25$. For $q\lesssim0.3$, the TDVP ratio $\kappa_2/\kappa_1$ exceeds 2, which the bound of App.~\ref{app:bound} excludes for free fermions, consistent with the domain-wall confinement of Eq.~\eqref{eq:Vm}. A related departure from independently formed defects was observed in Rydberg chains~\cite{Balewski25}. As an additional check for small system sizes, we evolved the full spin Hamiltonian, Eq.~\eqref{eq:H_Ising}, exactly on open chains of $L=12$--$20$ sites with all couplings at $q=0.2$ and $\tau_Q=50$, without any truncation of the state. Here, $\kappa_3/\kappa_2$ reaches $2.3$--$2.5$, above the free-fermion limit of 2 (App.~\ref{app:bound}), whereas free fermions at the same sizes give about $1.3$, and $\kappa_2/\kappa_1\simeq2.0$--$2.2$. At $q=0.1$ and $\tau_Q=100$, the kink number on the central half of the chain exceeds both free-fermion bounds from $L=16$ on, with $\kappa_2/\kappa_1\simeq2.2$--$2.4$ and $\kappa_3/\kappa_2\simeq2.3$--$2.6$ for $L=16$--$20$.

\subsection{Kink-kink correlations}\label{sec:kinkcorr}

In this subsection, we investigate the spatial structure of the kinks, which provides further support for the Gaussian form of the histograms in Sec.~\ref{sec:cumulants_num}. To this end, we extract the connected correlation function between the middle bond and the bond at distance $r$ from the time-evolved MPS of the TDVP simulations, using the same techniques as for the characteristic function (App.~\ref{app:char_function}),
\begin{equation}\label{eq:Ckink}
    C(r)=\langle\hat n_{L/2}\,\hat n_{L/2+r}\rangle-\langle\hat n_{L/2}\rangle\langle\hat n_{L/2+r}\rangle,
\end{equation}
with the kink projectors $\hat n_j=\bigl(1-\hat\sigma^x_j\hat\sigma^x_{j+1}\bigr)/2$ of Eq.~\eqref{eq:Ndef_app} and $r=-(L/2-1),\dots,L/2-1$. As the middle bond lies farthest from the edges, it serves as a representative reference, and boundary effects enter only for the largest values of $|r|$.

As for the cumulants, we also compare the results with those obtained within the free-fermion approximation, i.e., with the Hamiltonian of Eq.~\eqref{eq:H_MF} with all couplings $r\leq L-1$ and open boundary conditions. As the single-bond projectors remain quadratic in the fermionic operators (see App.~\ref{app:bound}),
\begin{equation}\label{eq:nj_fermions}
    \hat n_j=\frac12\Bigl[1-\bigl(\hat c_j^\dagger-\hat c_j\bigr)\bigl(\hat c_{j+1}^\dagger+\hat c_{j+1}\bigr)\Bigr],
\end{equation}
the correlation function is affected by the free-fermion approximation solely through the time-evolved wave function and can be expressed by the two-point functions $\langle\hat c_i^\dagger\hat c_j\rangle$ and $\langle\hat c_i\hat c_j\rangle$ via Wick's theorem. Using this, the correlations of a given bond with all other bonds are bounded by its local variance, $\bigl|\sum_{r\neq0}C(r)\bigr|\leq C(0)$, and every cumulant of the kink number grows at most linearly with $L$, as shown in App.~\ref{app:corr_bound}. Thus, for a finite kink density, the kink number distribution of free fermions becomes exactly Gaussian in the thermodynamic limit.

\begin{figure}[t]
    \centering
    \begin{tabular}{c c}
    \includegraphics[width=0.49\columnwidth]{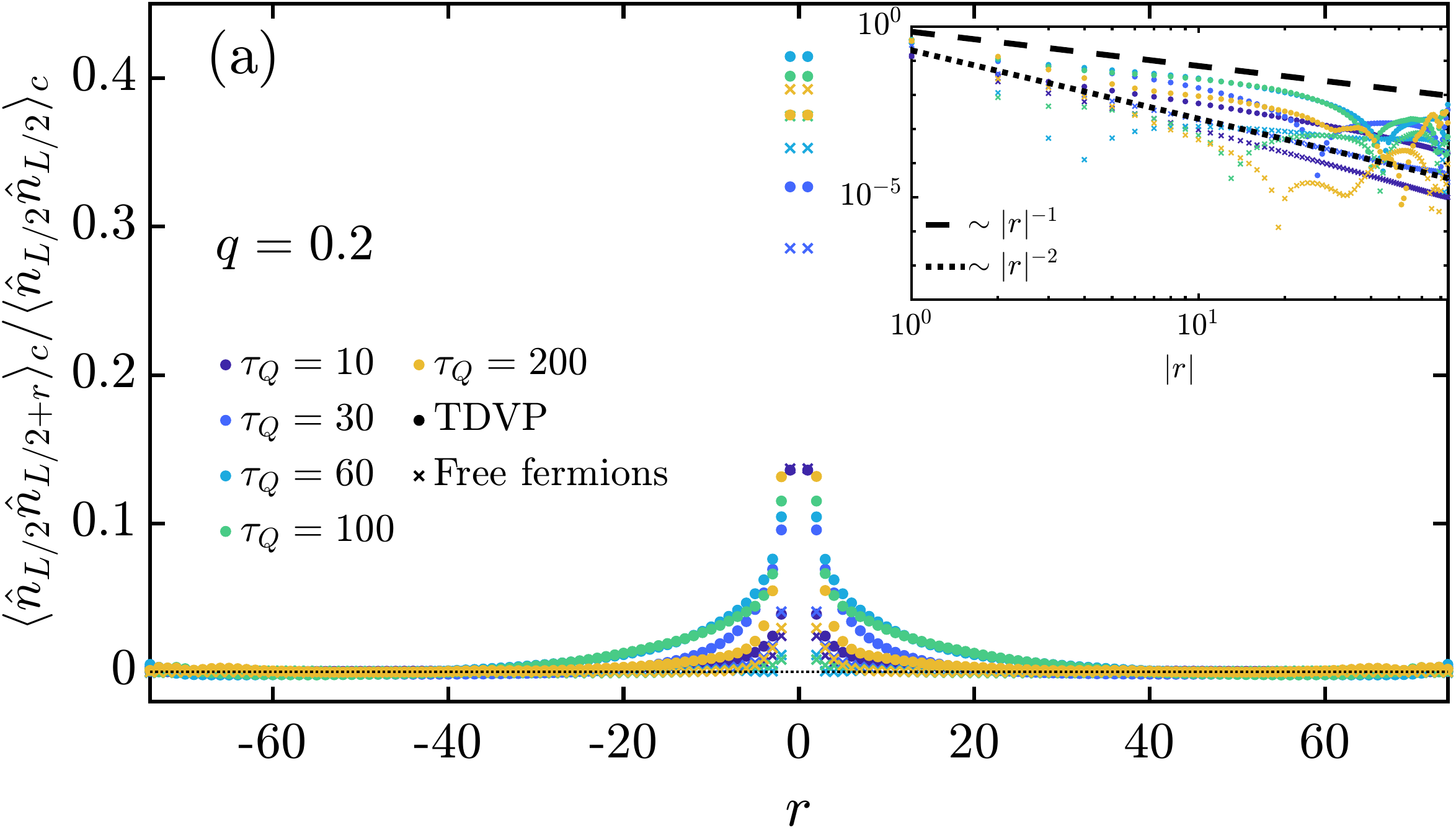} &
    \includegraphics[width=0.49\columnwidth]{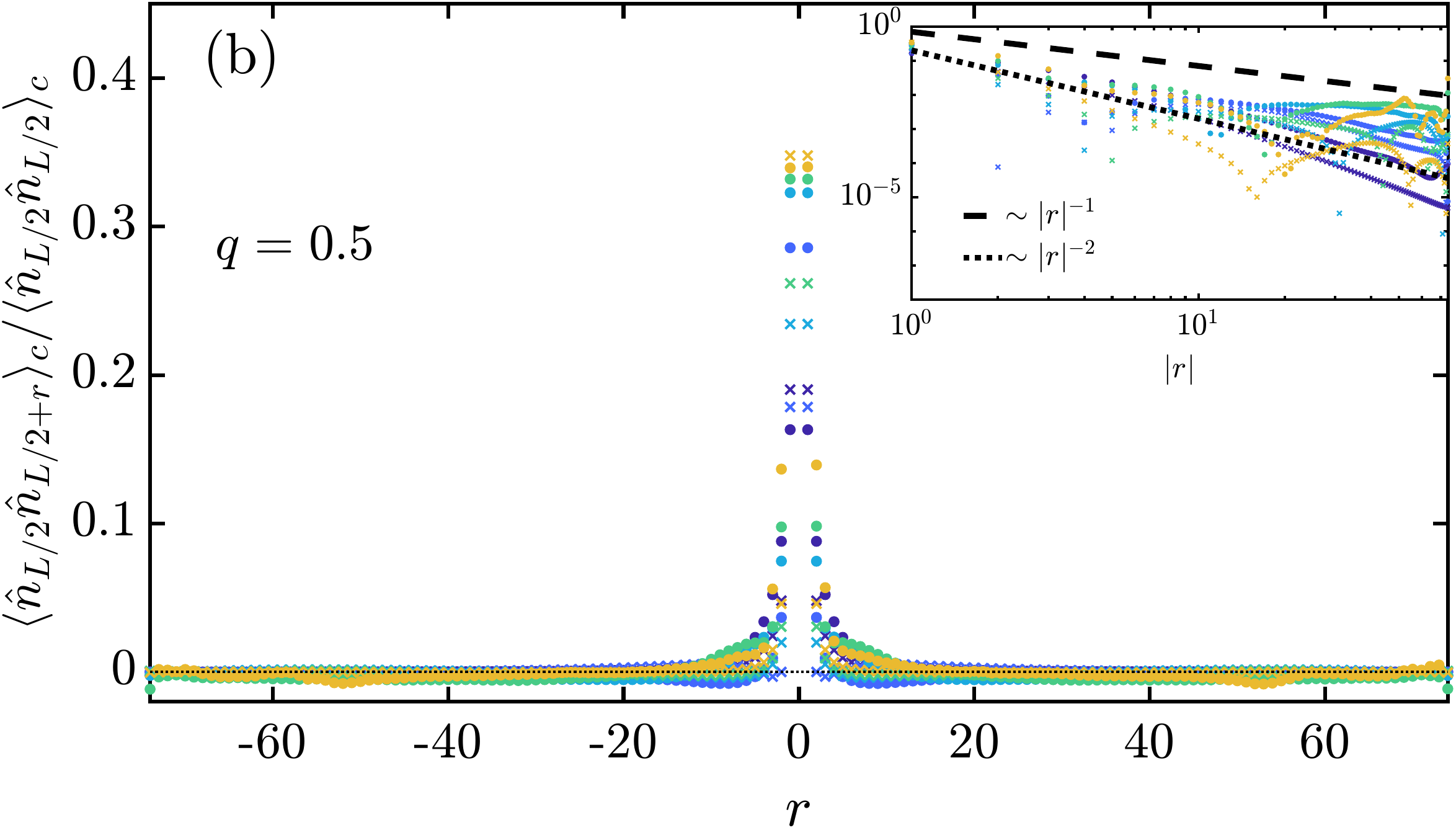} \\
    \includegraphics[width=0.49\columnwidth]{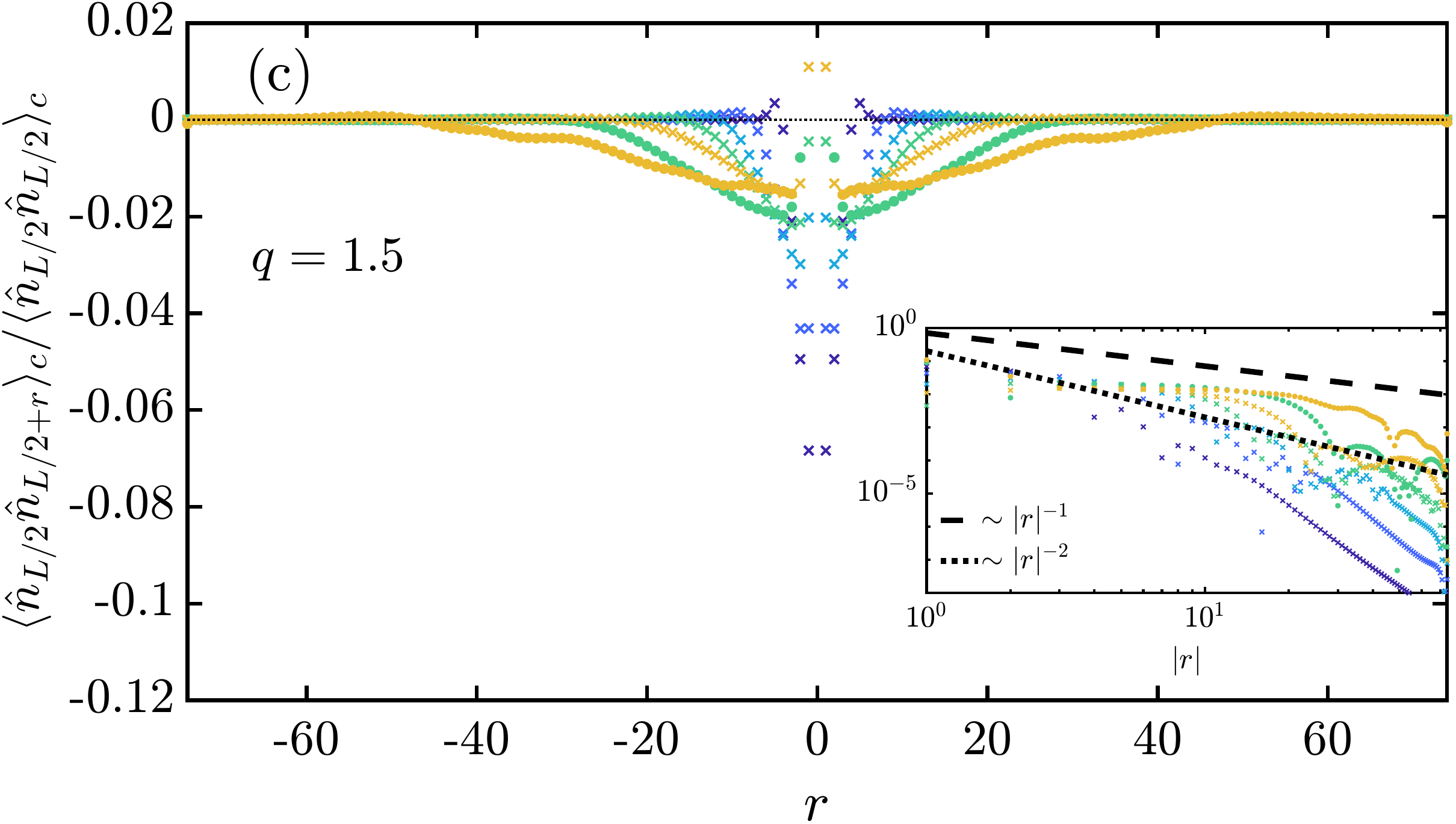} &
    \includegraphics[width=0.49\columnwidth]{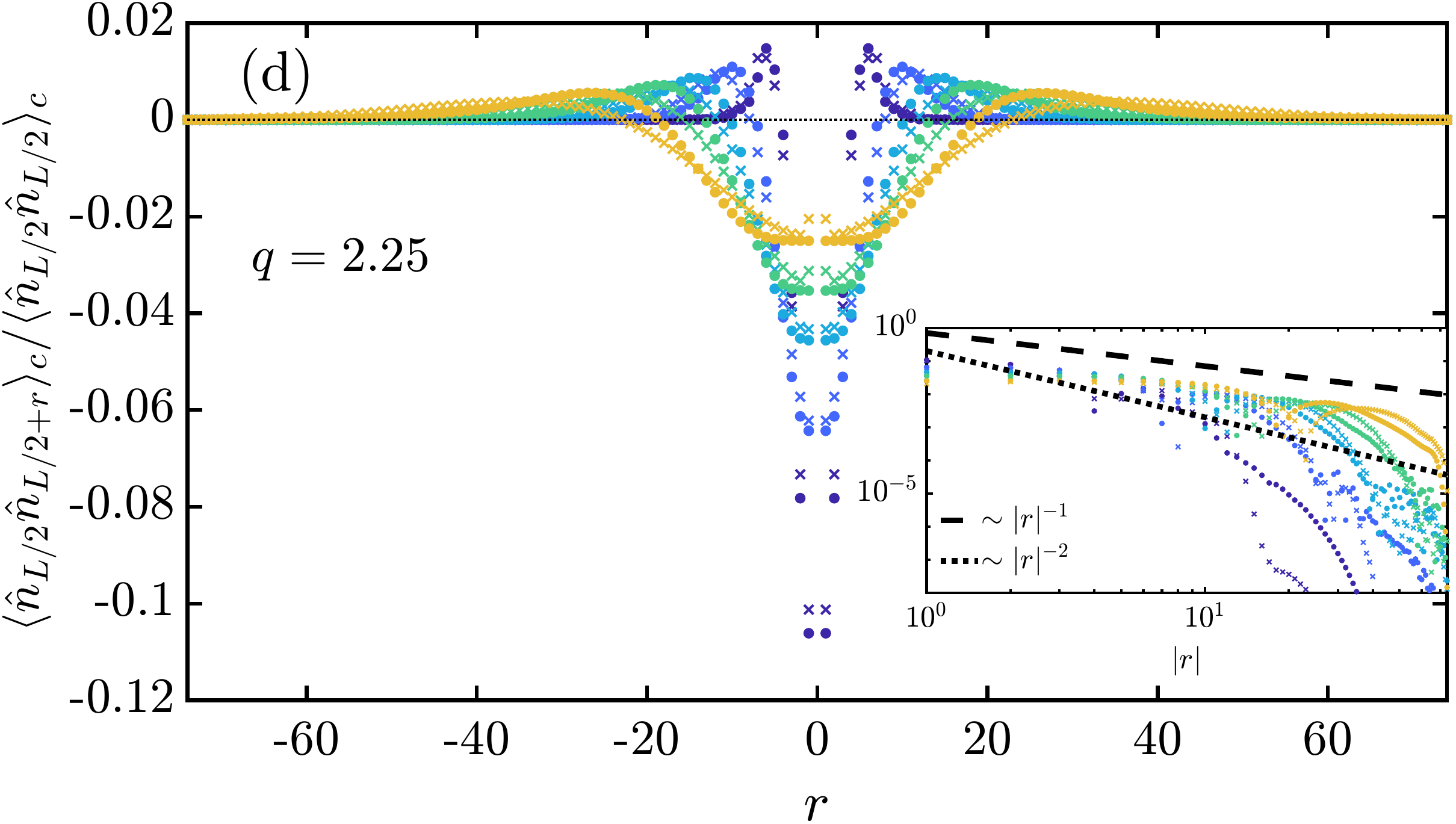} \\
    \end{tabular}
    \caption{Connected kink-kink correlation function of Eq.~\eqref{eq:Ckink}, normalized by the local variance $\langle\hat n_{L/2}\hat n_{L/2}\rangle_c$, for (a) $q=0.2$, (b) $q=0.5$, (c) $q=1.5$ and (d) $q=2.25$. Filled circles and crosses correspond to the TDVP simulations and to the free-fermion approximation on the same open chain, respectively, with the term $r=0$ omitted. The insets display the absolute values on logarithmic scales, together with the reference power laws $|r|^{-1}$ and $|r|^{-2}$ as guides to the eye. Positive correlations for $q\leq0.5$ turn into anticorrelations for $q\geq1.5$, while the fall-off with the distance is only slightly slower in the TDVP simulations.}
    \label{fig:kink_corr}
\end{figure}

\begin{figure}[t]
    \includegraphics[width=\columnwidth,trim={0 0 0cm 0cm},clip]{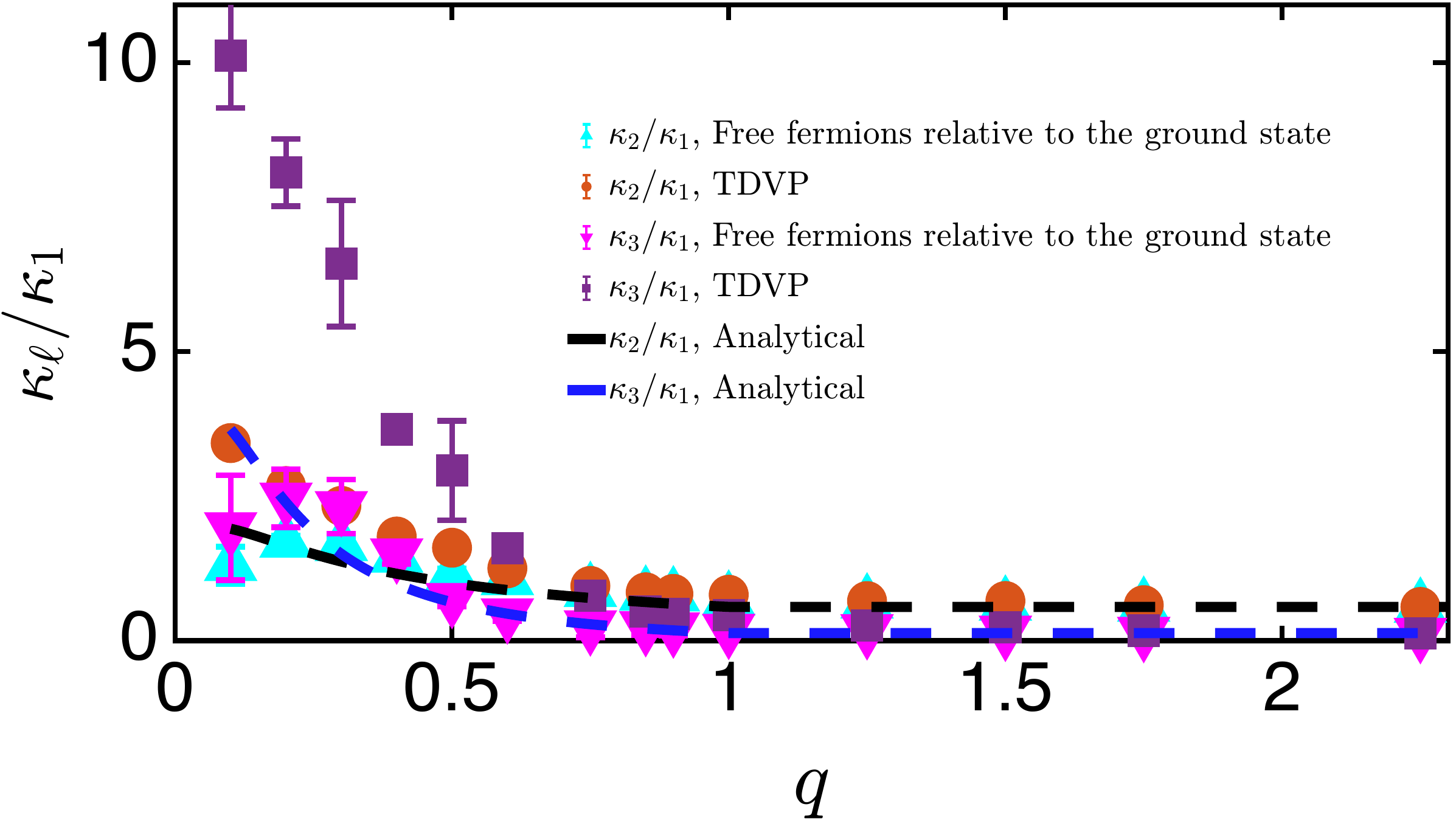} 
    \caption{Cumulant ratios $\kappa_2/\kappa_1$ and $\kappa_3/\kappa_1$, each averaged over the middle half of the sampled driving times, as a function of the fractional order $q$. The free-fermion kink ratios measured from the ground state on the open chain (cyan triangles) for $\kappa_2$ and (magenta triangles) for $\kappa_3$ and the analytical predictions of Eq.~\eqref{eq:kappa_scaling} on the ring (black and blue dashed lines) start converging to the TDVP ones (orange circles for $\kappa_2$ and purple squares for $\kappa_3$) for $q\gtrsim1.25$. For $q\lesssim0.3$ the TDVP results show a clear crossover to variances above twice the mean, while the free-fermion ratio $\kappa_2/\kappa_1$ stays below this limit throughout.}
    \label{fig:ratios_q}
\end{figure}

Figure~\ref{fig:kink_corr} shows that for all fractional orders the correlations decay within a few tens of bonds around the middle of the chain. For $q=1.5$ and $q=2.25$ neighboring kinks are anticorrelated, with close agreement between the TDVP and free-fermion data. In contrast, for $q=0.2$ and $q=0.5$ kinks tend to be created close to each other, with the many-body simulations showing stronger short-distance correlations, consistent with the enhanced variances of Fig.~\ref{fig:ratios_q} and with the binding of domain walls into pairs, Eq.~\eqref{eq:Vm}. As $\kappa_2>2\kappa_1\geq2\sum_jn_j(1-n_j)$ for $q\lesssim0.2$, the TDVP correlations of at least some bonds exceed the free-fermion limit of Eq.~\eqref{eq:rowsum}, and this holds for the middle bond itself: at $q=0.2$ its summed correlations reach $\sum_{r\neq0}C(r)\simeq1.5$--$2.3\,C(0)$ for $\tau_Q\geq30$, whereas the free-fermion values stay at $0.8$--$0.9\,C(0)$. The violation therefore arises in the bulk of the chain and not only near the edges. At intermediate distances both decay roughly between $|r|^{-2}$ and $|r|^{-1}$. Further results on the kink density profiles for both the TDVP and free-fermion results are shown in App.~\ref{app:profiles}.

\section{Summary and outlook}\label{sec:conclusions}

Long-range interactions provide a continuous parameter along which critical exponents vary. In the fractional Ising chain a single derivative order fixes the couplings at every scale, from the nearest-neighbor bond to the far tail, which makes the chain a testbed for the Kibble-Zurek mechanism in nonintegrable systems across the range of interactions.

We have characterized the kink number
statistics generated by driving the fractional Ising chain across its quantum critical point. In particular, we have reported the full counting statistics of kinks and the scaling of their cumulants with the driving time, obtained from TDVP simulations within the MPS and MPO
formalism, and benchmarked throughout against the exactly solvable fractional $p$-wave superconductor that follows from the truncated Jordan-Wigner transformation.

   First, we tested the KZ scaling of the mean kink number with the quench time, using the correlation-length and dynamical critical exponents extracted from the finite-size scaling of the gap minimum obtained by DMRG and, for $z$, from the light cone of the entanglement entropy spreading from a local perturbation. For fractional orders $q\lesssim1$, the measured decay is slower than the KZ prediction, while for $q\gtrsim1$ both agree at $L=150$ within 17\%, at the upper end of the systematic uncertainty of the gap fits. The kink number histograms converge to a Gaussian when the mean is large. The second and third cumulants remain proportional to the mean for $q\gtrsim1.25$, while the range of driving times where this proportionality holds shrinks substantially deep in the long-range regime.

We also compared our results with a free-fermion $p$-wave superconductor approach, dropping all interactions in the fermionic representation of the interacting spin Hamiltonian, including the kink-kink interaction that confines the domain walls into pairs for $q\leq1$. Within this description, we derived the kink and excitation statistics in closed form, the former relative to the free-fermion ground state, and identified a crossover at $q=1/4$, below which slow ramps mainly excite modes without avoided level crossings and all cumulants decay as $\tau_Q^{-2}$. We further proved that every free-fermion state limits the kink number variance to twice its mean and the total correlation of each bond with the rest of the chain to its local variance, which implies an exact Gaussian limiting kink number statistics for large enough mean. While the KZ exponents of the gap analysis match the TDVP decay already for $q\gtrsim1$, the free-fermion mean agrees precisely with the TDVP results only in the short-range regime, $q\gtrsim2$, and the open boundary accounts for about half of the remaining difference at $q=1.5$. However, the second and third cumulants approach the free-fermion results at smaller $q$, and their ratios agree with the free-fermion values for $q\gtrsim1.25$. For $q\lesssim0.3$, in contrast, the variance of the kink number in the TDVP simulations exceeds twice its mean and thus violates the free-fermion bound. For the kink-kink correlations, we found stronger short-distance correlations than for free fermions in this regime, consistent with domain-wall confinement, and their sum exceeds the free-fermion limit already at the center of the chain.

Fractional couplings are not an exotic choice. L\'evy statistics are as generic as Gaussian ones, and fractional operators are their continuum description in disordered and multiscale media~\cite{MetzlerKlafter00,Zaburdaev15,Berkowitz06}. As quantum simulators move toward realistic materials, how defects form under such couplings becomes a practical question. The Kibble-Zurek argument is at heart causal. Regions farther apart than the frozen correlation length order independently because no signal connects them before freeze-out. Fractional couplings weaken this premise, since correlations spread along L\'evy light cones that widen as $q$ decreases~\cite{LewisGongCarr25LevyLightCones}. The breakdown of KZ scaling found here lies in the same regime. Whether it reflects this altered causal structure, the binding of defects, or both is open, and deciding it requires a theory of defect formation with an algebraic rather than linear causal horizon.

At small $q$ a slow ramp prepares a dilute gas of bound kink-antikink pairs whose number and fluctuations are set by the ramp rate. The fractional order tunes the binding potential of Eq.~\eqref{eq:Vm} continuously, from growth as $m^{1-q}$ through logarithmic growth at $q=1$ to free domain walls for $1<q<2$. The chain is therefore a one-parameter setting for the nonequilibrium formation of bound states across a confinement threshold, which connects critical dynamics to the confinement physics now probed in quantum simulators~\cite{Liu19confined,Tan21confinement,bayocboc2026}.

Cumulant ratios take universal values when defects form independently and depart from them when defects interact. Counting statistics can therefore probe critical points that lack an equilibrium theory, using the snapshot measurements already available on Rydberg arrays, trapped ions and quantum annealers~\cite{Bernien17,Balewski25,LiWu23,king2022coherent}. Fractional derivatives extend directly to higher dimensions and, in time, to memory effects, which opens defect formation in multiscale quantum media to the same analysis.

\acknowledgements
AG and AdC acknowledge the Luxembourg National Research Fund under Grant No. C22/MS/17132054/AQCQNET and AG acknowledges the Luxembourg National Research Fund under Grant No. CC25/MS/19559370/FastQOPT. This material is based upon work supported by the U.S. National Science Foundation under Grant No. PHY-2515059. LDC thanks Maksym Serbyn for hosting him during the sabbatical in which he completed this work. Claude (Anthropic) was used to assist with checking calculations and references and with editing the text; the authors take full responsibility for the content.
\appendix

\section{Characteristic-function reconstruction of the defect distribution}\label{app:char_function}

For the MPS simulations of the fractional Ising chain, it is convenient to
characterize the full counting statistics of domain-wall defects through the
characteristic function rather than through high moments alone, which become
increasingly noisy once the corresponding MPOs are contracted.  To this end, we
define the defect-number operator
\begin{equation}\label{eq:Ndef_app}
    \hat N=\sum_{i=1}^{L-1}\hat n_i,
    \qquad
    \hat n_i=\frac{1-\hat\sigma_i^x\hat\sigma_{i+1}^x}{2},
\end{equation}
where $\hat n_i$ projects onto a kink on the bond $(i,i+1)$.  Since each
$\hat n_i$ has eigenvalues $0,1$, one has $\hat n_i^2=\hat n_i$.  Moreover, all
bond projectors commute, $[\hat n_i,\hat n_j]=0$, because they are diagonal in
the $\hat\sigma^x$ basis.  As a result, the characteristic function of the defect
number,
\begin{equation}\label{eq:char_app}
    \tilde P(\theta)\equiv\bra{\Psi(\tau_Q)}e^{i\theta\hat N}\ket{\Psi(\tau_Q)},
\end{equation}
factorizes exactly as
\begin{equation}\label{eq:char_fact_app}
    \tilde P(\theta)=\bra{\Psi(\tau_Q)}\prod_{i=1}^{L-1}e^{i\theta\hat n_i}\ket{\Psi(\tau_Q)}.
\end{equation}
This factorization makes the evaluation efficient on an MPS: rather than
constructing $\hat N^{\ell}$ for each cumulant separately, one applies a single
layer of two-site gates.

The local two-site gate associated with the bond $(i,i+1)$ is therefore
\begin{equation}\label{eq:gate_app}
    \hat U_i(\theta)\equiv e^{i\theta\hat n_i}
    =\mathbb I+\left(e^{i\theta}-1\right)\hat n_i .
\end{equation}
In the two-spin $\hat\sigma^x$ basis
$\{\ket{\uparrow\uparrow},\ket{\uparrow\downarrow},
\ket{\downarrow\uparrow},\ket{\downarrow\downarrow}\}$ this gate is diagonal,
\begin{equation}\label{eq:gate_diag_app}
    \hat U_i(\theta)=\mathrm{diag}\!\left(1,e^{i\theta},e^{i\theta},1\right),
\end{equation}
showing explicitly that aligned neighboring spins contribute phase $1$, whereas
a domain wall contributes the phase $e^{i\theta}$.

For a many-body state $\ket{\Psi(\tau_Q)}$ represented as an MPS, the characteristic
function is thus obtained as
\begin{equation}\label{eq:chi_mps_app}
    \tilde P(\theta)=
    \frac{\bra{\Psi(\tau_Q)}\prod_{i=1}^{L-1}\hat U_i(\theta)\ket{\Psi(\tau_Q)}}
         {\braket{\Psi(\tau_Q)}{\Psi(\tau_Q)}}.
\end{equation}
Because all $\hat U_i(\theta)$ commute and are local two-site operators, this
evaluation is particularly efficient within the MPS framework. As $e^{i\theta\hat n_i}=e^{i\theta/2}\bigl[\cos\frac\theta2-i\sin\frac\theta2\,\hat\sigma^x_i\hat\sigma^x_{i+1}\bigr]$, the product of all gates forms a matrix-product operator of bond dimension 2, so that the cost of Eq.~\eqref{eq:chi_mps_app} scales as that of the norm, $\mathcal O(L\chi^3)$ for each value of $\theta$. The gates are applied successively to the MPS with a truncation cutoff of $10^{-12}$. In
practice, one computes $\tilde P(\theta)$ on a discrete grid
\begin{equation}\label{eq:lambda_grid_app}
    \theta_m=\frac{2\pi m}{M},\qquad m=0,1,\dots,M-1,
\end{equation}
with $M\geq L$.  Since the spectrum of $\hat N$ is integer-valued and bounded,
$\hat N\in\{0,1,\dots,L-1\}$, the defect-number distribution
$P(n)=\mathrm{Prob}(\hat N=n)$ is reconstructed exactly, with no truncation in the
number of moments, from the inverse discrete Fourier transform,
\begin{equation}\label{eq:invDFT_app}
    P(n)=\frac{1}{M}\sum_{m=0}^{M-1}e^{-i\theta_m n}\,\tilde P(\theta_m),
\end{equation}
for $n=0,1,\dots,L-1$.

\section{Low-momentum expansion of the dispersions}\label{app:lowk}

Here we derive the small-$k$ forms of $j_q(k)$ and $d_q(k)$ from
Sec.~\ref{sec:JW}, together with the marginal case $q=1$ and the amplitude
$c_*$ of Eq.~\eqref{eq:cstar}.  The cosine sum closes in terms of the
generating identity of the centered generalized difference,
\begin{equation}\label{eq:app_ortig}
    \sum_{r=-\infty}^{\infty}(-1)^{r}\binom{q}{q/2+r}e^{ikr}
    =\left\lvert2\sin\frac{k}{2}\right\rvert^{q},
\end{equation}
so that, separating the $r=0$ term and using the parity of the coefficients,
one obtains Eq.~\eqref{eq:jqk}.  Expanding the sine for $k\to0$,
\begin{equation}\label{eq:app_jq_lowk}
    j_q(k)=J_q-k^{q}\left[1-\frac{q}{24}k^{2}+\mathcal O(k^{4})\right].
\end{equation}
No non-analytic contribution
beyond $k^{q}$ is generated, since Eq.~\eqref{eq:app_ortig} holds identically
on the Brillouin zone.

The sine sum does not terminate and is expressed through the ordinary
hypergeometric function, Eq.~\eqref{eq:jqk}.  The limit $k\to0$ corresponds to
$\zeta=e^{ik}\to1$, where ${}_2F_1$ is singular.  Applying the connection formula
\begin{equation}\label{eq:app_connection}
\begin{split}
    &{}_2F_1(a,b;c;\zeta)\\
    &=\frac{\Gamma(c)\Gamma(c-a-b)}{\Gamma(c-a)\Gamma(c-b)}
    \,{}_2F_1(a,b;a+b-c+1;1-\zeta)\\
    &\quad+\frac{\Gamma(c)\Gamma(a+b-c)}{\Gamma(a)\Gamma(b)}(1-\zeta)^{c-a-b}\\
    &\quad\times{}_2F_1(c-a,c-b;c-a-b+1;1-\zeta)
\end{split}
\end{equation}
with $(a,b,c)=(-q/2,1,q/2+1)$ gives $c-a-b=q$, so that the second term carries
the non-analytic factor $(1-\zeta)^{q}$ while the first is analytic in $1-\zeta$.
Taking the imaginary part with $1-e^{ik}=-ik+k^2/2+\mathcal O(k^3)$ separates
the two contributions,
\begin{equation}\label{eq:app_dq_lowk}
    d_q(k)=a_q k+c_q k^{q}+\mathcal O\!\left(k^{\min(3,q+2)}\right),
\end{equation}
with the amplitudes $a_q$ and $c_q$ of Eq.~\eqref{eq:aq_cq}.  Whichever
exponent is the smaller controls the limit, giving $d_q(k)\simeq c_*k^{q_*}$
with $q_*=\min(q,1)$ and $c_*$ as in Eq.~\eqref{eq:cstar}.

Two special cases follow directly.  At even integers $q=2,4,\dots$ the
hypergeometric series terminates and $c_q=0$, since $1/\Gamma(-q/2)=0$, while the zero of $\sin(\pi q/2)$ cancels the pole of $\Gamma(-q)$, so that
the expansion is purely analytic; for $q=2$ one recovers $d_2(k)=2\sin k$.  At
odd integers the non-analytic term is dressed by $\ln k$, which changes the
amplitude but not the exponent, except at $q=1$ where the analytic and
non-analytic powers coincide.  There the sum can be evaluated in closed form,
\begin{equation}\label{eq:app_d1}
    d_1(k)=\frac{4}{\pi}\sin\frac{k}{2}\ln\cot\frac{k}{4}
    \simeq\frac{2}{\pi}k\ln\frac{4}{k},
\end{equation}
consistent with the divergence of $a_q$ as $q\to1$: the pole of $a_q$ is
cancelled by the pole of $\Gamma(-q)$ in $c_q$, leaving the logarithm.

At $q=1$ the Landau-Zener exponent of Eq.~\eqref{eq:pk} therefore becomes
\begin{equation}\label{eq:app_pk_p1}
    p_k\simeq\exp\left[-\beta k^{2}\ln^{2}(4/k)\right],
    \qquad
    \beta=\frac{4\pi c_0^{2}\tau_Q}{\bigl(2+J_q\bigr)^{3}},
\end{equation}
with $c_0=2/\pi$ the amplitude of Eq.~\eqref{eq:app_d1}.
Writing $\mathcal L=\ln\tau_Q$, the integral in
Eq.~\eqref{eq:nex} is dominated by $k\lesssim k_*\sim\tau_Q^{-1/2}/\mathcal L$, and expanding the
logarithm about $k_*$ yields
\begin{equation}\label{eq:app_nex_p1}
    n_{\rm ex}\big|_{q=1}
    =\mathcal C\,\frac{\tau_Q^{-1/2}}{\mathcal L}
    -2\mathcal C\,\frac{\tau_Q^{-1/2}\ln\mathcal L}{\mathcal L^{2}}
    +\mathcal O\!\left(\frac{\tau_Q^{-1/2}}{\mathcal L^{2}}\right),
\end{equation}
with $\mathcal C=(2+J_q)^{3/2}/(2\pi\,c_0)\simeq1.481$.  The
Kibble-Zurek power $\tau_Q^{-1/2}$ is retained, dressed by the logarithmic
suppression, which makes the effective
finite-time exponent approach $1/2$ from above.

Finally, at $g=g_c^{\rm FF}$ of Eq.~\eqref{eq:gc} the kinetic term reduces to
\begin{equation}\label{eq:app_xi_gc}
    \xi_k(g_c^{\rm FF})=\bigl(1-g_c^{\rm FF}\bigr)\left\lvert2\sin\frac k2\right\rvert^{q}
    \simeq\frac{2}{2+J_q}\,k^{q},
\end{equation}
with the constant part canceling by the definition of $g_c^{\rm FF}$, while the pairing is
$\Delta_k(g_c^{\rm FF})\simeq2c_*k^{q_*}/(2+J_q)$.  Both appear in the
quasiparticle energy squared,
\begin{equation}\label{eq:app_eps_gc}
    \epsilon_k^{2}(g_c^{\rm FF})=\xi_k^{2}+\Delta_k^{2}\propto k^{2q}+k^{2q_*},
\end{equation}
so that $\epsilon_k(g_c^{\rm FF})\propto k^{q_*}$ and hence $z=\min(q,1)$, up to the
marginal logarithm at $q=1$.

As a check of the normalization, at $q=2$ one has $J_2=2$ and hence
$g_c^{\rm FF}=1/2$, the exact critical point of the transverse-field Ising chain, while
$c_*=a_2=2$ agrees with $d_2(k)=2\sin k$.  Equation~\eqref{eq:pk} then gives
$\lambda_2=4\pi c_*^2/(2+J_2)^3=\pi/4$, so that $p_k=e^{-\pi\tau_Qk^2/4}$
and $n_{\rm ex}=1/(\pi\sqrt{\tau_Q})$.  We have further verified
Eq.~\eqref{eq:pk} against direct integration of the Bogoliubov-de Gennes
equations~\eqref{eq:EOM}, obtaining agreement to within two percent over
$0.5\leq q\leq3$ at the momenta that dominate the defect density.

\section{Finite-size threshold for the low-momentum excitations}\label{app:finite_size}

In this section, we show how the time scale for the breakdown of the continuum approximation in the low-momentum regime changes with the fractional order. The continuum limit of Eq.~\eqref{eq:nex} requires a sufficient number of modes below the Landau-Zener momentum $k_*=(\lambda_q\tau_Q)^{-1/(2q_*)}$ for which the integral approximations with a finite range can be applied, where $\lambda_q=4\pi c_*^2/(2+J_q)^3$ and $q_*=\min(q,1)$. Equating $k_*$ with the lowest momentum $\pi/L$ gives the finite-size time scale
\begin{equation}\label{eq:app_fs_tauFS}
    \tau_{\rm FS}=\frac{1}{\lambda_q}\left(\frac{L}{\pi}\right)^{2q_*},
\end{equation}
beyond which only the pair $k=\pi/L$ is excited appreciably. The continuum limit persists, and the discreteness of the momenta is not resolved, as long as the time scale for the onset of adiabaticity, $\tau_\mathrm{ad}$, is shorter than $\tau_\mathrm{FS}$. The former is determined by $N_{\rm ex}=L\mathcal N_q\tau_Q^{-1/(2q_*)}\lesssim2$, i.e., when the number of kink pairs drops below unity, where $\mathcal N_q=\Gamma\bigl(1+\frac{1}{2q_*}\bigr)/\bigl(\pi\lambda_q^{1/(2q_*)}\bigr)$, and which yields in the continuum limit
\begin{equation}\label{eq:app_fs_tau2}
    \tau_\mathrm{ad}=\frac{1}{\lambda_q}\left[\frac{L\,\Gamma\bigl(1+\frac{1}{2q_*}\bigr)}{2\pi}\right]^{2q_*}.
\end{equation}
Both time scales grow as $(L/\pi)^{2q_*}/\lambda_q$, so that their ratio is independent of the system size,
\begin{equation}\label{eq:app_fs_ratio}
    \frac{\tau_\mathrm{ad}}{\tau_{\rm FS}}=\left[\frac{\Gamma\bigl(1+\frac{1}{2q_*}\bigr)}{2}\right]^{2q_*}.
\end{equation}
This function decreases monotonically, becomes unity at $q=1/4$ and saturates to the constant $\pi/16$ for $q\geq1$, where $\tau_\mathrm{ad}\simeq0.2\,\tau_\mathrm{FS}$. For $q>1/4$, the kink pair number thus drops below unity before the discreteness of the momenta becomes visible, $\tau_\mathrm{ad}<\tau_\mathrm{FS}$, while for $q<1/4$ the discreteness is resolved first, $\tau_\mathrm{FS}<\tau_\mathrm{ad}$. As the ratio does not depend on $L$, this ordering holds for every system size, and for the fractional orders $q\geq0.1$ the two time scales differ at most by the factor $16/\pi\simeq5$.
For $L=150$, Eqs.~\eqref{eq:app_fs_tauFS} and~\eqref{eq:app_fs_tau2} give $\tau_{\rm FS}\simeq1.5\times10^3$ and $\tau_\mathrm{ad}\simeq2.9\times10^2$ at $q=1.5$, and $\tau_{\rm FS}\simeq3.4\times10^3$ and $\tau_\mathrm{ad}\simeq6.6\times10^2$ at $q=2.25$, while $\tau_{\rm FS}\simeq90$--$190$ for $0.1\leq q\leq0.5$. Comparing the discrete Landau-Zener sum with the corresponding momentum integral at $L=150$, we find deviations below $5\%$ whenever more than one kink pair is excited. For $q<1/4$, where the slowest ramps resolve the discreteness, the modes near $k=\pi$ dominate, and their excitation probability varies with $k$ on a scale of order one, so that their contribution is unaffected (App.~\ref{app:noncrossing}).

\section{Adiabatic perturbation theory for high-momentum modes at small $q$}\label{app:noncrossing}

In this section, we derive the leading-order formula, $p_k\simeq\frac{d_q^2(k)}{4\tau_Q^2\,\epsilon_k^{6}(g=0)}$, for the excitation probability of high-momentum modes in which $\xi_k$ in Eq.~\eqref{eq:Hk} does not change sign during the ramp. As $\xi_k=2g-(1-g)j_q(k)$ decreases from $2$ at $g=1$ to $-j_q(k)$ at $g=0$, it changes sign only for $j_q(k)=J_q-|2\sin(k/2)|^q>0$. As a result, for momenta $k_c<k<\pi$ above the threshold defined by
\begin{equation}\label{eq:app_kc}
    2\sin\frac{k_c}{2}=J_q^{1/q},
\end{equation}
no avoided crossing emerges, and the Landau-Zener form of $p_k=\exp\left[-\frac{4\pi\tau_Q\,d_q^2(k)}{\bigl(2+j_q(k)\bigr)^3}\right]$ does not hold. 
However, this regime can still allow for complementary excitations, most dominantly around $k=\pi$, where the gap is the smallest at $g=0$. Here $d_q(\pi)=0$ and $\epsilon_\pi(g=0)=|j_q(\pi)|=2^q-J_q\simeq q\ln2$ for small $q$, allowing for an adiabatic perturbation theory around that point.
Each pair of modes $(k,-k)$ evolves independently under the momentum-space Hamiltonian $\hat\psi_k^\dagger H_k(t)\hat\psi_k$, with $\hat\psi_k=(\hat c_k,\hat c_{-k}^\dagger)^T$ containing the creation and annihilation operators, and $H_k(t)=\xi_k(t)\tau^z+\Delta_k(t)\tau^x$ with $\tau^{x,z}$ denoting the Pauli operators within the $k$-th mode. Starting from the fermionic vacuum of the two modes, $\ket{0}_k$, the dynamics remains in the subspace spanned by $\ket{0}_k$ and the doubly occupied state $\ket{2}_k\equiv\hat c_k^\dagger\hat c_{-k}^\dagger\ket{0}_k$.
The corresponding instantaneous ground and excited states of the $k$-th mode are given by
\begin{eqnarray}\label{eq:app_apt_states}
    \ket{\Omega_k(t)}&=&\cos\vartheta_k\ket{0}_k-\sin\vartheta_k\ket{2}_k,\nonumber\\
    \ket{\Omega^+_k(t)}&=&\sin\vartheta_k\ket{0}_k+\cos\vartheta_k\ket{2}_k ,
\end{eqnarray}
with the energies $\mp\epsilon_k$ and $\vartheta_k(t)$ of Eq.~\eqref{eq:theta}. The time-evolved state of Eq.~\eqref{eq:EOM} is written as
\begin{equation}\label{eq:app_apt_ansatz}
    \ket{\Psi_k(t)}=a_k(t)\,e^{i\varphi_k(t)}\ket{\Omega_k(t)}+b_k(t)\,e^{-i\varphi_k(t)}\ket{\Omega^+_k(t)},
\end{equation}
with $\varphi_k(t)=\int_0^t\epsilon_k(t')\,dt'$, $a_k(0)=1$, and $b_k(0)=0$, since at $g=1$ one has $\vartheta_k=0$ and $\ket{\Omega_k}=\ket{0}_k$. Using that $\partial_t\ket{\Omega_k}=-\dot\vartheta_k\ket{\Omega^+_k}$ and $\partial_t\ket{\Omega^+_k}=\dot\vartheta_k\ket{\Omega_k}$, the Schr\"odinger equation translates to
\begin{equation}\label{eq:app_apt_eom}
    \dot a_k=-\dot\vartheta_k\,e^{-2i\varphi_k}\,b_k,\qquad
    \dot b_k=\dot\vartheta_k\,e^{2i\varphi_k}\,a_k .
\end{equation}
To the leading order in $\tau_Q$ one finds $\dot\vartheta_k\propto\tau_Q^{-1}$ and $a_k\simeq1$, and integrating by parts with $e^{2i\varphi_k}=(2i\epsilon_k)^{-1}\,d e^{2i\varphi_k}/dt$, one finds
\begin{eqnarray}\label{eq:app_apt_b}
    b_k(\tau_Q)&=&\int_0^{\tau_Q}\dot\vartheta_k\,e^{2i\varphi_k}\,dt\\
    &=&\left[\frac{\dot\vartheta_k}{2i\epsilon_k}\,e^{2i\varphi_k}\right]_0^{\tau_Q}
    -\int_0^{\tau_Q}\frac{d}{dt}\left(\frac{\dot\vartheta_k}{2i\epsilon_k}\right)e^{2i\varphi_k}\,dt,\nonumber
\end{eqnarray}
where the remaining integral is of order $\tau_Q^{-2}$, so that the boundary terms dominate. Using that $\tan2\vartheta_k=(1-g)d_q(k)/[2g-(1-g)j_q(k)]$ and $\dot g=-1/\tau_Q$, one has $\xi_k=2$, $\Delta_k=0$ and $d(\tan2\vartheta_k)/dg=-d_q(k)/2$ at $g=1$, while $d(\tan2\vartheta_k)/dg=-2d_q(k)/j_q^2(k)$ and $\cos^22\vartheta_k=j_q^2(k)/\epsilon_k^2(g=0)$ at $g=0$. By these relations the angle derivatives at the endpoints become up to the leading order
\begin{equation}\label{eq:app_apt_thetadot}
    \dot\vartheta_k\big|_{g=1}=\frac{d_q(k)}{4\tau_Q},\qquad
    \dot\vartheta_k\big|_{g=0}=\frac{d_q(k)}{\tau_Q\,\epsilon_k^2(g=0)},
\end{equation}
which yields for the excitation probability
\begin{equation}\label{eq:app_apt_pk_full}
    p_k=|b_k(\tau_Q)|^2\simeq\frac{d_q^2(k)}{4\tau_Q^2}\left|\frac{e^{2i\varphi_k(\tau_Q)}}{\epsilon_k^3(g=0)}-\frac18\right|^2 .
\end{equation}
As for small $q$ the first term in the absolute value square dominates through the leading order expression $\epsilon_k(g=0)\propto q$, one finds
\begin{equation}\label{eq:app_apt_pk}
    p_k\simeq\frac{d_q^2(k)}{4\tau_Q^2\,\epsilon_k^{6}(g=0)} .
\end{equation}
Neglecting the remaining integral in Eq.~\eqref{eq:app_apt_b} requires the gap to vary slowly on the time scale $1/\epsilon_k(g=0)$ near the end of the ramp, i.e., $\tau_Q\,\epsilon_k^2(g=0)\gg1$, which near $k=\pi$ reads $\tau_Q\gg(2^q-J_q)^{-2}\simeq(q\ln2)^{-2}$ in the leading order for small $q$.

Summing Eq.~\eqref{eq:app_apt_pk} over the modes of Eq.~\eqref{eq:app_kc} in the continuum limit, modes above the threshold $k_c<k<\pi$ give
\begin{equation}\label{eq:app_apt_Nnc}
    N_{\rm nc}=L\,B_q\,\tau_Q^{-2},\qquad
    B_q=\frac1\pi\int_{k_c}^{\pi}\frac{d_q^2(k)}{4\,\epsilon_k^{6}(g=0)}\,dk.
\end{equation}
First assuming that the continuum approximation can be applied for the low-momentum excitations, $N_{\rm ex}=\frac{L}{\pi}\int_0^\infty e^{-\lambda_q\tau_Qk^{2q_*}}dk=L\,\Gamma\bigl(1+\frac{1}{2q_*}\bigr)\big/\bigl[\pi(\lambda_q\tau_Q)^{1/(2q_*)}\bigr]$, the competition captured by the ratio of the two types of contributions reads
\begin{equation}\label{eq:app_apt_ratio}
    \frac{N_{\rm nc}}{N_{\rm ex}}=\frac{\pi\,B_q\,\lambda_q^{1/(2q_*)}}{\Gamma\!\left(1+\frac{1}{2q_*}\right)}\,\tau_Q^{\frac{1}{2q_*}-2}.
\end{equation}
As it grows with $\tau_Q$ for $q<1/4$ and decays for $q>1/4$, the low-momentum excitations dominate for $q>1/4$, while for $q<1/4$ the modes near $k=\pi$ dominate the excitations of slow ramps. For $q>1/4$, the onset of adiabaticity precedes the finite-size time scale, $\tau_\mathrm{ad}<\tau_\mathrm{FS}$, so that the continuum approximation applies up to $\tau_\mathrm{ad}$, as shown in more detail in App.~\ref{app:finite_size}.

In the complementary regime, where $\tau_Q\gg\tau_{\rm FS}$, the lowest mode determines the excitation number at leading order, $N_{\rm ex}\simeq2e^{-\tau_Q/\tau_{\rm FS}}$. At the same time, $N_{\rm nc}$ of Eq.~\eqref{eq:app_apt_Nnc} remains unchanged, as the excitation probability of Eq.~\eqref{eq:app_apt_pk} varies with $k$ on a scale of order one. Introducing $x=\tau_Q/\tau_{\rm FS}$, the ratio becomes
\begin{equation}\label{eq:app_apt_ratio_single}
    \frac{N_{\rm nc}}{N_{\rm ex}}=\frac{L\,B_q}{2\,\tau_Q^2}\,e^{\tau_Q/\tau_{\rm FS}}
    =\frac{L\,B_q}{2\,\tau_{\rm FS}^2}\,\frac{e^{x}}{x^2},
\end{equation}
which grows exponentially with $\tau_Q$ for every $q$, so that the modes near $k=\pi$ dominate for any $q$ once $x$ exceeds the crossover $x_*$, given by $e^{x_*}/x_*^2=2\tau_{\rm FS}^2/(L B_q)$. For small $q$ the couplings behave as $J_{r,q}\simeq q/(2r)$, so that $j_q(k)\simeq-q\ln\bigl(2\sin\frac k2\bigr)$, $d_q(k)\simeq\frac q2(\pi-k)$, $J_q\simeq1$, $c_q\simeq\pi q/2$ and $k_c\simeq\pi/3$. By this, at leading order one finds the expressions $\lambda_q\simeq\pi^3q^2/27$, $\tau_{\rm FS}\simeq\frac{27}{\pi^3q^2}\left(\frac L\pi\right)^{2q}$ and
\begin{equation}\label{eq:app_Bq_smallq}
    B_q\!\simeq\!\frac{I_0}{4\pi q^4},\,\,
    I_0\!=\!\int_{\pi/3}^{\pi}\frac{\left(\frac{\pi-k}{2}\right)^2}{\Bigl[\ln^2\!\bigl(2\sin\frac k2\bigr)+\left(\frac{\pi-k}{2}\right)^2\Bigr]^3}\,dk\simeq2.10 .
\end{equation}

Reassuringly, in this regime $\tau_Q\gg\tau_\mathrm{FS}$ implies immediately the original $\tau_Q\gg q^{-2}$ condition.
The factor $q^{-4}$ of $B_q$ cancels against $\tau_{\rm FS}^{-2}\propto q^4$, and Eq.~\eqref{eq:app_apt_ratio_single} becomes
\begin{equation}\label{eq:app_apt_ratio_smallq}
    \frac{N_{\rm nc}}{N_{\rm ex}}\simeq\frac{\pi^5I_0}{5832}\,\pi^{4q}L^{1-4q}\,\frac{e^{x}}{x^2},\qquad \frac{\pi^5I_0}{5832}\simeq0.11 .
\end{equation}
The modes near $k=\pi$ therefore dominate the excitations for $q\lesssim\ln\bigl(0.11\,L\,e^{x}/x^2\bigr)/[4\ln(L/\pi)]$, which for $x\simeq1$ becomes $q\lesssim\ln(0.30\,L)/[4\ln(L/\pi)]$, i.e., $q\lesssim0.246$ for $L=150$, approaching $q=1/4$ for $L\to\infty$, in agreement with the threshold of the continuum limit, Eq.~\eqref{eq:app_apt_ratio}. 

\section{Kink number cumulants and their relation to excitations}\label{app:kinks}

In this section, we provide the analytical details for the kink number cumulants and show that, at leading order in $\tau_Q$, the cumulants of the kink number measured from its value in the final ground state become identical to the cumulants of the excitation number of Sec.~\ref{sec:defects} whenever the low-momentum excitations dominate, while close to $k=\pi$ the two differ by a momentum-dependent weight in the mean and by the overlap with the ground-state kinks in the higher cumulants. With periodic boundary conditions the kink number operator reads
\begin{equation}\label{eq:app_Nring}
    \hat N=\frac12\sum_{i=1}^{L}\left(1-\hat\sigma^x_i\hat\sigma^x_{i+1}\right),\qquad \hat\sigma^x_{L+1}\equiv\hat\sigma^x_1 .
\end{equation}
The Jordan-Wigner transformation
\begin{equation}\label{eq:app_JW}
    \hat\sigma_i^z=1-2\hat c_i^\dagger\hat c_i,\qquad
    \hat\sigma_i^x=\prod_{j<i}\bigl(1-2\hat c_j^\dagger\hat c_j\bigr)\bigl(\hat c_i^\dagger+\hat c_i\bigr),
\end{equation}
together with $(\hat c_i^\dagger+\hat c_i)(1-2\hat c_i^\dagger\hat c_i)=\hat c_i^\dagger-\hat c_i$, maps the nearest-neighbor bonds $i=1,\dots,L-1$ exactly, without any truncation, to
\begin{equation}\label{eq:app_JW_bond}
    \hat\sigma^x_i\hat\sigma^x_{i+1}
    =\hat c_i^\dagger\hat c_{i+1}+\hat c_{i+1}^\dagger\hat c_i+\hat c_i^\dagger\hat c_{i+1}^\dagger+\hat c_{i+1}\hat c_i .
\end{equation}
For the closing bond, which carries the string of all other sites, using $(1-2\hat c_L^\dagger\hat c_L)(\hat c_L^\dagger+\hat c_L)=\hat c_L-\hat c_L^\dagger$ one finds
\begin{equation}\label{eq:app_JW_closing}
    \hat\sigma^x_L\hat\sigma^x_1=-\hat P\bigl(\hat c_L^\dagger-\hat c_L\bigr)\bigl(\hat c_1^\dagger+\hat c_1\bigr),\qquad
    \hat P=\prod_{j=1}^{L}\bigl(1-2\hat c_j^\dagger\hat c_j\bigr),
\end{equation}
where $\hat P$ is the fermion parity. The Hamiltonian conserves $\hat P$, and the ramp starts from the paramagnet, so the state stays in the even-parity sector $\hat P=1$, translating the original periodic boundary condition to an antiperiodic one for the fermionic operators, $\hat c_{L+1}=-\hat c_1$. The kink-number operator in terms of the fermionic operators thus becomes
\begin{equation}\label{eq:app_N_fermions}
    \hat N=\frac L2-\frac12\sum_{i=1}^{L}\left(\hat c_i^\dagger\hat c_{i+1}+\hat c_{i+1}^\dagger\hat c_i+\hat c_i^\dagger\hat c_{i+1}^\dagger+\hat c_{i+1}\hat c_i\right).
\end{equation} 
The fermionic operators are decomposed in Fourier space as
\begin{equation}\label{eq:app_fourier}
    \hat c_j=\frac{e^{i\pi/4}}{\sqrt L}\sum_k e^{-ijk}\,\hat c_k,\qquad k=\frac{\pi(2m-1)}{L},
\end{equation}
with $m=-\frac L2+1,\dots,\frac L2$, which automatically satisfies the antiperiodic boundary condition. Inserting Eq.~\eqref{eq:app_fourier} into the four terms of Eq.~\eqref{eq:app_N_fermions}, using $\frac1L\sum_{j=1}^{L}e^{ij(k-k')}=\delta_{kk'}$, grouping the contributions of $k$ and $-k$ for $k>0$, and using $\hat c_{-k}^\dagger\hat c_k^\dagger=-\hat c_k^\dagger\hat c_{-k}^\dagger$, one obtains
\begin{equation}\label{eq:app_N_kpairs}
\begin{split}
    \sum_{i=1}^{L}\hat\sigma^x_i\hat\sigma^x_{i+1}=\sum_{k>0}\Bigl[&2\cos k\left(\hat c_k^\dagger\hat c_k+\hat c_{-k}^\dagger\hat c_{-k}\right)\\
    &-2\sin k\left(\hat c_k^\dagger\hat c_{-k}^\dagger+\hat c_{-k}\hat c_k\right)\Bigr].
\end{split}
\end{equation}
With $\hat\psi_k=(\hat c_k,\hat c_{-k}^\dagger)^T$ containing the creation and annihilation operators, the kink number operator becomes
\begin{equation}\label{eq:app_Nk}
    \hat N=\frac L2+\frac12\sum_{k>0}\hat\psi_k^\dagger\bigl(-2\cos k\,\tau^z+2\sin k\,\tau^x\bigr)\hat\psi_k=\sum_{k>0}\hat n_k,
\end{equation}
with $\hat n_k=1+\hat\psi_k^\dagger\bigl(-\cos k\,\tau^z+\sin k\,\tau^x\bigr)\hat\psi_k$ and with $\tau^{x,z}$ denoting the Pauli operators in the two-level systems.
At $q=2$, where $J_{r,2}=\delta_{r,1}$, $j_2(k)=2\cos k$ and $d_2(k)=2\sin k$, Eq.~\eqref{eq:app_Nk} reduces to the transverse-field Ising result, $\hat N=\frac L2+\frac12\hat H_{\rm FF}(g=0)$. The dynamics of each pair of modes takes place in the subspace spanned by $\ket{0}_k$ and $\ket{2}_k\equiv\hat c_k^\dagger\hat c_{-k}^\dagger\ket{0}_k$, where the time-evolved state is written as $\ket{\Psi_k}=u_k\ket{0}_k+v_k\ket{2}_k$ of Eq.~\eqref{eq:EOM} with initial condition $u_k(0)=1,\,v_k(0)=0$. In this subspace $\hat n_k$ has the eigenvalues $0$ and $2$, corresponding to no kink and a kink pair, respectively. The corresponding kink probability in the $k$-th mode is then given by
\begin{equation}\label{eq:app_ptilde}
    \tilde p_k=\frac12\langle \Psi_k|\hat n_k|\Psi_k\rangle=\Bigl|\cos\tfrac k2\,u_k+\sin\tfrac k2\,v_k\Bigr|^2 .
\end{equation}
For relating it to the excitation probability, we use again the instantaneous basis of $H_k$ at the end of the ramp, $g=0$, Eq.~\eqref{eq:app_apt_states},
\begin{eqnarray}\label{eq:app_eigen_g0}
    \ket{\Omega_k}&=&\cos\vartheta_k\ket{0}_k-\sin\vartheta_k\ket{2}_k,\nonumber\\
    \ket{\Omega^+_k}&=&\sin\vartheta_k\ket{0}_k+\cos\vartheta_k\ket{2}_k ,
\end{eqnarray}
where, by Eqs.~\eqref{eq:xiDelta} and~\eqref{eq:theta} with $\xi_k=-j_q(k)$ and $\Delta_k=d_q(k)$ at $g=0$, the angle $\vartheta_k$ is given by $\tan2\vartheta_k=-d_q(k)/j_q(k)$, or equivalently $\cos2\vartheta_k=-j_q(k)/\epsilon_k(g=0)$ and $\sin2\vartheta_k=d_q(k)/\epsilon_k(g=0)$ with $\epsilon_k(g=0)=\sqrt{j_q^2(k)+d_q^2(k)}$. The final state thus can be expressed as
\begin{equation}\label{eq:app_final_state}
    \ket{\Psi_k}=\alpha_k\ket{\Omega_k}+\beta_k\ket{\Omega^+_k}.
\end{equation}
Comparing with $\ket{\Psi_k}=u_k\ket{0}_k+v_k\ket{2}_k$, the two sets of amplitudes are related by
\begin{equation}\label{eq:app_uv_ab}
    u_k=\alpha_k\cos\vartheta_k+\beta_k\sin\vartheta_k,\quad
    v_k=-\alpha_k\sin\vartheta_k+\beta_k\cos\vartheta_k .
\end{equation}
Inserting Eq.~\eqref{eq:app_uv_ab} into Eq.~\eqref{eq:app_ptilde} and using the angle addition formulas gives
\begin{equation}\label{eq:app_amplitude}
    \cos\tfrac k2\,u_k+\sin\tfrac k2\,v_k=\alpha_k\cos\Bigl(\vartheta_k+\frac k2\Bigr)+\beta_k\sin\Bigl(\vartheta_k+\frac k2\Bigr).
\end{equation}
Measuring $\vartheta_k$ from its transverse-field Ising value $(\pi-k)/2$ through the angle
\begin{equation}\label{eq:app_delta}
    \delta_k=\vartheta_k-\frac{\pi-k}{2},\qquad
    \cos2\delta_k=\frac{j_q(k)\cos k+d_q(k)\sin k}{\epsilon_k(g=0)},
\end{equation}
one has $\cos(\vartheta_k+k/2)=-\sin\delta_k$ and $\sin(\vartheta_k+k/2)=\cos\delta_k$, so that the kink probability becomes
\begin{equation}\label{eq:app_ptilde_ab}
    \tilde p_k=\bigl|\beta_k\cos\delta_k-\alpha_k\sin\delta_k\bigr|^2 .
\end{equation}
The squared amplitude of the excited state is the excitation probability of Sec.~\ref{sec:defects}, $|\beta_k|^2=p_k$, while $|\alpha_k|^2=1-p_k$ and $\alpha_k^*\beta_k=\sqrt{p_k(1-p_k)}\,e^{i\Phi_k}$, with $\Phi_k$ the relative phase of $\alpha_k$ and $\beta_k$. Expanding Eq.~\eqref{eq:app_ptilde_ab} then relates the kink probability to the excitation probability,
\begin{equation}\label{eq:app_ptilde_split}
    \tilde p_k=\sin^2\delta_k+p_k\cos2\delta_k-\sqrt{p_k(1-p_k)}\,\sin2\delta_k\cos\Phi_k .
\end{equation}
For small momenta the hopping and pairing dispersions behave as $j_q(k)\simeq J_q-k^q\simeq J_q>0$ and $d_q(k)\simeq c_*k^{q_*}>0$. Hence $\cos2\vartheta_k=-j_q(k)/\epsilon_k(g=0)<0$ and $\sin2\vartheta_k=d_q(k)/\epsilon_k(g=0)>0$, so that $2\vartheta_k$ lies in the second quadrant and $\tan2\vartheta_k=-d_q(k)/j_q(k)$ gives
\begin{equation}\label{eq:app_theta_lowk}
    2\vartheta_k=\pi-\arctan\frac{d_q(k)}{j_q(k)}\simeq\pi-\frac{c_*}{J_q}\,k^{q_*}.
\end{equation}
Using that $2\delta_k=2\vartheta_k-\pi+k$, one also finds
\begin{equation}\label{eq:app_delta_lowk}
    2\delta_k\simeq k-\frac{c_*}{J_q}\,k^{q_*}\xrightarrow{k\to0}0 .
\end{equation}
As a result, for the low-energy, Landau-Zener type excitations only the second term of Eq.~\eqref{eq:app_ptilde_split} contributes to the kinks above the ground state,
\begin{equation}\label{eq:app_ptilde_LO}
    \tilde p_k\simeq p_k+\mathcal O\bigl(k^{q_*}\bigr),
\end{equation}
where the correction stems from the interference term of Eq.~\eqref{eq:app_ptilde_split}, whose phase $\Phi_k$ varies rapidly with $k$, so that it averages out in the sum over modes. For the remaining modes with avoided crossings, where $p_k$ is exponentially small, Eq.~\eqref{eq:app_ptilde_split} instead gives $\tilde p_k=\sin^2\delta_k$, which is the kink probability of the ground state at $g=0$, while the excitations for momenta close to $k=\pi$ are treated separately below. Given that $p_k\sin^2\delta_k\simeq0$, at leading order for small $k$ and through the exponential suppression of $p_k$ for larger ones, the cumulant generating function of the kink number decomposes into the separate contributions of the ground-state kinks and the ones stemming from the excitations,
\begin{eqnarray}\label{eq:app_charf_split}
    \ln\langle e^{i\theta\hat N}\rangle&\simeq&\sum_{k>0}\ln\left[1+\bigl(e^{2i\theta}-1\bigr)\sin^2\delta_k\right]\nonumber\\
    &&+\sum_{k>0}\ln\left[1+\bigl(e^{2i\theta}-1\bigr)p_k\right].
\end{eqnarray}
Here the first sum gives the $\tau_Q$-independent cumulants of the kink number in the ground state at $g=0$, and the second sum is identical to Eq.~\eqref{eq:charf}. As a result, kinks relative to the ground state follow the same statistics as those of the excitations as long as they are limited to the low-energy region, i.e., for fractional orders $q>1/4$.

Next we consider the modes without an avoided crossing close to $k=\pi$ for small $q$, for which $j_q(k)<0$ keeps $\xi_k=2g-(1-g)j_q(k)$ positive throughout the ramp. The derivation proceeds analogously to the case of the excitations in App.~\ref{app:noncrossing}. Expanding the time-evolved state in the instantaneous eigenbasis of $H_k(t)$ as $\ket{\Psi_k}=a_ke^{i\varphi_k}\ket{\Omega_k}+b_ke^{-i\varphi_k}\ket{\Omega^+_k}$, with $\varphi_k(t)=\int_0^t\epsilon_k\,dt'$, $a_k(0)=1$ and $b_k(0)=0$, the Schr\"odinger equation yields $\dot b_k=\dot\vartheta_k\,e^{2i\varphi_k}a_k$. For slow ramps $a_k\simeq1$, and integrating twice by parts gives
\begin{eqnarray}\label{eq:app_pi_parts}
    b_k(\tau_Q)&=&\int_0^{\tau_Q}\dot\vartheta_k\,e^{2i\varphi_k}\,dt\\
    &=&\left[\left(-i\frac{\dot\vartheta_k}{2\epsilon_k}+\frac{1}{4\epsilon_k}\frac{d}{dt}\frac{\dot\vartheta_k}{\epsilon_k}\right)e^{2i\varphi_k}\right]_0^{\tau_Q}+\mathcal O\bigl(\tau_Q^{-3}\bigr).\nonumber
\end{eqnarray}
As each further integration by parts brings an additional factor $1/(\tau_Q\epsilon_k^2)$, the neglected terms remain small under the same condition as for the excitations, $\tau_Q\epsilon_\pi^2\gg1$, i.e., $\tau_Q\gg(q\ln2)^{-2}$ for small $q$.
Since $\Delta_k'\xi_k-\Delta_k\xi_k'=-2d_q(k)$ for every $g$, with $(\dots)'=\mathrm d(\dots)/\mathrm dg$, the angle derivative becomes $\dot\vartheta_k=d_q(k)/(\tau_Q\epsilon_k^2)$. The lower limit acquires the phase $e^{2i\varphi_k(\tau_Q)}$ relative to the upper one and averages out in the momentum sum. From the same general form of the time-evolved wave function in Eq.~\eqref{eq:app_final_state} at the end of the ramp, one has $\alpha_k=a_k(\tau_Q)e^{i\varphi_k(\tau_Q)}$ and $\beta_k=b_k(\tau_Q)e^{-i\varphi_k(\tau_Q)}$, so that evaluating the upper limit at $g=0$ leads to $\alpha_k\simeq e^{i\varphi_k(\tau_Q)}$ and
\begin{eqnarray}\label{eq:app_pi_ab}
    \beta_k&\simeq& e^{i\varphi_k(\tau_Q)}\bigl(-iX_k+Y_k\bigr),\qquad
    X_k=\frac{d_q(k)}{2\tau_Q\,\epsilon_k^3},\nonumber\\
    Y_k&=&-\frac{3\,d_q(k)\bigl[2j_q(k)+\epsilon_k^2\bigr]}{4\tau_Q^2\,\epsilon_k^6},
\end{eqnarray}
where $\epsilon_k\equiv\epsilon_k(g=0)$ from here on. The first term gives the excitation probability $p_k\simeq X_k^2=d_q^2(k)/(4\tau_Q^2\epsilon_k^6)$. Substituting both amplitudes into $\tilde p_k=|\beta_k\cos\delta_k-\alpha_k\sin\delta_k|^2$ and using $\sin2\delta_k=[j_q(k)\sin k-d_q(k)\cos k]/\epsilon_k$, one finds
\begin{eqnarray}\label{eq:app_pi_ptilde}
    \tilde p_k&\simeq&\sin^2\delta_k+w_k\,p_k,\\
    w_k&=&\cos^2\delta_k+\frac{3\bigl[2j_q(k)+\epsilon_k^2\bigr]\bigl[j_q(k)\sin k-d_q(k)\cos k\bigr]}{d_q(k)\,\epsilon_k}.\nonumber
\end{eqnarray}
In contrast to small momenta, the second-order amplitude $Y_k$ matters here. For $k=\pi-\kappa$ one has $d_q(k)\simeq D_q\kappa$ with $D_q=-d_q'(\pi)$, and $j_q(k)\simeq-\epsilon_\pi$ with $\epsilon_\pi=2^q-J_q$, so that both $2\delta_k\simeq\kappa(D_q/\epsilon_\pi-1)$ and $X_k\simeq D_q\kappa/(2\tau_Q\epsilon_\pi^3)$ are linear in $\kappa$, and the interference of $Y_k$ with $\sin\delta_k$ is of the same order $\kappa^2/\tau_Q^2$ as $p_k$. The weight then approaches
\begin{equation}\label{eq:app_pi_wpi}
    w_\pi=1+\frac{3(\epsilon_\pi-D_q)(2-\epsilon_\pi)}{D_q},
\end{equation}
which reaches $1+6(2\ln2-1)\simeq3.32$ as $q\to0$, where $D_q/\epsilon_\pi\to1/(2\ln2)$, and reduces to unity at $q=2$, where $D_2=\epsilon_\pi=2$. As a result, close to $k=\pi$ the mean number of kinks relative to the ground state decays as $\tau_Q^{-2}$, differing only by the $w_k$ prefactor from the mean excitation number.

For the higher cumulants, the overlap with the ground-state kinks no longer drops out. The kink number of each pair of modes takes the values $0$ and $2$ with probabilities $1-\tilde p_k$ and $\tilde p_k$, so that its first three cumulants are $2\tilde p_k$, $4\tilde p_k(1-\tilde p_k)$ and $8\tilde p_k(1-\tilde p_k)(1-2\tilde p_k)$. Expanding them to first order in $\tilde p_k-\sin^2\delta_k=w_kp_k$ around the ground-state value $\sin^2\delta_k$ gives
\begin{eqnarray}\label{eq:app_pi_cumulants}
    \kappa_1-\kappa^{\rm GS}_1&\simeq&2\sum_{k>0}w_kp_k,\qquad
    \kappa_2-\kappa^{\rm GS}_2\simeq4\sum_{k>0}w_kp_k\cos2\delta_k,\nonumber\\
    \kappa_3-\kappa^{\rm GS}_3&\simeq&8\sum_{k>0}w_kp_k\Bigl(1-\frac32\sin^22\delta_k\Bigr),
\end{eqnarray}
which corresponds to replacing the second sum of Eq.~\eqref{eq:app_charf_split} by $\sum_{k>0}(e^{2i\theta}-1)w_kp_k/[1+(e^{2i\theta}-1)\sin^2\delta_k]$. The angle $\delta_k$ vanishes at $k=\pi$ but reaches $\sin^2\delta_k\simeq0.07$ at $k_c$, which follows from Eq.~\eqref{eq:app_delta} with $j_q(k_c)=0$ and $k_c\to\pi/3$ for small $q$, $\sin^2\delta_{k_c}=(1-\sin k_c)/2\simeq(1-\sqrt3/2)/2$. As a result, the ratios $\tilde R_\ell$ of Sec.~\ref{sec:defects} increase monotonically from $\tilde R_2\simeq1.89$ and $\tilde R_3\simeq3.36$ for $q\to0$ to $\tilde R_2\simeq1.93$ and $\tilde R_3\simeq3.57$ at $q=1/4$, below the values $2$ and $4$ of the excitations. The terms quadratic in $w_kp_k$ add the corrections $-2\sum_kw_k^2p_k^2/\sum_kw_kp_k=-\tilde J_0/(2\tilde I_0q^4\tau_Q^2)$ and $-12\sum_kw_k^2p_k^2\cos2\delta_k/\sum_kw_kp_k\simeq-3\tilde J_0/(\tilde I_0q^4\tau_Q^2)$ to these ratios, with $\tilde I_0$ of Eq.~\eqref{eq:app_pi_Bq_smallq} and $\tilde J_0$ of Sec.~\ref{sec:defects}.

Summing over the modes in the continuum limit, $2\sum_{k>0}\to\frac{L}{\pi}\int dk$, the kinks above the ground state from the low-momentum region coincide with the excitations, $N_{\rm LZ}=L\,\mathcal N_q\,\tau_Q^{-1/(2q_*)}$ with $\mathcal N_q=\Gamma\bigl(1+\frac{1}{2q_*}\bigr)/\bigl(\pi\lambda_q^{1/(2q_*)}\bigr)$. The modes without an avoided crossing, $k_c<k\leq\pi$ with $2\sin(k_c/2)=J_q^{1/q}$, contribute in the regime $\tau_Q\epsilon_\pi^2\gg1$ as
\begin{equation}\label{eq:app_pi_Nnc}
    N_{\rm nc}=\frac{L}{\pi}\int_{k_c}^{\pi}w_k\,p_k\,dk=L\,\tilde B_q\,\tau_Q^{-2},\,\,
    \tilde B_q=\frac1\pi\int_{k_c}^{\pi}\frac{w_k\,d_q^2(k)}{4\,\epsilon_k^6}\,dk .
\end{equation}
Thus the competition between them is identical to that of the excitations, Eq.~\eqref{eq:app_apt_ratio}, up to a different, $\tau_Q$-independent prefactor,
\begin{equation}\label{eq:app_pi_ratio}
    \frac{N_{\rm nc}}{N_{\rm LZ}}=\frac{\tilde B_q}{\mathcal N_q}\,\tau_Q^{\frac{1}{2q_*}-2},
\end{equation}
growing with $\tau_Q$ for $q<1/4$ and decaying for $q>1/4$. As $w_k$ alters only the prefactor of the mean, slow ramps with $\tau_Q\gg(q\ln2)^{-2}$ produce kinks mainly near $k=\pi$ below the threshold $q=1/4$ found for the excitations. Similarly to the excitations, for $q>1/4$ the low-momentum kinks above the ground state follow the continuum limit up to the onset of adiabaticity, $\tau_\mathrm{ad}<\tau_\mathrm{FS}$, since at leading order they are identical to the excitations.

Finally, we also provide a cross-check in the complementary regime, where the slowest ramps resolve individual momenta, i.e., for $\tau_Q\gtrsim\tau_{\rm FS}=\lambda_q^{-1}(L/\pi)^{2q_*}$, where the Landau-Zener threshold momentum $(\lambda_q\tau_Q)^{-1/(2q_*)}$ drops below $\pi/L$. The derivation proceeds analogously to the case of the excitations with $N_{\rm LZ}\simeq2\,e^{-\tau_Q/\tau_{\rm FS}}$, as $\delta_k$ vanishes for $k\to0$. The kinks near $k=\pi$ remain unchanged, resulting in the same expression for the ratio governing the competition between the low- and high-momentum contributions as for the excitations, Eq.~\eqref{eq:app_apt_ratio_single}, up to the prefactor $\tilde B_q$,
\begin{equation}\label{eq:app_pi_ratio_single}
    \frac{N_{\rm nc}}{N_{\rm LZ}}=\frac{L\,\tilde B_q}{2\,\tau_{\rm FS}^2}\,\frac{e^{x}}{x^2},\,x=\tau_Q/\tau_\mathrm{FS}.
\end{equation}
The exponential growth in $x$ holds irrespective of $q$, and the kinks near the zone boundary take over beyond $x_*$, the solution of $e^{x_*}/x_*^2=2\tau_{\rm FS}^2/(L\tilde B_q)$. To make the $q$ dependence explicit, the small-$q$ forms $j_q(k)\simeq-q\ln\bigl(2\sin\frac k2\bigr)$ and $d_q(k)\simeq\frac q2(\pi-k)$, which follow from $J_{r,q}\simeq q/(2r)$, leave $w_k$ and $\delta_k$ independent of $q$ and yield
\begin{equation}\label{eq:app_pi_Bq_smallq}
    \tilde B_q\!\simeq\!\frac{\tilde I_0}{4\pi q^4},\,\,
    \tilde I_0\!=\!\int_{\pi/3}^{\pi}\frac{w_k\left(\frac{\pi-k}{2}\right)^2}{\Bigl[\ln^2\!\bigl(2\sin\frac k2\bigr)+\left(\frac{\pi-k}{2}\right)^2\Bigr]^3}\,dk\simeq4.89 .
\end{equation}
Together with $\tau_{\rm FS}\simeq\frac{27}{\pi^3q^2}\left(\frac L\pi\right)^{2q}$, the second-order expansion is automatically justified in this regime, as $\tau_Q\gg\tau_{\rm FS}\propto q^{-2}$. The dependence on $q^{-4}$ drops out of the ratio,
\begin{equation}\label{eq:app_pi_ratio_smallq}
    \frac{N_{\rm nc}}{N_{\rm LZ}}\simeq\frac{\pi^5\tilde I_0}{5832}\,\pi^{4q}L^{1-4q}\,\frac{e^{x}}{x^2},\qquad \frac{\pi^5\tilde I_0}{5832}\simeq0.26,
\end{equation}
so that the kinks near $k=\pi$ outnumber the low-momentum ones for $q<\ln\bigl(0.26\,L\,e^{x}/x^2\bigr)/[4\ln(L/\pi)]$. In agreement with the breakdown of the continuum approximation, $\tau_Q\gtrsim\tau_\mathrm{FS}$, i.e., $x\gtrsim1$, the bound becomes $q<\ln(0.70\,L)/[4\ln(L/\pi)]$, which converges for large $L$ to the value $q=1/4$ obtained in the continuum.

\section{Free-fermion bounds on the kink-number cumulants}\label{app:bound}

In this section we show that for every Gaussian state of fermions the cumulants of the kink number $\hat N$ obey the inequalities $\kappa_2<2\kappa_1$ and $|\kappa_3|\leq2\kappa_2$, provided that $\kappa_1>0$. Using the Jordan-Wigner representation of Eq.~\eqref{eq:xxJW}, which for $r=1$ contains no string, and expressing it in terms of the Majorana operators $\hat\eta_j=\hat c_j^\dagger+\hat c_j$ and $\hat\varphi_j=i(\hat c_j^\dagger-\hat c_j)$, one finds 
\begin{equation}
    \hat\sigma^x_j\hat\sigma^x_{j+1}=(\hat c_j^\dagger-\hat c_j)(\hat c_{j+1}^\dagger+\hat c_{j+1})=-i\hat\varphi_j\hat\eta_{j+1}.  
\end{equation}
Introducing the bond fermions
\begin{equation}\label{eq:bondf}
    \hat f_j=\tfrac12\bigl(\hat\varphi_j+i\hat\eta_{j+1}\bigr),\qquad j=1,\dots,L-1,
\end{equation}
the $j$-th term in the kink number operator, Eq.~\eqref{eq:Nhat}, can be expressed as
\begin{equation}\label{eq:kinkproj}
    \tfrac12\bigl(1-\hat\sigma^x_j\hat\sigma^x_{j+1}\bigr)=\hat f_j^\dagger\hat f_j,
    \qquad
    \hat N=\sum_{j=1}^{L-1}\hat f_j^\dagger\hat f_j .
\end{equation}
As the $\hat f_j$ are linear in the $\hat c_j$ and satisfy the canonical anticommutation relations, a Gaussian state of the $\hat c_j$ remains Gaussian in terms of the $\hat f_j$. As a result, the kink statistics is entirely encoded in the $(L-1)\times(L-1)$ correlation matrices $G_{jl}=\langle\hat f_l^\dagger\hat f_j\rangle$ and $F_{jl}=\langle\hat f_l\hat f_j\rangle$. As for any fermionic state it holds that $0\leq G\leq1$, and that the eigenvalues of the full correlation matrix of the $\hat f_j$ and $\hat f_j^\dagger$ lie between 0 and 1~\cite{BachLiebSolovej94}, one finds the following inequality
\begin{equation}\label{eq:BLS}
    FF^\dagger\leq G-G^2 .
\end{equation}
Within this setup Wick's theorem gives the first three cumulants as
\begin{align}\label{eq:wick}
    \kappa_1&=\Tr G,\qquad
    \kappa_2=\Tr\bigl(G-G^2\bigr)+\Tr\bigl(FF^\dagger\bigr),\\
    \kappa_3&=\Tr\bigl(G-3G^2+2G^3\bigr)\nonumber\\
    &\quad+3\Tr\bigl[(1-G)FF^\dagger\bigr]-3\Tr\bigl(FG^{T}F^\dagger\bigr).\label{eq:wick3}
\end{align}
Inserting Eq.~\eqref{eq:BLS} into $\kappa_2$ and using $\Tr G^2\geq(\Tr G)^2/(L-1)$, the variance is bounded by
\begin{equation}\label{eq:bound}
    \kappa_2\leq2\Tr\bigl(G-G^2\bigr)\leq2\kappa_1\left(1-\frac{\kappa_1}{L-1}\right)<2\kappa_1 .
\end{equation}
For the third cumulant, Eq.~\eqref{eq:BLS} together with $0\leq G\leq1$, which gives $0\leq\Tr(GFF^\dagger)\leq\Tr(FF^\dagger)$ and $0\leq\Tr(FG^TF^\dagger)\leq\Tr(FF^\dagger)$, leads to $2\kappa_2-\kappa_3\geq2\Tr G^2(1-G)\geq0$ and $2\kappa_2+\kappa_3\geq2\Tr G(1-G)^2\geq0$, so that $|\kappa_3|\leq2\kappa_2$. Without pairing, $F=0$, the bounds reduce to $\kappa_2\leq\kappa_1$ and $|\kappa_3|\leq\kappa_2$, which hold for independently occupied modes. Including pairing correlations as well, which create kink pairs in the bulk, the bounds are doubled, in agreement with the pair creation of Sec.~\ref{sec:defects}. As Eq.~\eqref{eq:BLS} holds for any fermionic state and Eqs.~\eqref{eq:wick} and~\eqref{eq:wick3} only require Wick's theorem, the bounds apply to pure and mixed Gaussian states for any couplings, boundary conditions and driving protocol, and in particular to the free-fermion approximation of Sec.~\ref{sec:JW}. As the derivation only uses that $\hat N$ is a sum of number operators of canonical fermions, the same bounds hold for the excitation number $\hat N_{\rm ex}$ of Sec.~\ref{sec:defects}, with $L$ modes in place of $L-1$. For the translation-invariant chain of Sec.~\ref{sec:defects}, where $\hat N$ is a sum of independent variables, analogous bounds hold for every order $\ell$ with constants that grow with $\ell$, while for general free-fermion states we have established them for $\ell=2,3$. Put differently, whenever the kink number variance exceeds twice the mean, or the magnitude of the third cumulant exceeds twice the variance, kink-kink interactions are present, and no free-fermion description can be applied.

\section{Free-fermion bounds on the kink-kink correlations and exact Gaussian limit}\label{app:corr_bound}

In this section we extend the bounds of App.~\ref{app:bound} to the connected kink-kink correlations $C_{ij}=\langle\hat n_i\hat n_j\rangle-\langle\hat n_i\rangle\langle\hat n_j\rangle$, with the kink projectors $\hat n_j=\hat f_j^\dagger\hat f_j$ of Eq.~\eqref{eq:kinkproj}, which enter the correlation function of Eq.~\eqref{eq:Ckink} as $C(r)=C_{L/2,L/2+r},\, r=-L/2+1,\dots,L/2-1$. Within the same setup of App.~\ref{app:bound}, Wick's theorem gives
\begin{equation}\label{eq:Cij_wick}
    C_{ij}=\delta_{ij}G_{ii}-|G_{ij}|^2+|F_{ij}|^2 .
\end{equation}
By this, the local and total variances read $C_{ii}=n_i(1-n_i)$ with $n_i=G_{ii}$, and $\kappa_2=\sum_{i,j}C_{ij}$, respectively, in agreement with Eq.~\eqref{eq:wick}. For a single pair of bonds $i\neq j$, the Cauchy-Schwarz inequality gives $|F_{ij}|^2\leq n_i(1-n_j)$ and $|G_{ij}|^2\leq\min[n_in_j,(1-n_i)(1-n_j)]$, and thus
\begin{eqnarray}\label{eq:Cij_pair}
    &&-\min\bigl[n_in_j,(1-n_i)(1-n_j)\bigr]\leq C_{ij}\nonumber\\
    &&\leq\min\bigl[n_i(1-n_j),n_j(1-n_i)\bigr].
\end{eqnarray}
These bounds, however, hold for any two commuting projectors in any state, since $\max(0,n_i+n_j-1)\leq\langle\hat n_i\hat n_j\rangle\leq\min(n_i,n_j)$, so that a single correlation cannot distinguish a free-fermion state from an interacting one.

The key part where the free-fermion structure becomes essential is the total correlation of a single bond. Summing Eq.~\eqref{eq:Cij_wick} over $j$ gives $\sum_jC_{ij}=(G-G^2)_{ii}+(FF^\dagger)_{ii}$. As Eq.~\eqref{eq:BLS} also holds for the diagonal elements, $(FF^\dagger)_{ii}\leq(G-G^2)_{ii}$, and as $(G^2)_{ii}=\sum_j|G_{ij}|^2\geq n_i^2$, one finds $0\leq\sum_jC_{ij}\leq2n_i(1-n_i)$, or equivalently
\begin{equation}\label{eq:rowsum}
    \Bigl|\sum_{j\neq i}C_{ij}\Bigr|\leq C_{ii}=n_i(1-n_i).
\end{equation}
Thus the total correlation of a bond with all other bonds cannot exceed its own variance, and summing Eq.~\eqref{eq:rowsum} over $i$ recovers Eq.~\eqref{eq:bound}. Far from the edges of the chain, where the kink density is approximately constant, Eq.~\eqref{eq:rowsum} reads $\bigl|\sum_{r\neq0}C(r)\bigr|\leq C(0)$ for the correlation function of Eq.~\eqref{eq:Ckink}. As a result, the kink correlations of any free-fermion state are summable for every system size, so that the variance grows at most linearly with $L$, which is the condition underlying the central limit theorem for weakly dependent variables~\cite{Newman80,Bolthausen82}. Regarding the higher-order cumulants, as $\hat N$ is quadratic in the fermionic operators, all its cumulants are traces of products of bounded correlation matrices, $|\kappa_\ell|\leq c_\ell L$. For a finite kink density, where $\kappa_2$ grows linearly with $L$, these higher-order cumulants vanish as $\kappa_\ell/\kappa_2^{\ell/2}\sim L^{1-\ell/2}$ for $\ell\geq3$, and $P(n)$ becomes Gaussian for $L\to\infty$. Conversely, in any state, a bond whose total correlation with all other bonds exceeds its own variance signals kink correlations beyond any free-fermion description, analogously to a variance exceeding twice the mean in App.~\ref{app:bound}.

\section{Finite-size scaling of the gap}\label{app:gapfits}

In this appendix, we provide further numerical details on the fitting parameters and their errors in the finite-size scaling of the minimal energy gap and its position $g_c(L)$ of Sec.~\ref{sec:gapscaling}. For each fractional order and system size, the minimum $\Delta_\mathrm{min}$ of the gap $\Delta=E_2-E_0$ was first bracketed on a coarse grid of field values around the free-fermion critical point $g_c^{\rm FF}$ of Eq.~\eqref{eq:gc} and then located by a golden-section search, in which the bracketing interval is reduced by the factor $(\sqrt5-1)/2$ in each step until its relative width falls below $2\times10^{-4}$. The minimum was then refined by computing the gap by DMRG on a uniform grid of 41 field values within $\pm10\%$ around it, and the position $g_c(L)$ and the value $\Delta(g_c,L)$ follow from a parabola through the three lowest points of this grid. The system sizes range from $L=20$ to $L=200$, as listed for each fractional order in Table~\ref{tab:gapfit}.

The scaling forms of Eqs.~\eqref{eq:gap_L} and~\eqref{eq:gc_L} are fitted by nonlinear least squares with all parameters free. The correction terms are retained only for $\omega,\omega'\geq0.2$ and when they remain smaller than the leading term for all sizes, since otherwise they become degenerate with the leading power law. For the drift of $g_c(L)$ we use one more test, in which the correction is kept only if it improves the fit enough to justify its two extra parameters. As a measure we use the Akaike information criterion~\cite{Akaike74} $n\ln(\mathrm{RSS}/n)+2k$, where $n$ is the number of system sizes, $k$ the number of fit parameters and $\mathrm{RSS}$ the sum of the squared residual errors. The correction is kept if it lowers this value by more than $2$, as smaller reductions are not significant~\cite{Akaike74}.

The uncertainties of $z$, $\nu$ and $\nu/(z\nu+1)$ are estimated in two ways. First, we repeat the fit $n$ times, each time leaving out one system size, and obtain the jackknife error~\cite{Efron82,Young15} $\sigma=\bigl[\frac{n-1}{n}\sum_{i=1}^{n}(\theta_i-\bar\theta)^2\bigr]^{1/2}$, where $\theta_i$ is the value of the parameter without the $i$-th size and $\bar\theta$ the mean of the $\theta_i$. Second, we repeat the fit without the one or two smallest sizes and take the change of the fitted values. We quote the larger of the two estimates.

A dash in Tables~\ref{tab:gapfit} and~\ref{tab:gcfit} marks a correction term that is not included, either because it is rejected by the above criteria or because it stays below $1\%$ of the leading term for all sizes, and the data then follow the leading power law alone.

\begin{figure*}[t]
    \centering
    \includegraphics[width=0.48\textwidth]{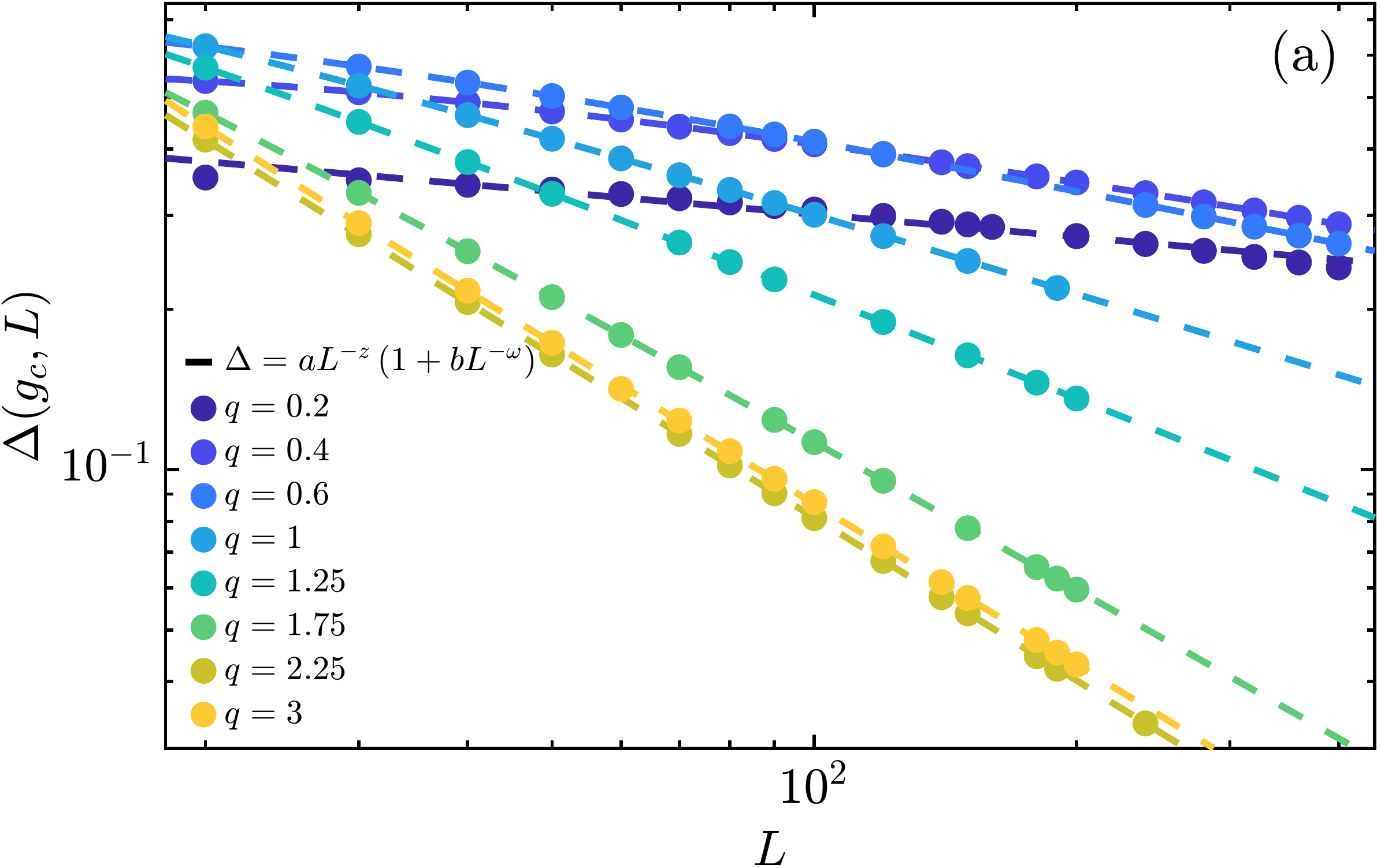}\hfill
    \includegraphics[width=0.48\textwidth]{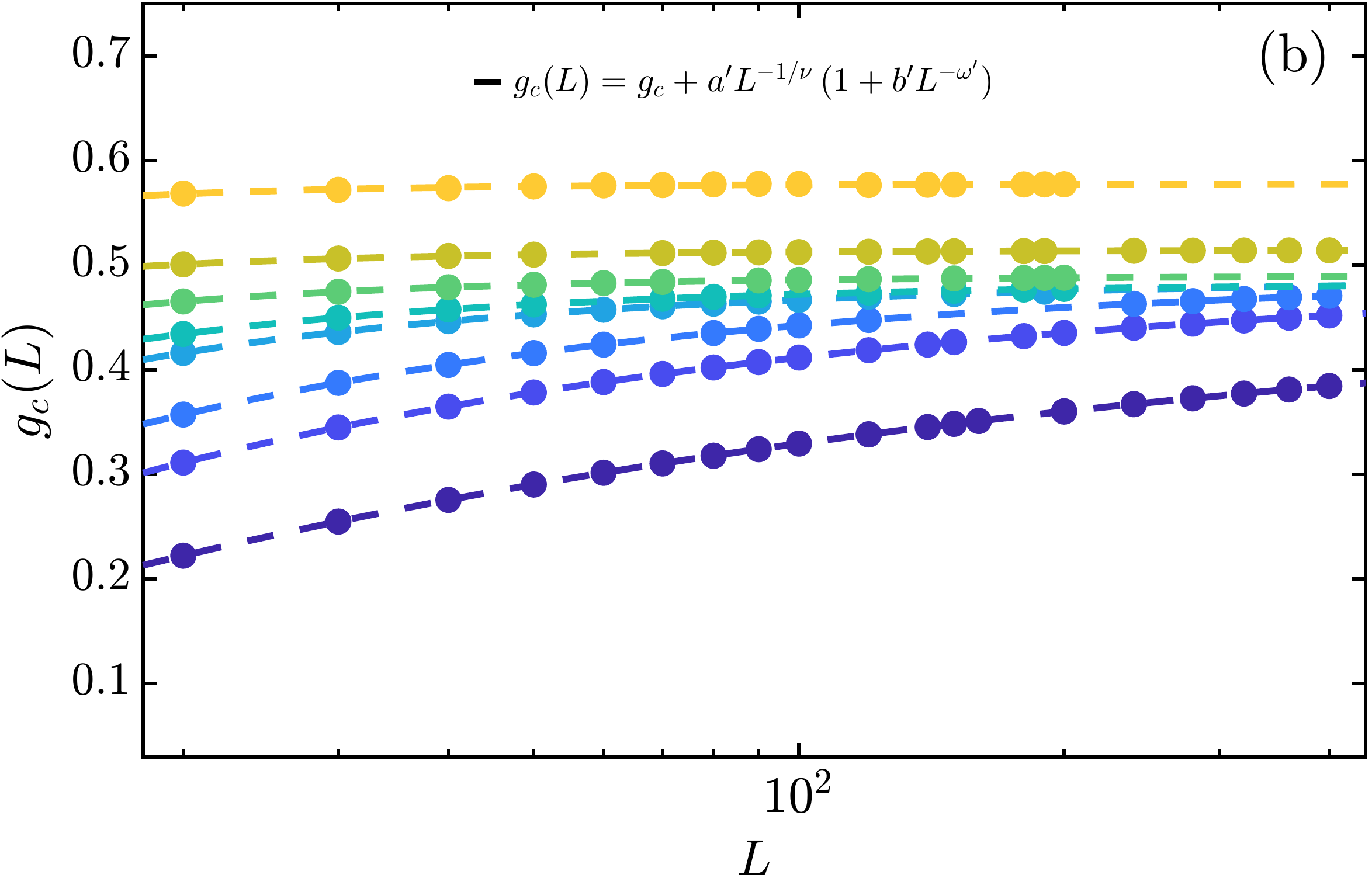}
    \caption{(a) Minimum gap $\Delta(g_c,L)$ and (b) its position $g_c(L)$ as functions of the system size for selected fractional orders. The dashed lines show the fits of Eqs.~\eqref{eq:gap_L} and~\eqref{eq:gc_L} with the parameters of Tables~\ref{tab:gapfit} and~\ref{tab:gcfit}.}
    \label{fig:gapfits}
\end{figure*}

The minimum gap and its position are shown for selected fractional orders in Fig.~\ref{fig:gapfits}, and the fitted parameters are listed in Tables~\ref{tab:gapfit} and~\ref{tab:gcfit}.

\begin{table*}[t]
\caption{Parameters of the fit of the minimum gap, Eq.~\eqref{eq:gap_L}, over the system sizes $L$. A dash marks a correction term that is not included.}
\label{tab:gapfit}
\begin{ruledtabular}
\begin{tabular}{cccccc}
$q$ & $L$ & $z$ & $a$ & $b$ & $\omega$ \\
\hline
$0.1$ & 20--200 & $0.101(24)$ & $0.299$ & -- & -- \\ 
$0.2$ & 20--200 & $0.141(25)$ & $0.578$ & -- & -- \\ 
$0.3$ & 20--190 & $0.177(22)$ & $0.828$ & -- & -- \\ 
$0.4$ & 20--200 & $0.3392(65)$ & $2.717$ & $-1.092$ & $0.29$ \\
$0.5$ & 20--120 & $0.3458(48)$ & $2.488$ & $-1.131$ & $0.41$ \\
$0.6$ & 20--120 & $0.3618(34)$ & $2.445$ & $-1.186$ & $0.52$ \\
$0.7$ & 20--200 & $0.3867(22)$ & $2.515$ & $-1.272$ & $0.64$ \\
$0.8$ & 20--160 & $0.4237(19)$ & $2.743$ & $-1.315$ & $0.71$ \\
$0.9$ & 20--190 & $0.471(18)$ & $3.07$ & $-1.28$ & $0.73$ \\
$1$ & 20--190 & $0.517(23)$ & $3.397$ & $-1.247$ & $0.74$ \\
$1.25$ & 20--200 & $0.6624(27)$ & $4.613$ & $-1.406$ & $0.87$ \\
$1.5$ & 20--200 & $0.8157(8)$ & $6.675$ & $-1.321$ & $0.82$ \\
$1.75$ & 20--200 & $0.9395(27)$ & $8.812$ & $-1.171$ & $0.78$ \\
$2.25$ & 20--190 & $1.0179(4)$ & $8.816$ & -- & -- \\
$2.5$ & 20--200 & $1.034(12)$ & $8.755$ & -- & -- \\
$2.75$ & 20--200 & $1.009(10)$ & $8.481$ & -- & -- \\
$3$ & 20--200 & $0.9300(51)$ & $6.69$ & $-1.25$ & $0.74$ \\
\end{tabular}
\end{ruledtabular}
\end{table*}

\begin{table*}[t]
\caption{Parameters of the fit of the position of the gap minimum, Eq.~\eqref{eq:gc_L}, over the system sizes of Table~\ref{tab:gapfit} and the resulting KZ exponent $\nu/(z\nu+1)$. The extrapolated critical fields $g_c$ are given in the convention of Eq.~\eqref{eq:H_Ising}, while $a'$ and $b'$ refer to the drift of $g_c(L)/[1-g_c(L)]$, the transverse field in units of the interaction used in the DMRG simulations. A dash marks a correction term that is not included.}
\label{tab:gcfit}
\begin{ruledtabular}
\begin{tabular}{ccccccc}
$q$ & $\nu$ & $a'$ & $b'$ & $\omega'$ & $g_c$ & $\nu/(z\nu+1)$ \\
\hline
$0.1$ & $6.87(21)$ & $-1.120$ & $-0.252$ & $0.73$ & $0.4604$ & $4.06(34)$ \\
$0.2$ & $4.0(16)$ & $-1.526$ & $-0.396$ & $0.62$ & $0.4887$ & $2.55(53)$ \\
$0.3$ & $2.843(27)$ & $-1.926$ & $-0.514$ & $0.53$ & $0.4948$ & $1.890(68)$ \\
$0.4$ & $2.240(11)$ & $-2.374$ & $-0.607$ & $0.47$ & $0.4955$ & $1.273(14)$ \\
$0.5$ & $1.895(27)$ & $-2.752$ & $-0.673$ & $0.45$ & $0.4942$ & $1.145(16)$ \\
$0.6$ & $1.6754(25)$ & $-3.012$ & $-0.721$ & $0.47$ & $0.4924$ & $1.0431(47)$ \\
$0.7$ & $1.5244(40)$ & $-3.167$ & $-0.756$ & $0.50$ & $0.4903$ & $0.9591(36)$ \\
$0.8$ & $1.415(16)$ & $-3.239$ & $-0.782$ & $0.54$ & $0.4884$ & $0.8845(48)$ \\
$0.9$ & $1.402(25)$ & $-2.95$ & $-0.80$ & $0.58$ & $0.4878$ & $0.844(16)$ \\
$1$ & $1.389(20)$ & $-2.058$ & -- & -- & $0.4874$ & $0.808(18)$ \\
$1.25$ & $1.1424(71)$ & $-2.318$ & -- & -- & $0.4833$ & $0.6503(23)$ \\
$1.5$ & $1.000(20)$ & $-2.440$ & -- & -- & $0.4844$ & $0.5508(59)$ \\
$1.75$ & $0.893(25)$ & $-2.563$ & -- & -- & $0.4897$ & $0.4856(65)$ \\
$2.25$ & $0.787(75)$ & $-2.495$ & -- & -- & $0.5142$ & $0.437(22)$ \\
$2.5$ & $1.02(9)$ & $-2.879$ & -- & -- & $0.5320$ & $0.496(22)$ \\
$2.75$ & $1.14(21)$ & $-0.840$ & -- & -- & $0.5562$ & $0.530(45)$ \\
$3$ & $0.835(34)$ & $-1.99$ & -- & -- & $0.5782$ & $0.470(15)$ \\
\end{tabular}
\end{ruledtabular}
\end{table*}

\section{Further results on the cumulants}\label{app:cumulants}
In this section we provide further results on the kink number cumulants. We first show the third cumulant and the cumulant ratios as functions of the driving time, and then the first three cumulants multiplied by the power of $\tau_Q$ expected for slow ramps, which shows directly over which range of driving times the analytical scaling holds.

\begin{figure*}[t]
    \centering
    \begin{tabular}{c  c  c c}\includegraphics[width=0.255\textwidth,trim={0 0 0 0cm},clip]{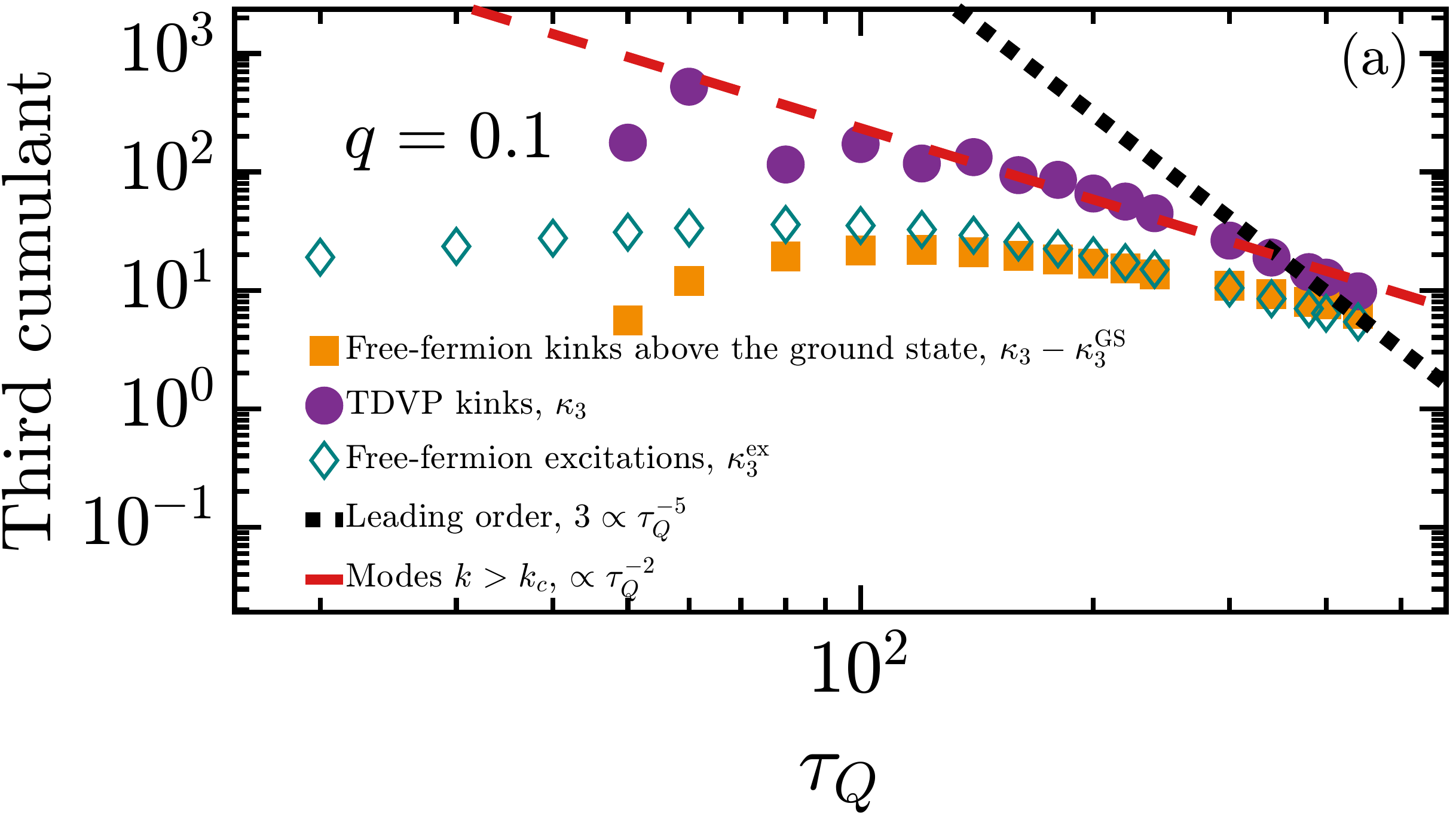} &
    \includegraphics[width=0.235\textwidth,trim={2.75cm 0 0 0},clip]{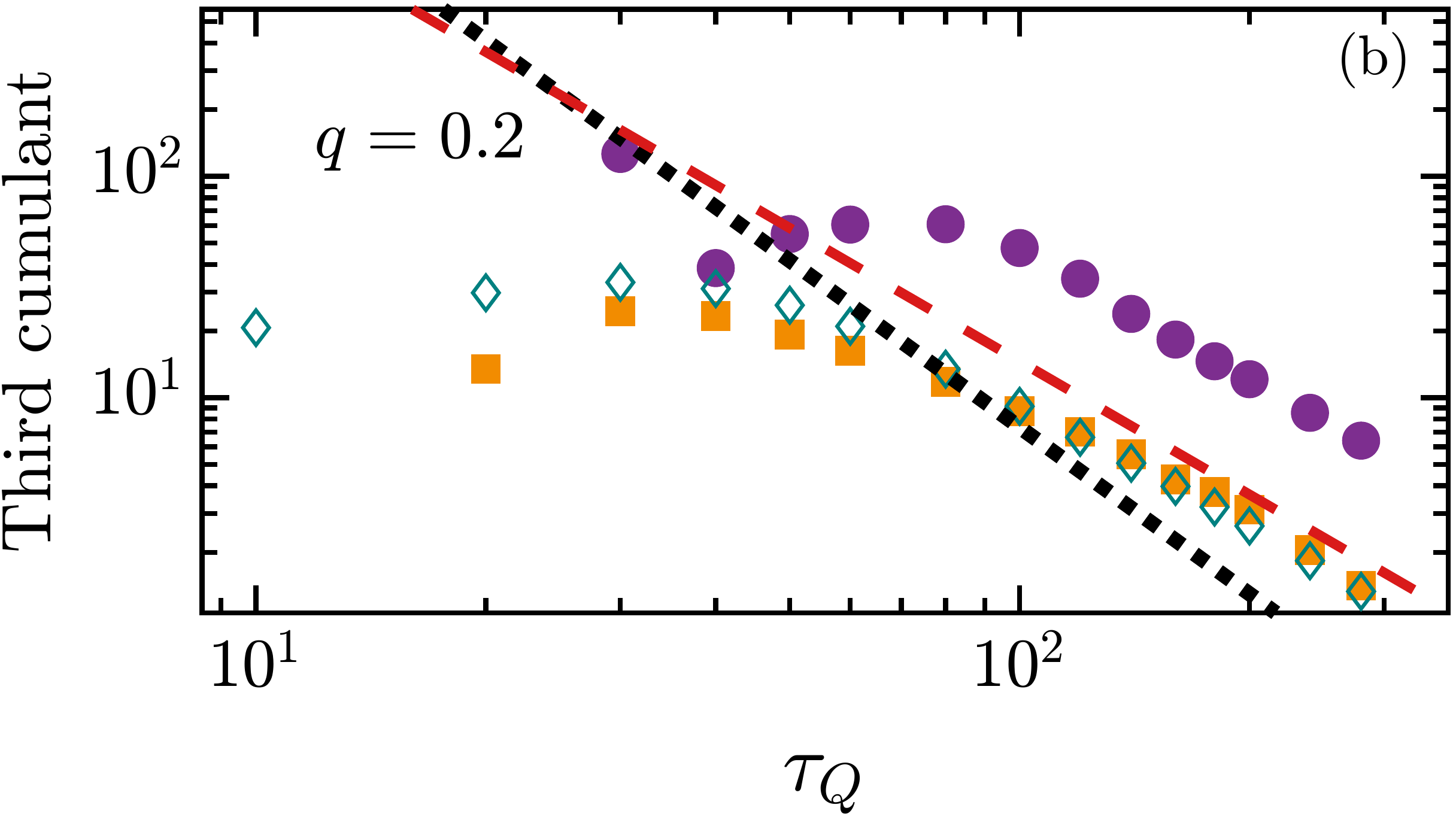} &
    \includegraphics[width=0.235\textwidth,trim={2.75cm 0 0cm 0},clip]{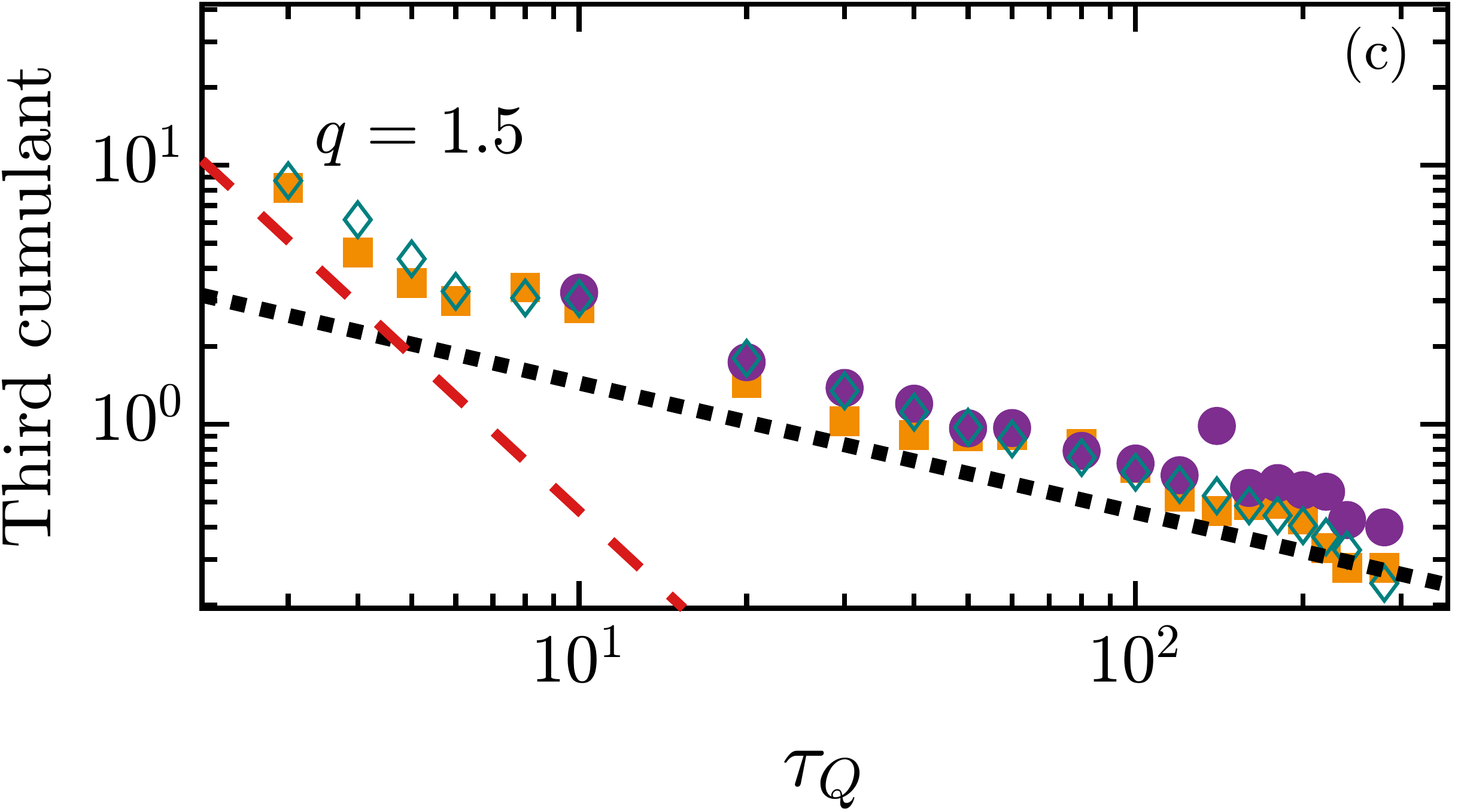} &
      \includegraphics[width=0.235\textwidth,trim={2.75cm 0 0cm 0},clip]{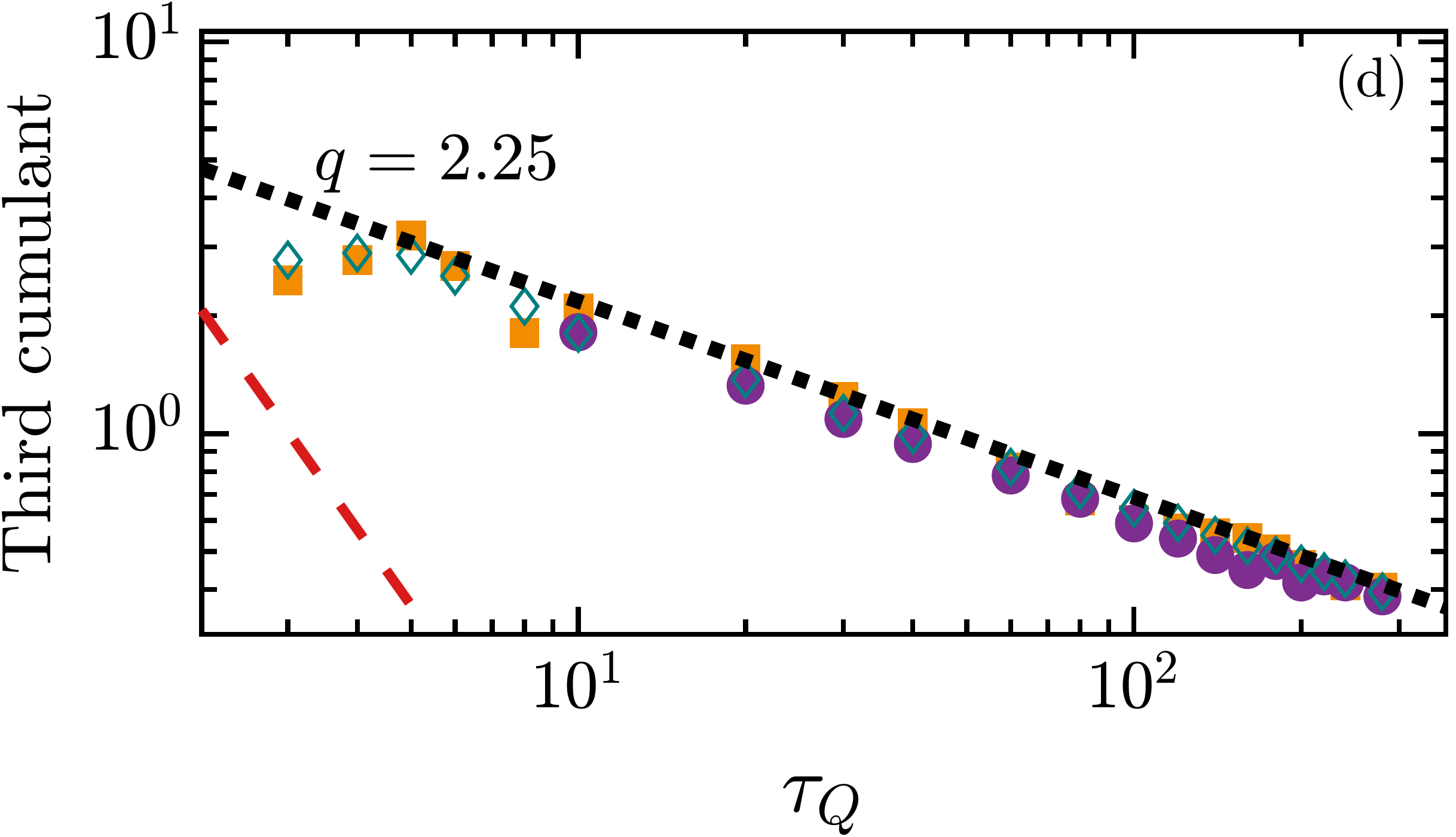} \\
    \end{tabular}
    \caption{Third cumulant $\kappa_3$ of the kink number in the (a), (b) long-range regime $q=0.1,0.2$, respectively, (c) intermediate regime $q=1.5$, and (d) short-range regime $q=2.25$, with the same conventions as in Fig.~\ref{fig:kappa_1}.}
    \label{fig:kappa_3}
\end{figure*}

The third cumulant is shown in Fig.~\ref{fig:kappa_3} for $q=0.1$ and $q=0.2$ in the long-range regime, $q=1.5$ in the intermediate regime and $q=2.25$ in the short-range regime. As for the variance, the free-fermion results match the TDVP data for $q\gtrsim1.25$, i.e., slightly earlier than for the mean. Strong discrepancies remain at $q=0.1$ and $q=0.2$, while the matching is almost perfect for $q=1.5$ and $q=2.25$.

\begin{figure*}[t]
    \centering
    \begin{tabular}{c c c}
      \includegraphics[width=0.33\textwidth]{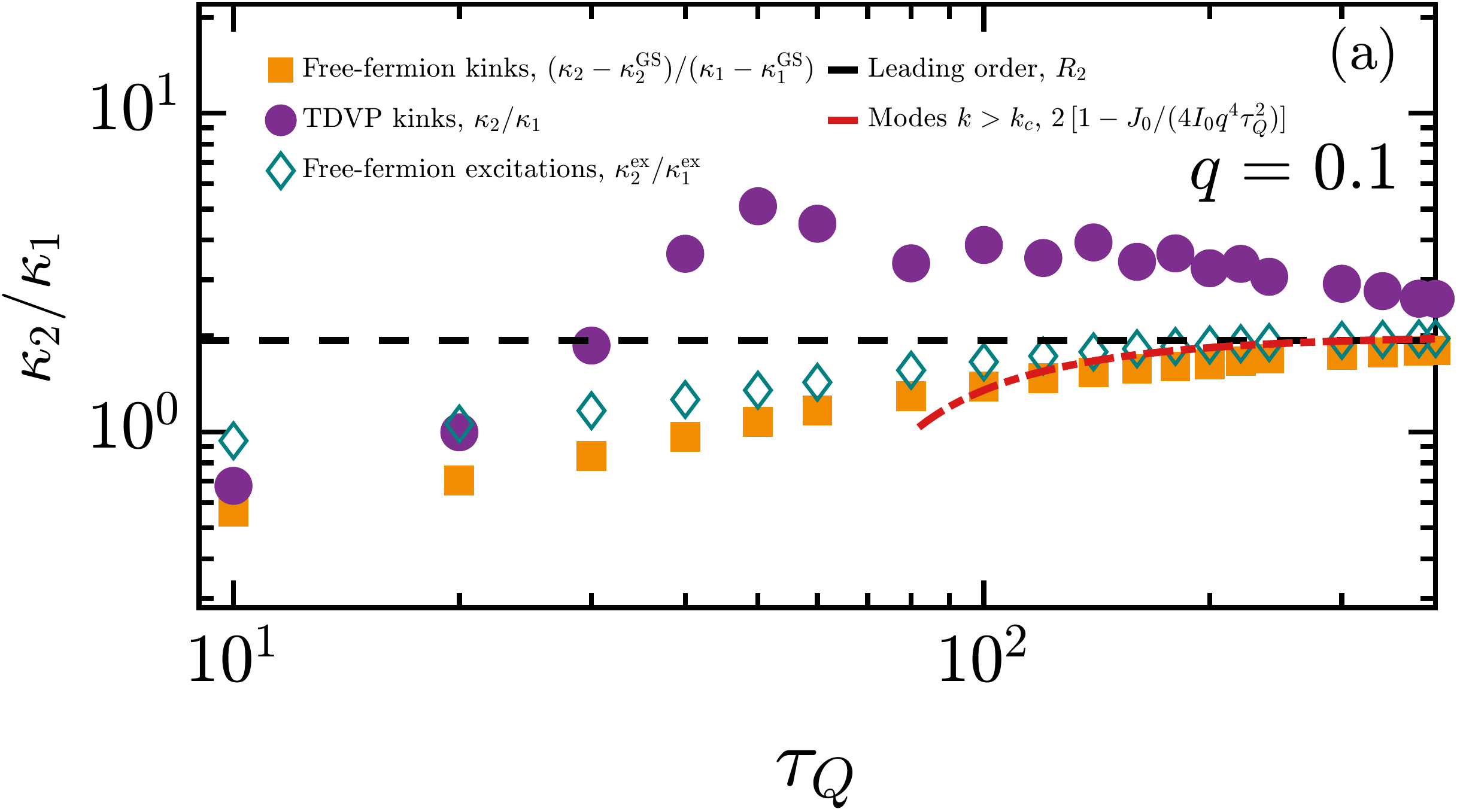} &
      \includegraphics[width=0.315\textwidth]{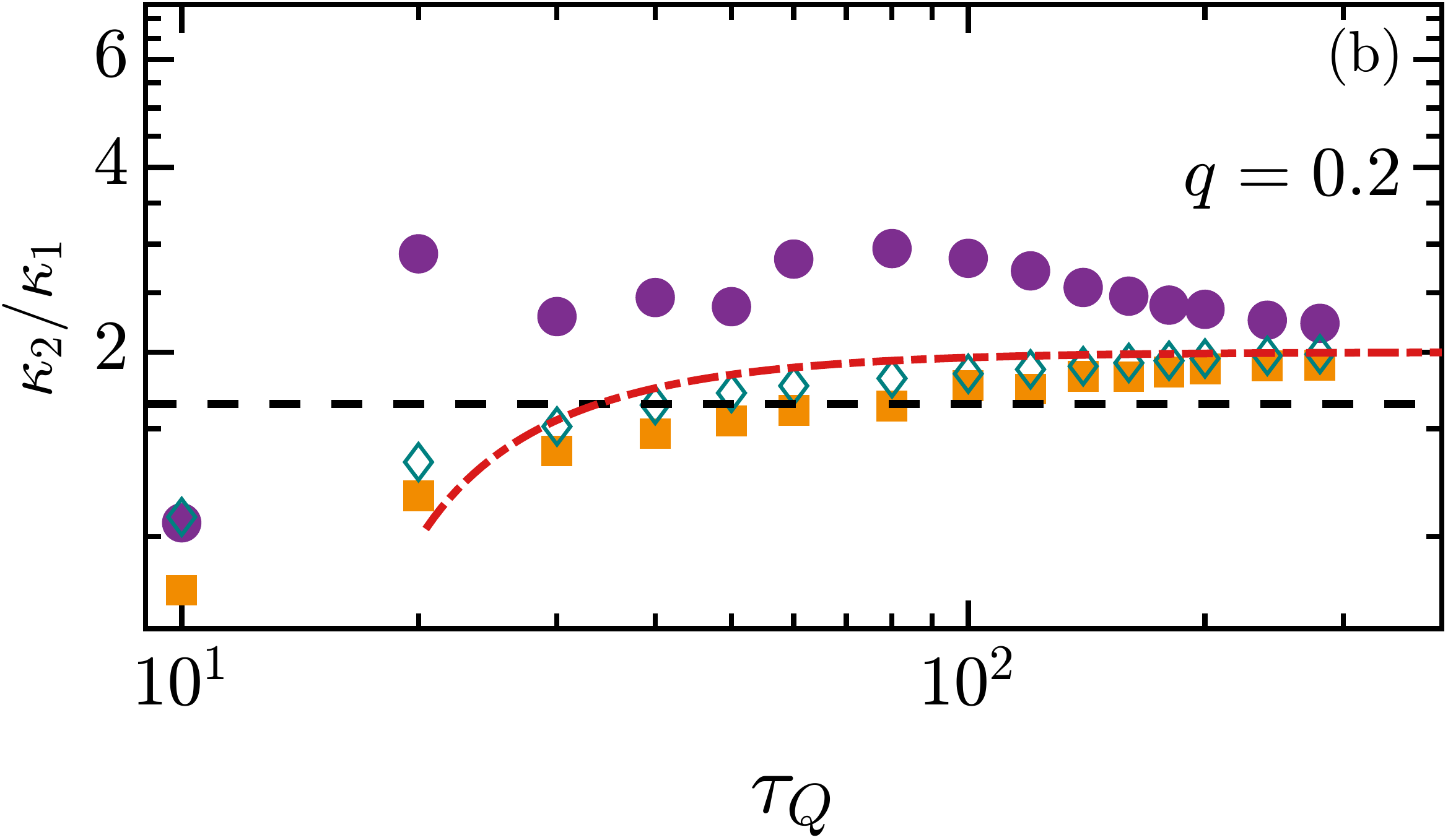} &
      \includegraphics[width=0.315\textwidth]{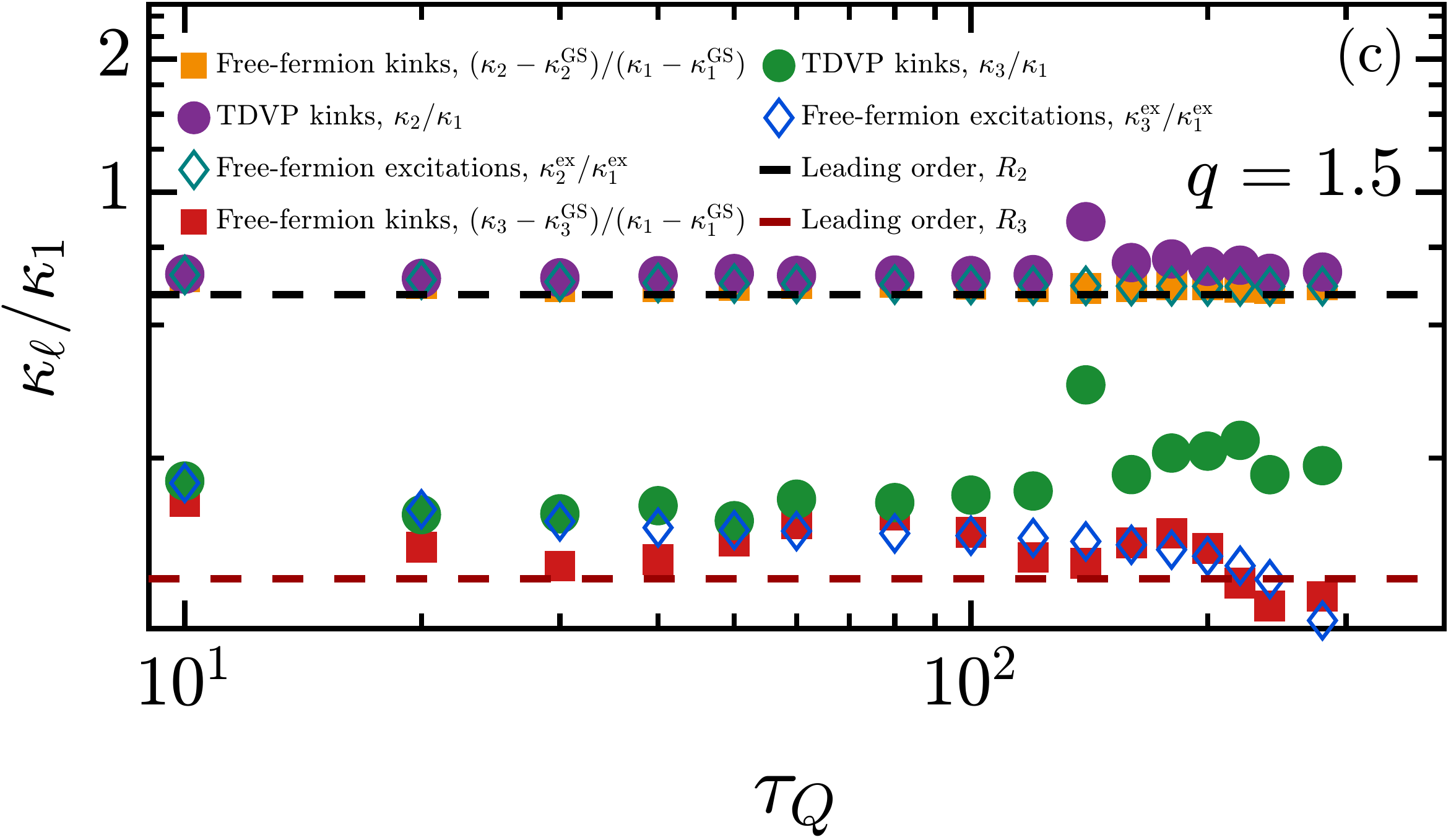} \\
      \includegraphics[width=0.3\textwidth]{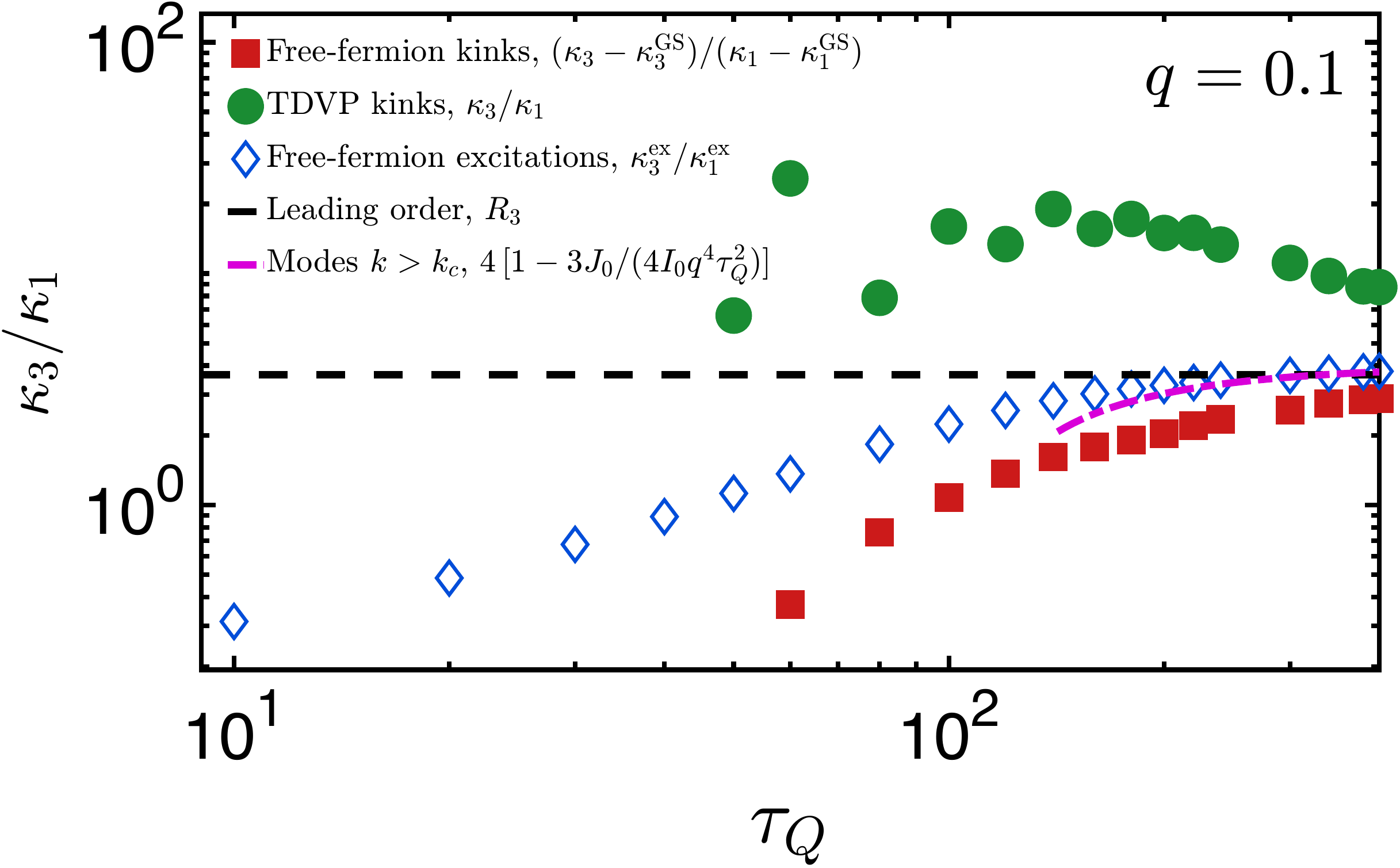} &
      \includegraphics[width=0.3325\textwidth]{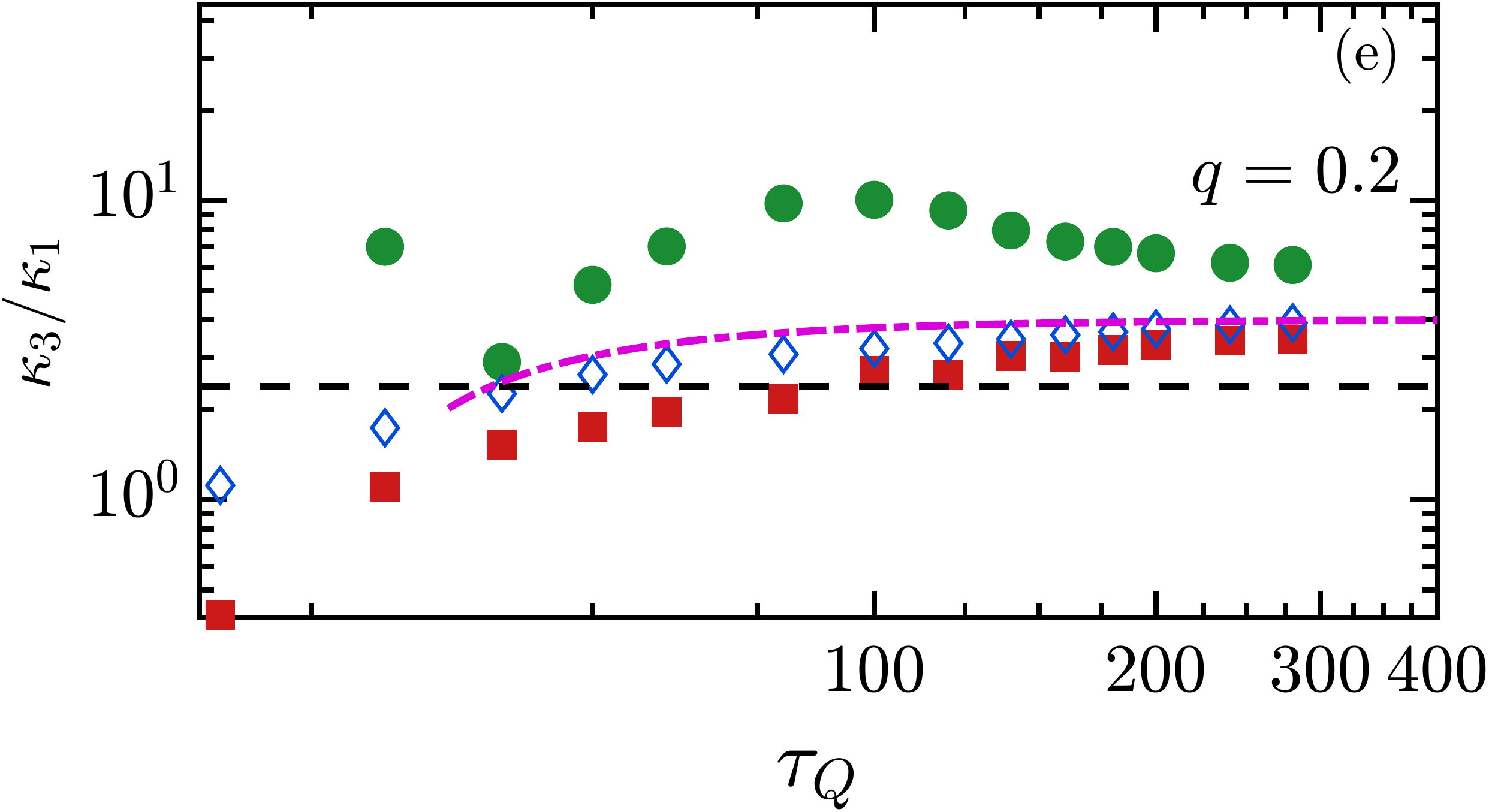} &
      \includegraphics[width=0.33\textwidth]{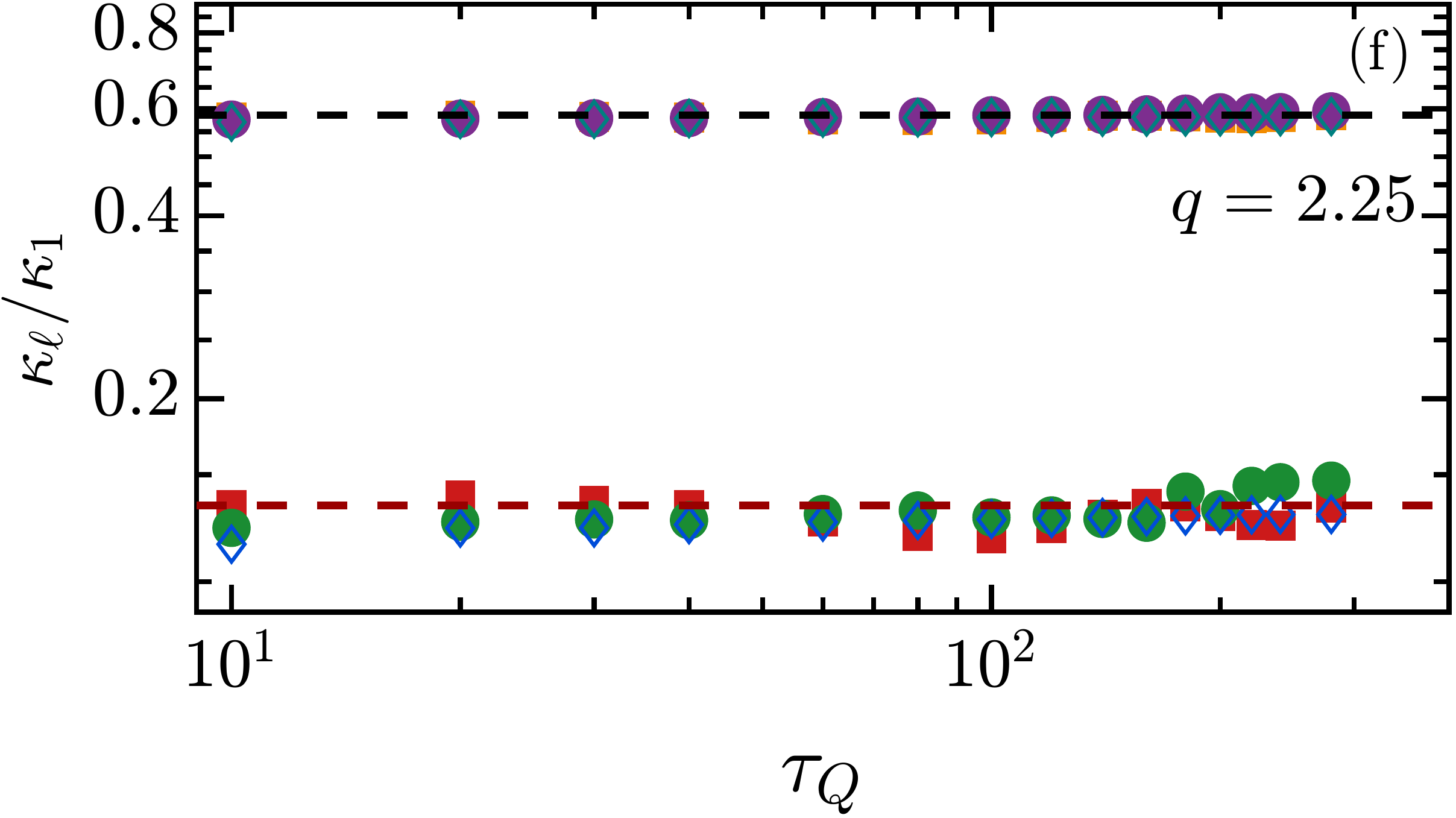} \\
    \end{tabular}
    \caption{Cumulant ratios as functions of the driving time. Panels (a) and (b) show $\kappa_2/\kappa_1$ and panels (d) and (e) show $\kappa_3/\kappa_1$ in the long-range regime, $q=0.1$ and $q=0.2$, while panels (c) and (f) show both ratios at $q=1.5$ and $q=2.25$. Filled squares mark the free-fermion kinks measured from the ground state at $g=0$ for the open chain, $(\kappa_\ell-\kappa^{\rm GS}_\ell)/(\kappa_1-\kappa^{\rm GS}_1)$, open diamonds the free-fermion excitations on the ring, $\kappa^{\rm ex}_\ell/\kappa^{\rm ex}_1$, and filled circles the TDVP kinks, $\kappa_\ell/\kappa_1$. The dashed lines mark the rate-independent values $R_2=2[1-2^{-1/(2q_*)}]$ and $R_3=4[1-3\cdot2^{-1/(2q_*)}+2\cdot3^{-1/(2q_*)}]$ of Eq.~\eqref{eq:kappa_scaling}, with $q_*=\min(q,1)$. For $q<1/4$ the dash-dotted lines show the excitation ratios of the modes without an avoided crossing, $\kappa^{\rm ex}_2/\kappa^{\rm ex}_1=2[1-J_0/(4I_0q^4\tau_Q^2)]$ and $\kappa^{\rm ex}_3/\kappa^{\rm ex}_1=4[1-3J_0/(4I_0q^4\tau_Q^2)]$, with $I_0$ of Eq.~\eqref{eq:app_Bq_smallq} and $J_0$ of Sec.~\ref{sec:defects}, which approach the values $2$ and $4$ of independently created pairs for slow ramps, while the kink ratios above the ground state approach $\tilde R_2\simeq1.91$ and $\tilde R_3\simeq3.45$ at $q=0.1$, and $\tilde R_2\simeq1.92$ and $\tilde R_3\simeq3.53$ at $q=0.2$ (Sec.~\ref{sec:defects}). In panels (c) and (f) all ratios sit on the predicted plateaus over the whole range of driving rates, so that the higher cumulants remain proportional to the mean. For $q\leq0.2$ the TDVP ratio $\kappa_2/\kappa_1$ exceeds 2, while the free-fermion ratios stay below this bound. All panels use logarithmic scales.}
    \label{fig:ratios}
\end{figure*}

The cumulant ratios as functions of $\tau_Q$ are shown in Fig.~\ref{fig:ratios} for the same fractional orders.     Although the mean kink numbers of the TDVP simulations and of free fermions on the ring differ for $q\lesssim2$, the ratios of the variance and of the third cumulant to the mean agree closely for $q=1.5$ and $q=2.25$. At $q=0.1$ and $q=0.2$ the free-fermion ratios $\kappa_2/\kappa_1$ stay below $2$, which the excitations approach only for slow ramps, as expected for independently created pairs, while the kinks above the ground state approach the lower plateau $\tilde R_2\simeq1.91$--$1.92$ of Sec.~\ref{sec:defects}, as their overlap with the ground-state kinks reduces the higher cumulants (App.~\ref{app:kinks}). At the same time, the TDVP ratios exceed 2, in agreement with Fig.~\ref{fig:ratios_q} of the main text.

To better characterize the range of driving times where the cumulants follow a power law, we multiply them by $\tau_Q^{\alpha}$, with the exponent $\alpha$ of the free-fermion analysis. For $q>1/4$ we use $\alpha=1/(2q_*)$ of the Landau-Zener modes, Eq.~\eqref{eq:kappa_scaling}, while for $q<1/4$, $\alpha=2$ is employed from the calculations of adiabatic perturbation theory, Eq.~\eqref{eq:nex_nc}. An expected power law then appears as a horizontal line, at the free-fermion values $C_\ell$ for $q>1/4$ and at $\tilde R_\ell L\tilde B_q$, with $\tilde R_1=1$, for the kinks above the ground state at $q<1/4$ (Sec.~\ref{sec:defects}).

\begin{figure*}[t]
    \centering
    \begin{tabular}{c  c  c c}\includegraphics[width=0.25\textwidth,trim={0 0 0 0cm},clip]{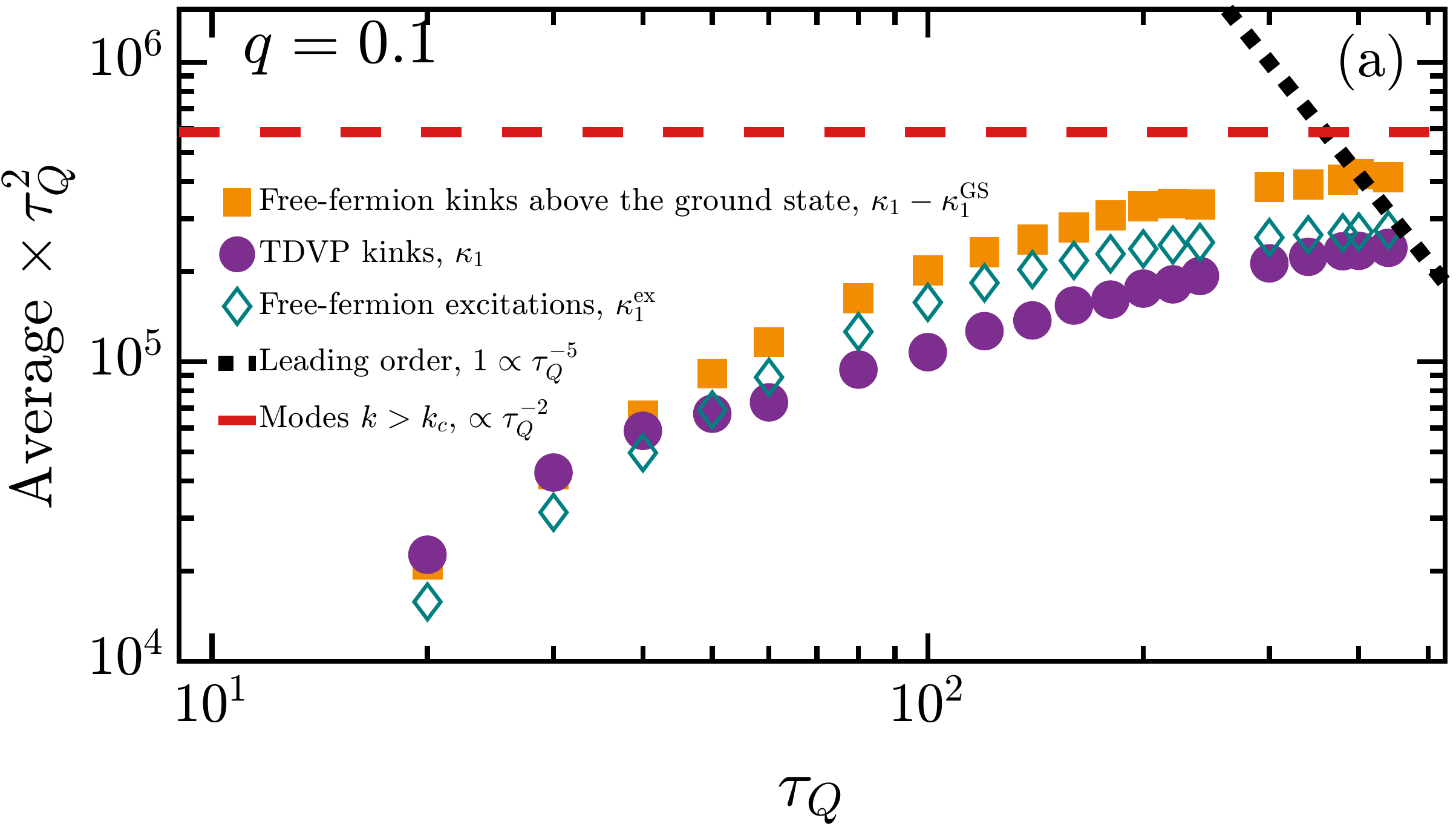} &
    \includegraphics[width=0.2365\textwidth,trim={2.04cm 0 0 0},clip]{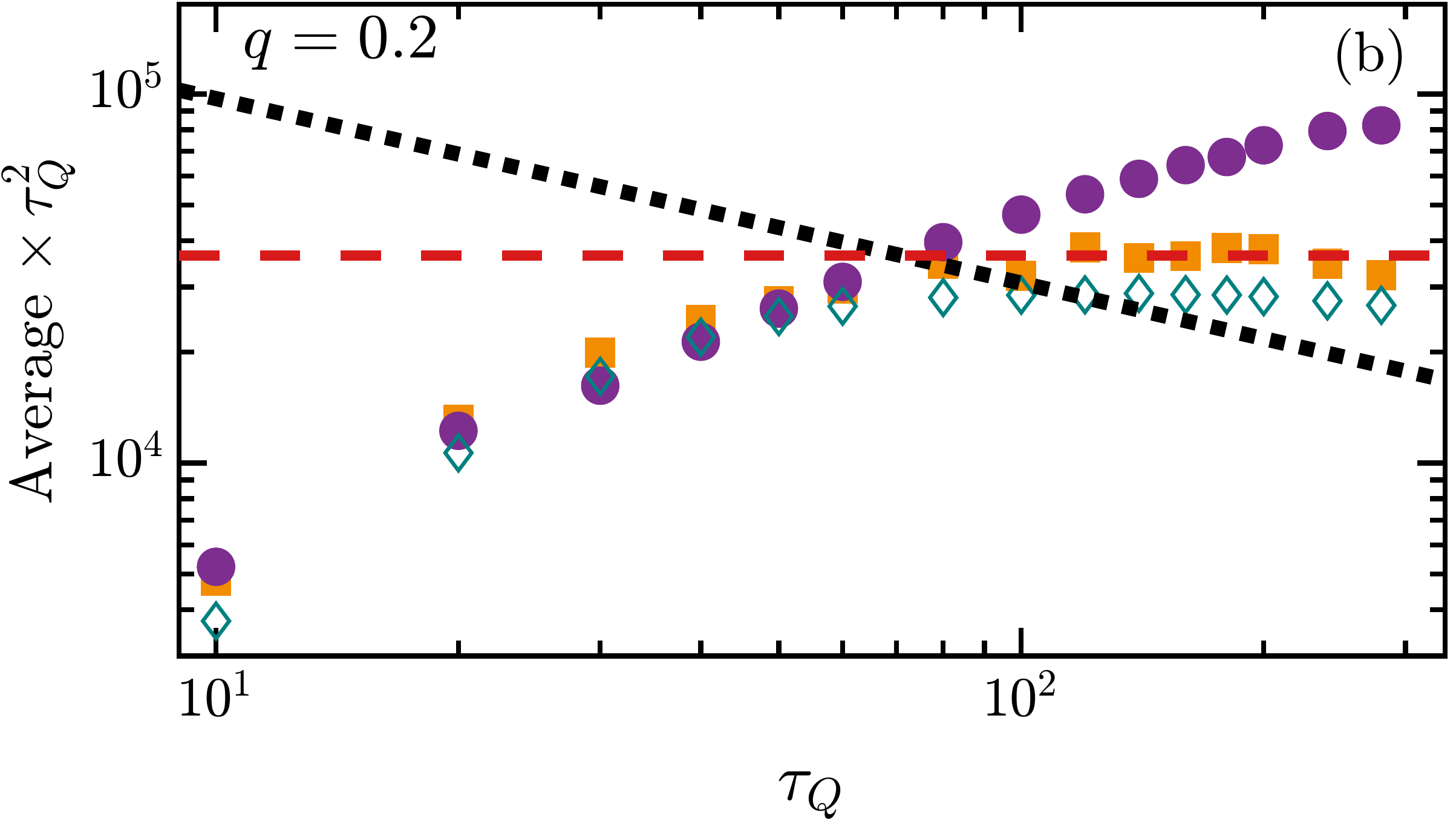} &
    \includegraphics[width=0.2365\textwidth,trim={2.04cm 0 0cm 0},clip]{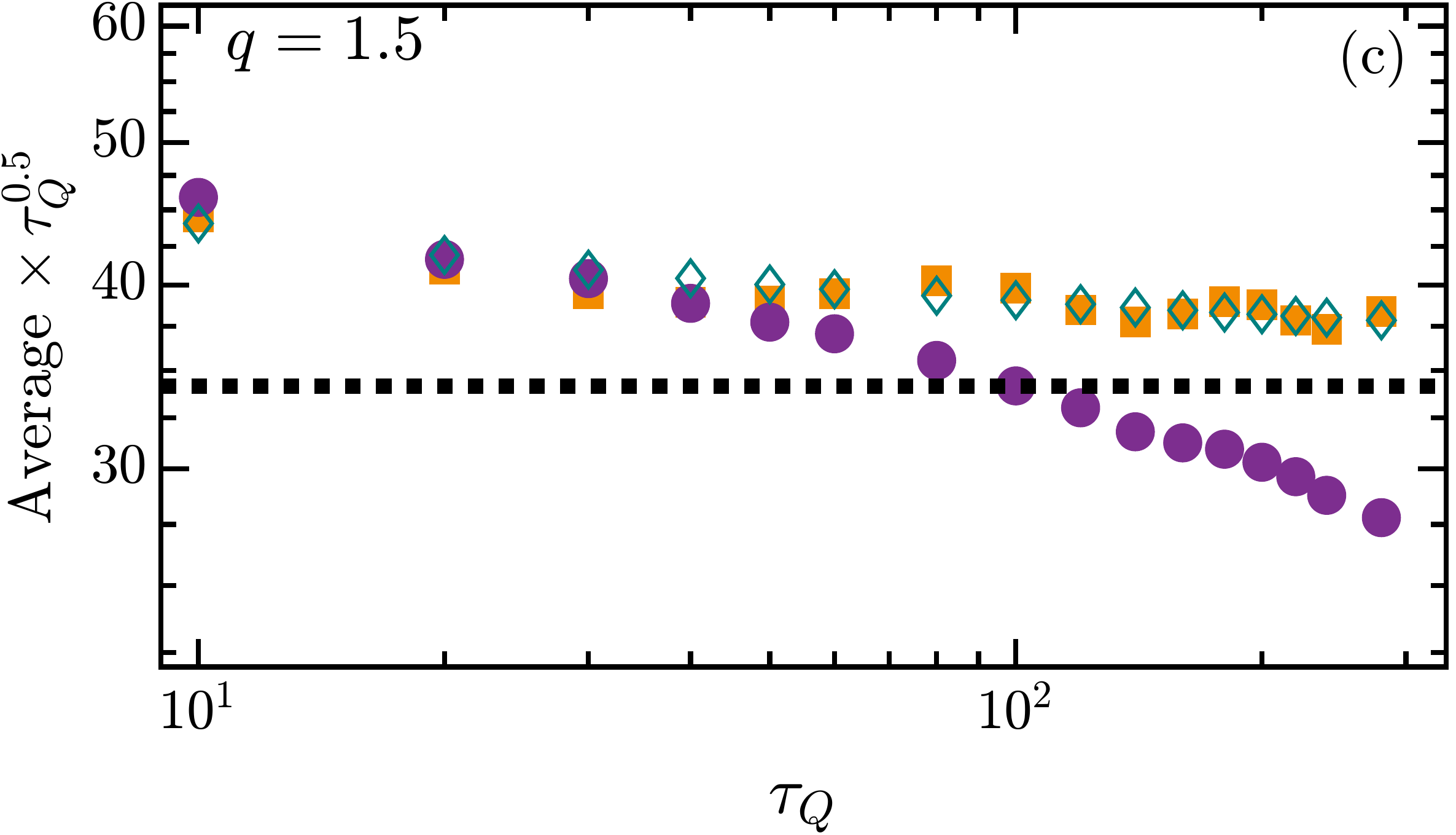} &
      \includegraphics[width=0.2365\textwidth,trim={2.04cm 0 0cm 0},clip]{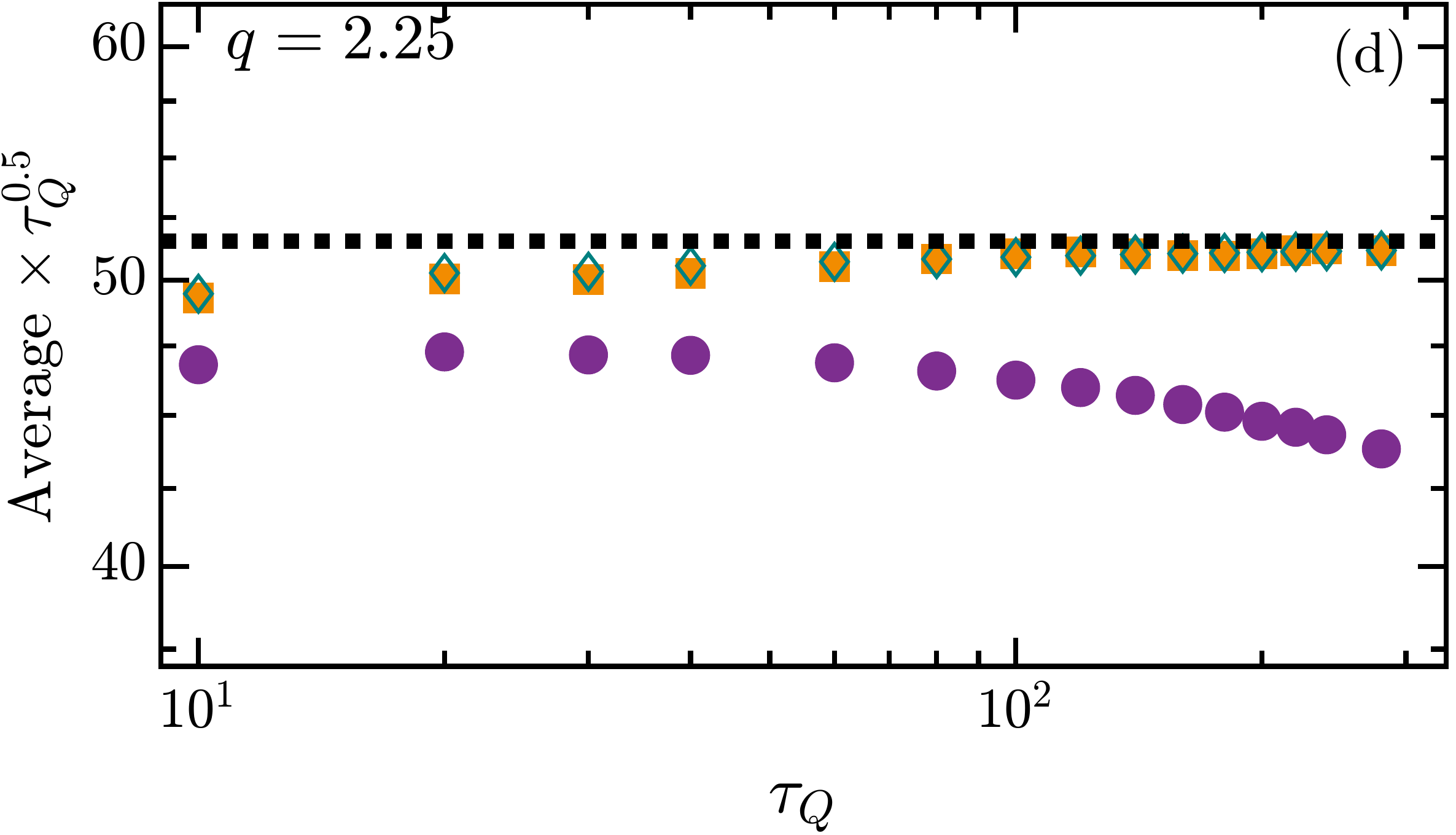} \\
    \end{tabular}
    \caption{Mean kink number multiplied by $\tau_Q^{\alpha}$ for (a) $q=0.1$, (b) $q=0.2$, (c) $q=1.5$ and (d) $q=2.25$. We use $\alpha=2$ for $q<1/4$, where slow ramps excite mainly the modes without an avoided crossing, and $\alpha=1/(2q_*)$ otherwise, so that the expected power law appears as a horizontal line. Purple circles show the TDVP data, orange squares the free-fermion kinks counted from the ground state at $g=0$, and open diamonds the free-fermion excitations. The black dotted and red dashed lines are the leading-order prediction of Eq.~\eqref{eq:kappa_scaling} and the contribution $L\tilde B_q\tau_Q^{-2}$ of the modes with $k>k_c$, multiplied by the same factor.}
    \label{fig:kappa1_comp}
\end{figure*}

For the mean, Fig.~\ref{fig:kappa1_comp}, the free-fermion kinks and excitations at $q=2.25$ lie on the leading-order line over the whole range of $\tau_Q$, as $\tau_{\rm FS}$ exceeds the simulated driving times at this order (App.~\ref{app:finite_size}). The TDVP data lie slightly below it and decrease slowly with $\tau_Q$, in agreement with the steeper decay on the open chain discussed in Sec.~\ref{sec:cumulants_num}. At $q=1.5$ the free-fermion data are nearly constant and are slightly above the leading-order line, whereas the TDVP data decrease slowly with $\tau_Q$, dropping by less than a factor of two over two decades of $\tau_Q$, which corresponds to a decay faster than $\tau_Q^{-1/2}$ by about $0.13$ in the exponent, consistent with the steeper decay on the open chain. 
For $q<1/4$ the free-fermion data grow at fast ramps and level off once the modes with $k>k_c$ dominate. At $q=0.2$ this happens for $\tau_Q\gtrsim80$, close to the value $L\tilde B_q$, while at $q=0.1$ the data approach it only at the slowest ramps, since the condition $\tau_Q\gg(q\ln2)^{-2}\approx200$ is fulfilled only there. The TDVP mean keeps increasing at $q=0.2$ over the whole range, i.e., it decays more slowly than $\tau_Q^{-2}$, and at $q=0.1$ it stays below the free-fermion data without reaching a plateau either. As an accidental consequence of these finite-size effects, the free-fermion and TDVP results stay close to each other before the former reaches the expected $\tau^{-2}_Q$ scaling, as also seen in Figs.~\ref{fig:kappa_1} and~\ref{fig:alpha}.

\begin{figure*}[t]
    \centering
    \begin{tabular}{c  c  c c}\includegraphics[width=0.24\textwidth,trim={0 0 0 0cm},clip]{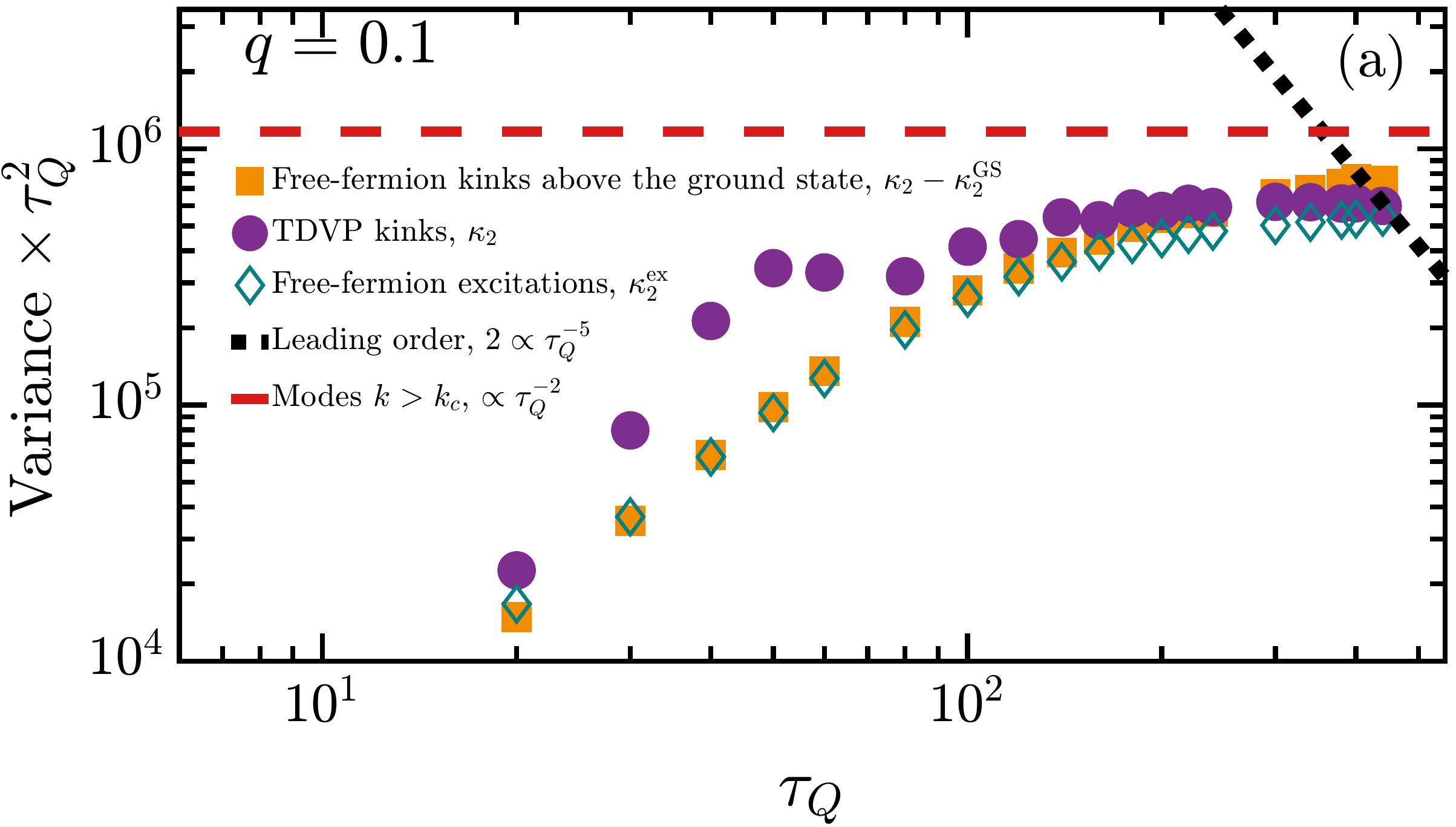} &
    \includegraphics[width=0.24\textwidth,trim={0cm 0 0 0},clip]{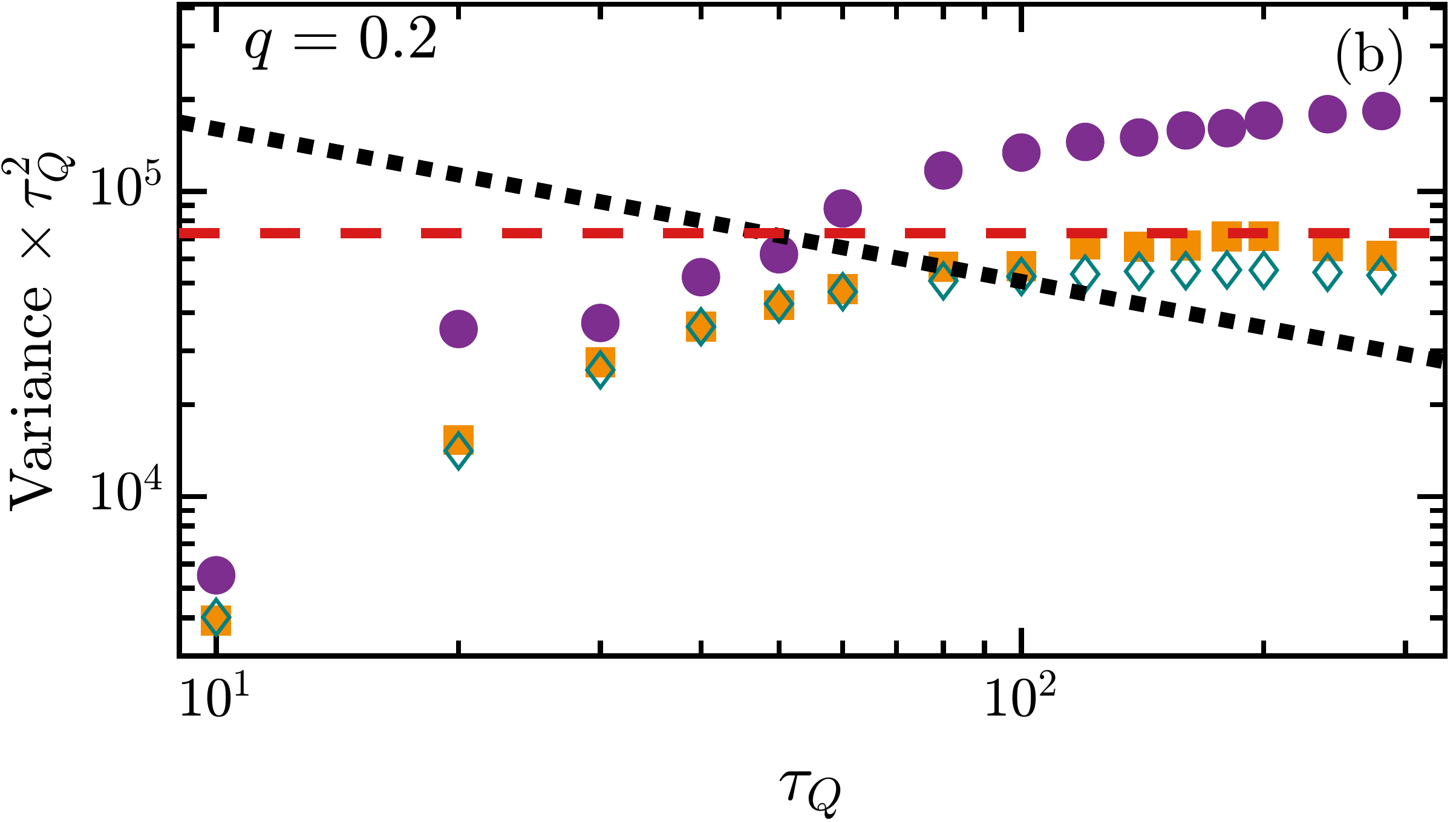} &
    \includegraphics[width=0.24\textwidth,trim={0cm 0 0cm 0},clip]{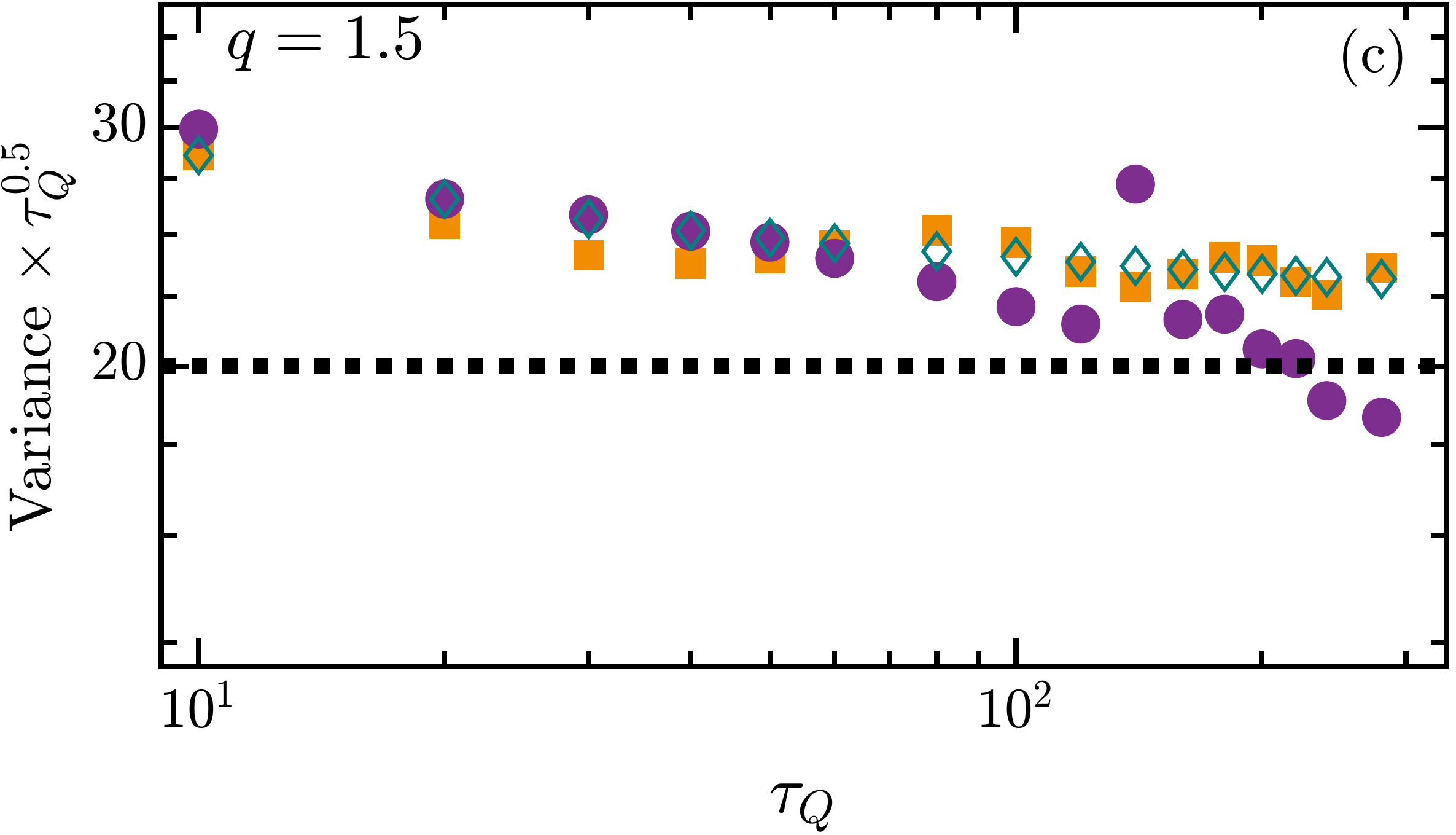} &
      \includegraphics[width=0.24\textwidth,trim={0cm 0 0cm 0},clip]{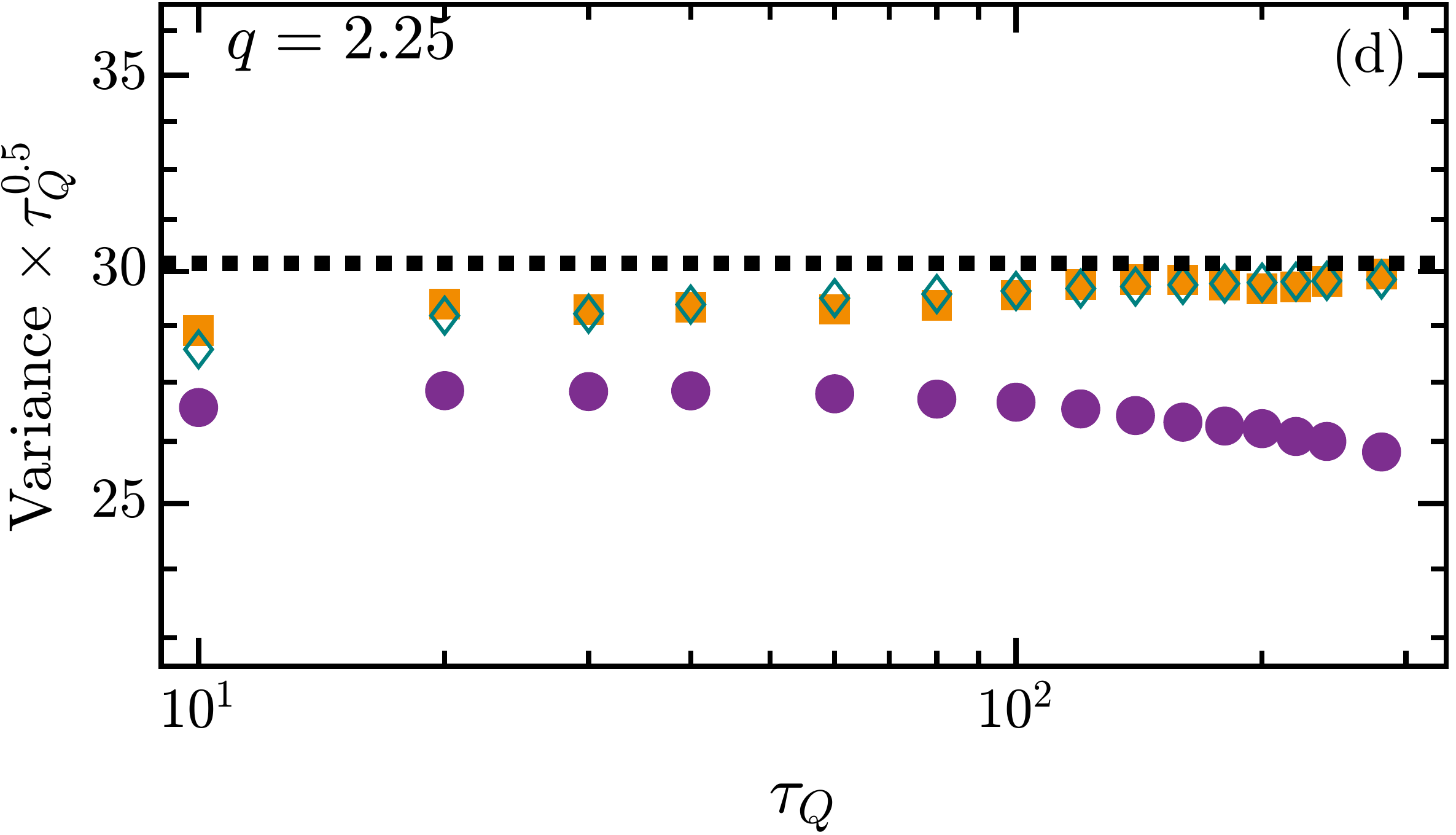} \\
    \end{tabular}
     \caption{Variance of the kink number multiplied by $\tau_Q^{\alpha}$ for (a) $q=0.1$, (b) $q=0.2$, (c) $q=1.5$ and (d) $q=2.25$, with the exponents $\alpha$ and symbols of Fig.~\ref{fig:kappa1_comp}. The black dotted lines show the leading-order prediction $C_2\tau_Q^{-1/(2q_*)}$ of Eq.~\eqref{eq:kappa_scaling} and the red dashed lines the contribution $\tilde R_2L\tilde B_q\tau_Q^{-2}$ of the modes with $k>k_c$ to the kinks above the ground state, multiplied by the same factor.}
    \label{fig:kappa2_comp}
\end{figure*}

The variance in Fig.~\ref{fig:kappa2_comp} behaves similarly to the mean at $q=1.5$ and $q=2.25$. The free-fermion data stay nearly constant, while the TDVP data decrease slowly with $\tau_Q$ and remain close to them. At $q=0.2$ the free-fermion variance levels off close to the value $\tilde R_2L\tilde B_q$ of the modes with $k>k_c$ for $\tau_Q\gtrsim100$, whereas the TDVP variance keeps growing and exceeds it by about a factor of three at the slowest ramps. At $q=0.1$ the TDVP variance lies above the free-fermion one for fast ramps and approaches it for the slowest ramps, with slightly better matching between them than for the mean.

\begin{figure*}[t]
    \centering
    \begin{tabular}{c  c  c c}\includegraphics[width=0.245\textwidth,trim={0 0 0 0cm},clip]{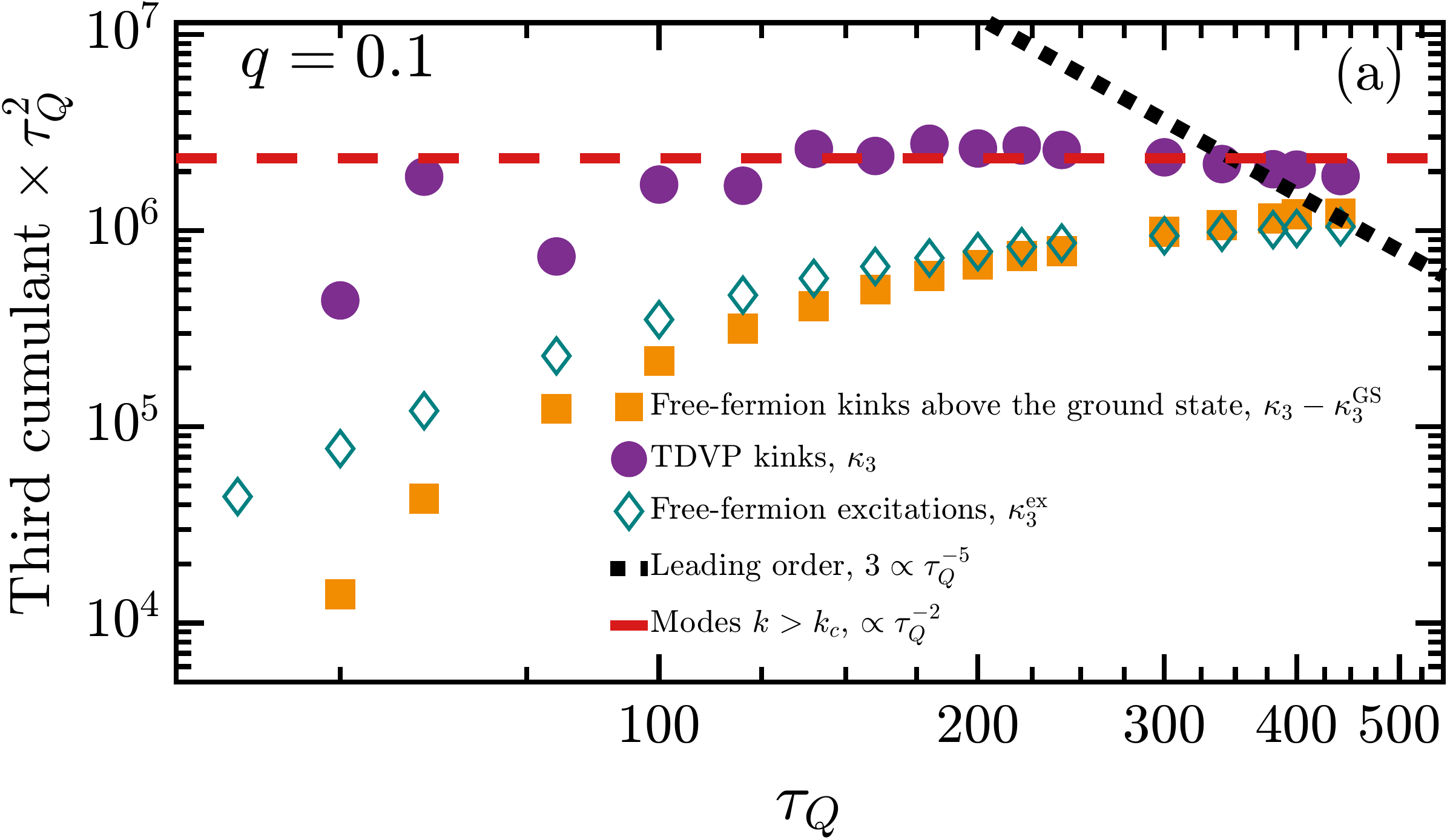} &
    \includegraphics[width=0.242\textwidth,trim={.25cm 0 0 0},clip]{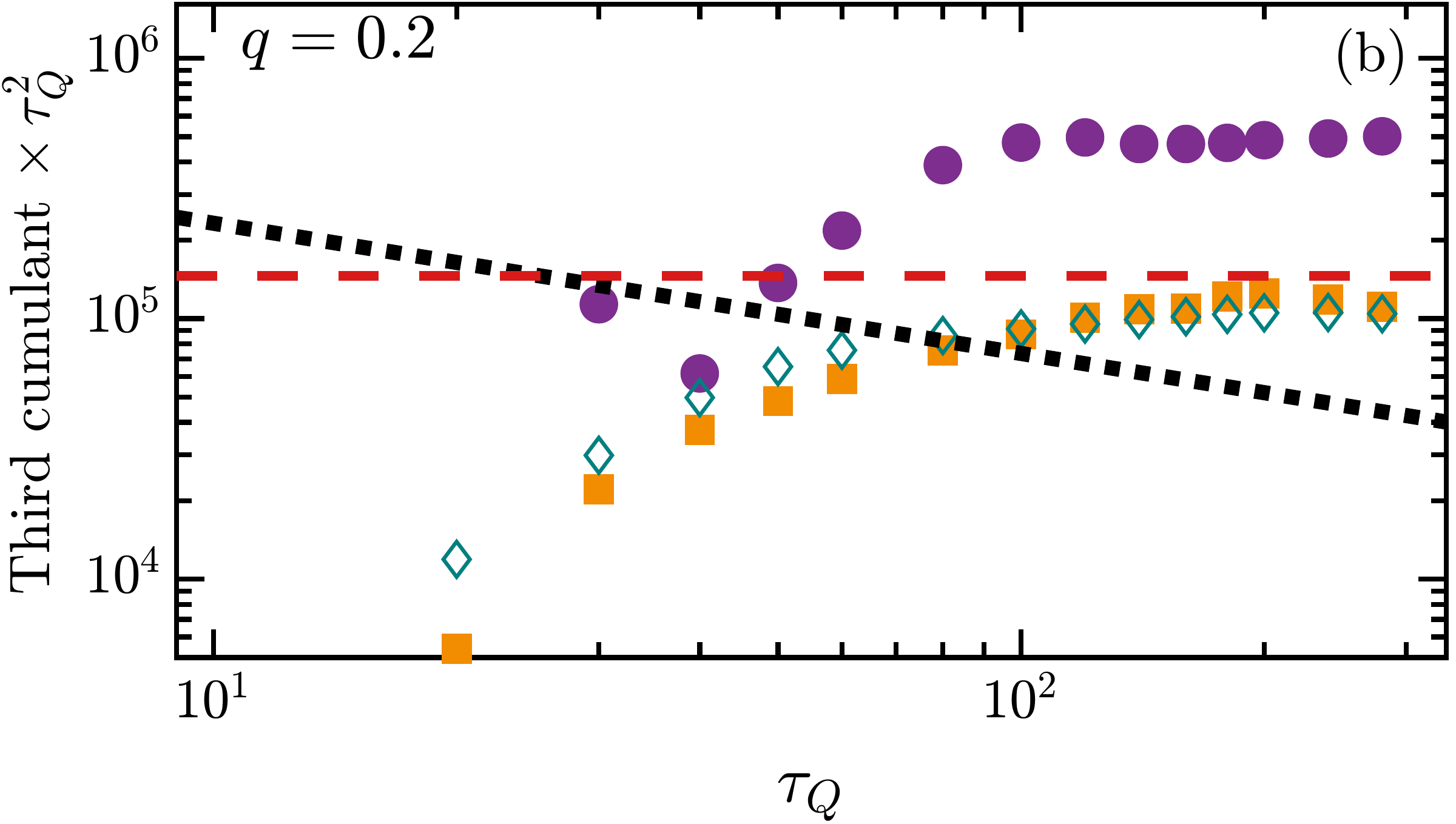} &
    \includegraphics[width=0.241\textwidth,trim={.25cm 0 0cm 0},clip]{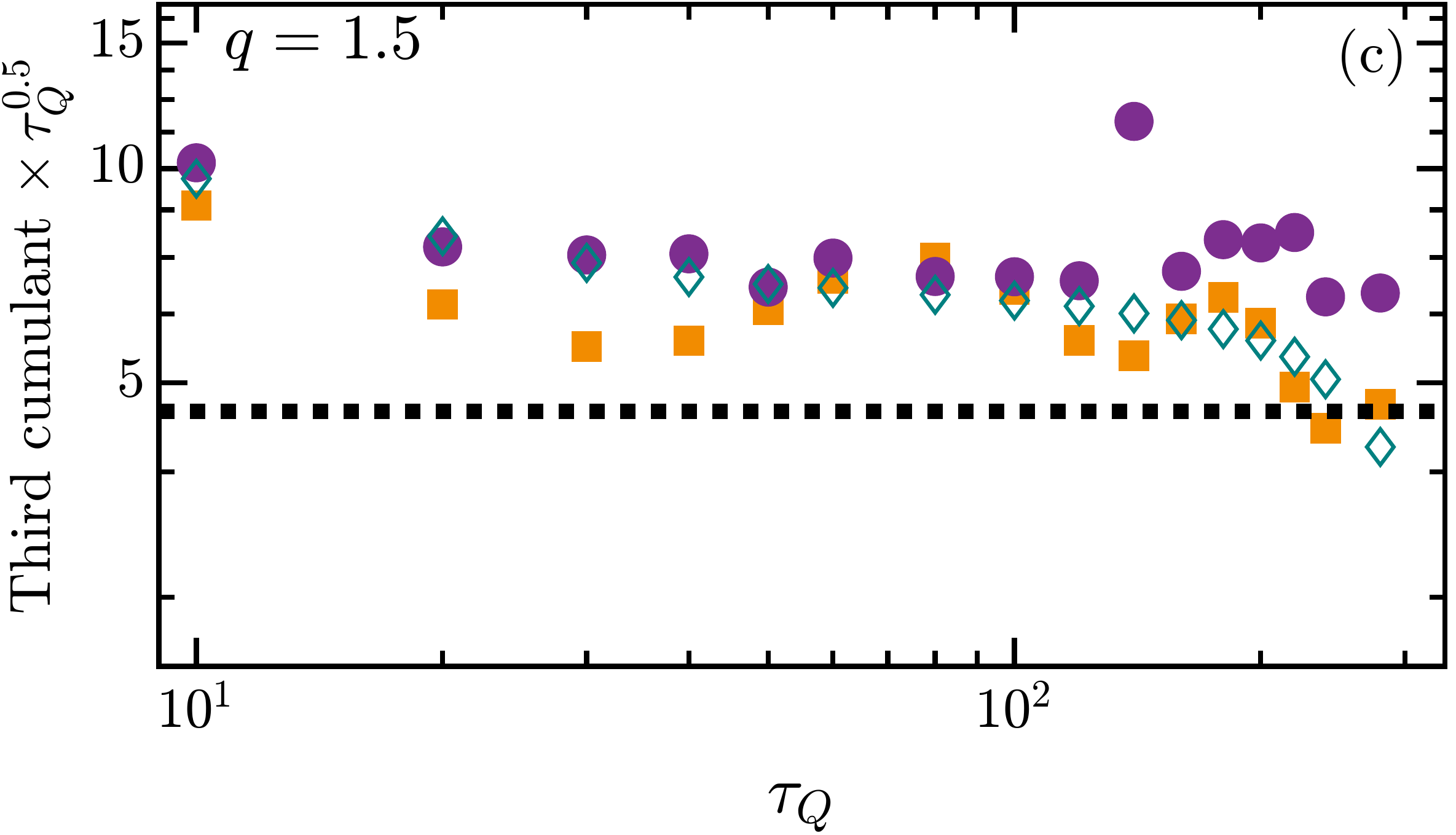} &
      \includegraphics[width=0.24\textwidth,trim={.25cm 0 0cm 0},clip]{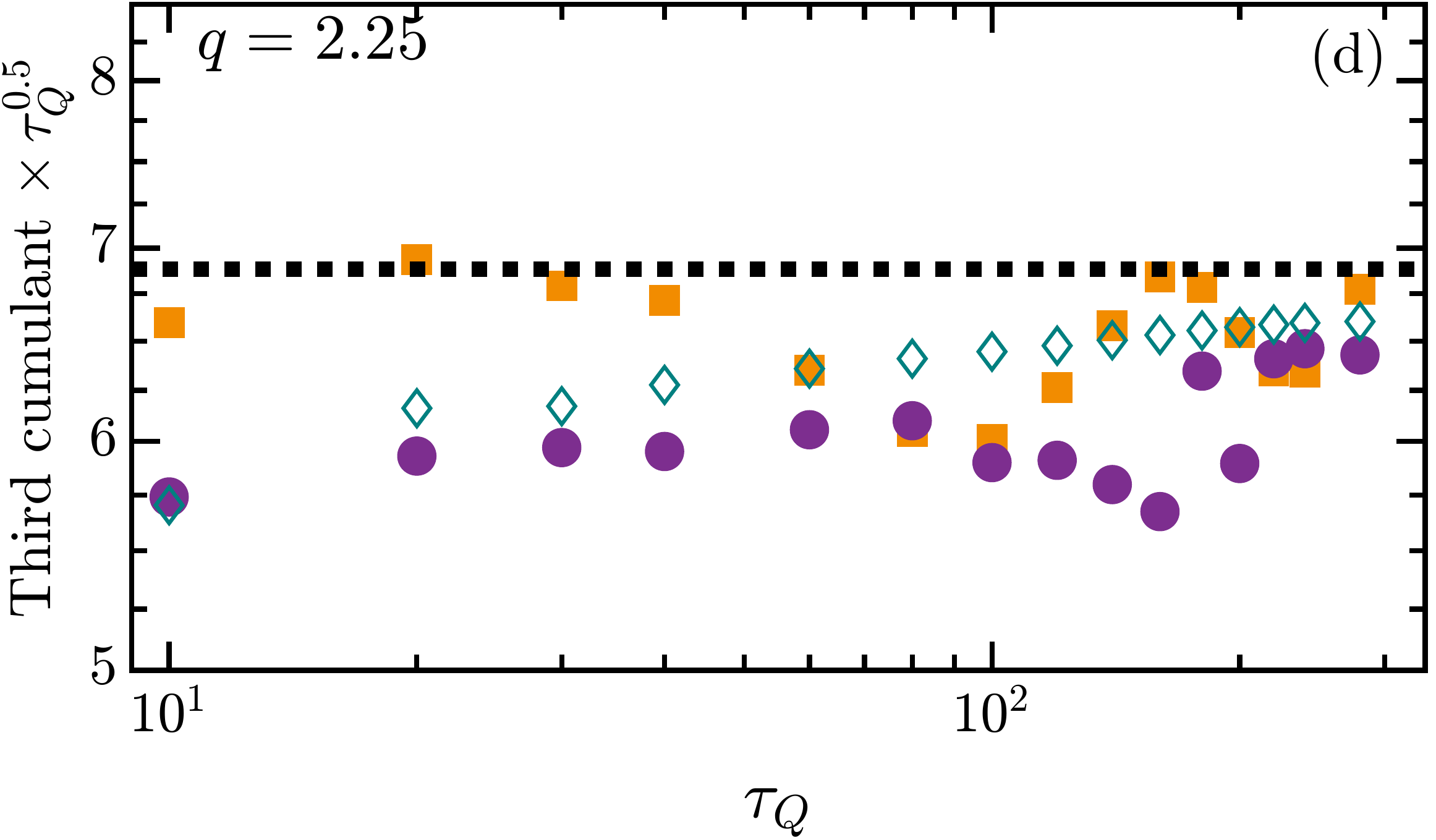} \\
    \end{tabular}
    \caption{Third cumulant of the kink number multiplied by $\tau_Q^{\alpha}$ for (a) $q=0.1$, (b) $q=0.2$, (c) $q=1.5$ and (d) $q=2.25$, with the exponents $\alpha$ and symbols of Fig.~\ref{fig:kappa1_comp}. The black dotted lines show the leading-order prediction $C_3\tau_Q^{-1/(2q_*)}$ of Eq.~\eqref{eq:kappa_scaling} and the red dashed lines the contribution $\tilde R_3L\tilde B_q\tau_Q^{-2}$ of the modes with $k>k_c$ to the kinks above the ground state, multiplied by the same factor.}
    \label{fig:kappa3_comp}
\end{figure*}

For the third cumulant, Fig.~\ref{fig:kappa3_comp}, the TDVP data at $q=0.2$ reach a plateau for $\tau_Q\gtrsim100$, so that $\kappa_3$ decays as $\tau_Q^{-2}$ while the mean decays more slowly. The plateau lies about four times above the free-fermion data, which approach the line of the modes with $k>k_c$ from below.
At $q=0.1$ the TDVP data happen to match precisely the line of the modes with $k>k_c$ already for $\tau_Q\gtrsim150$, while the free-fermion data approach it only at the slowest ramps. The former agreement appears accidental, while the latter agrees with the behavior of the first two cumulants. At $q=1.5$ and $q=2.25$ the third cumulant is small, and the TDVP and free-fermion data scatter around the leading-order value by about $10$--$20\%$, approaching it at the slowest ramps.

\section{Kink density profiles}\label{app:profiles}

\begin{figure*}[t]
    \centering
    \begin{tabular}{c c}
    \includegraphics[width=0.45\textwidth]{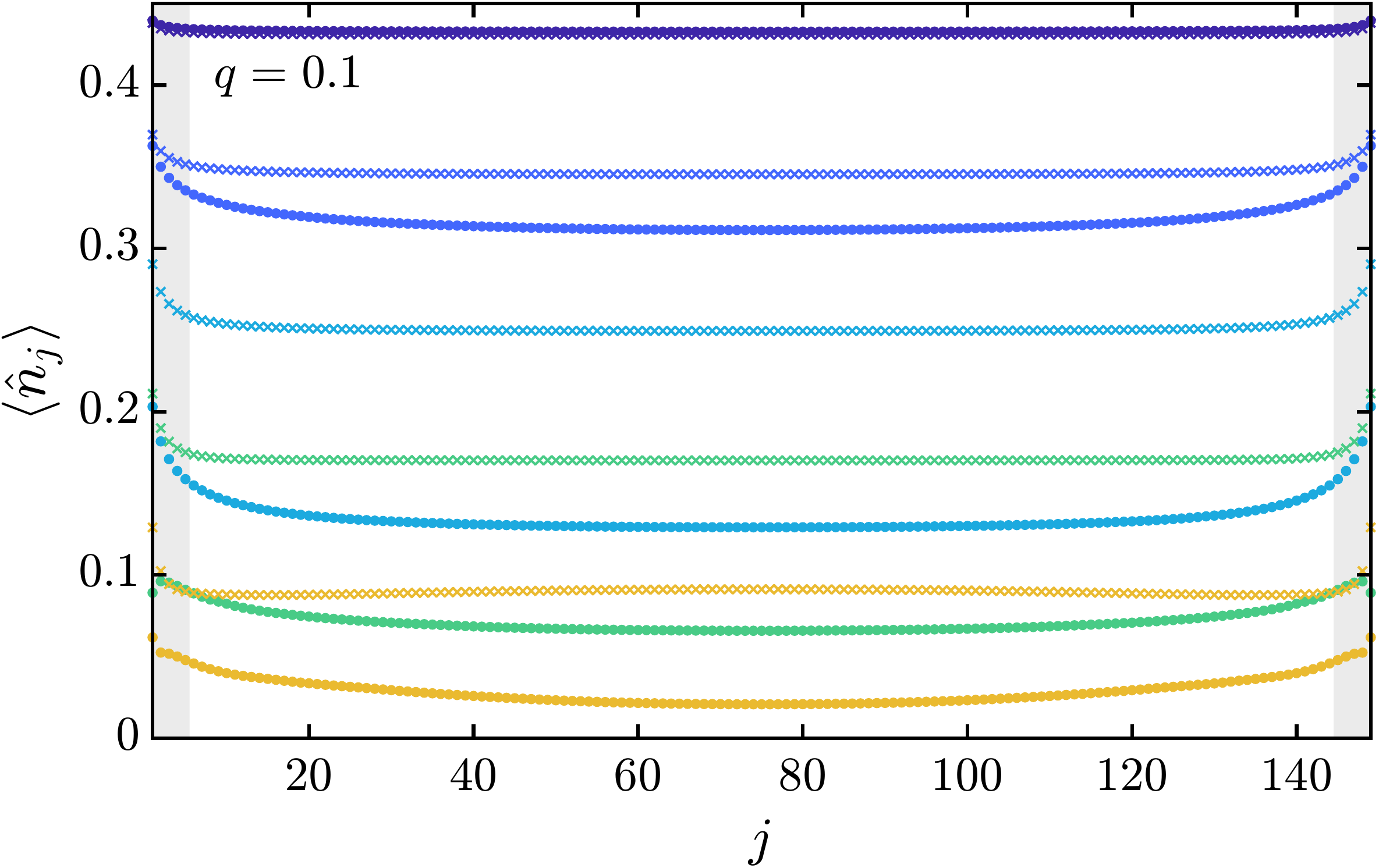} &
    \includegraphics[width=0.45\textwidth]{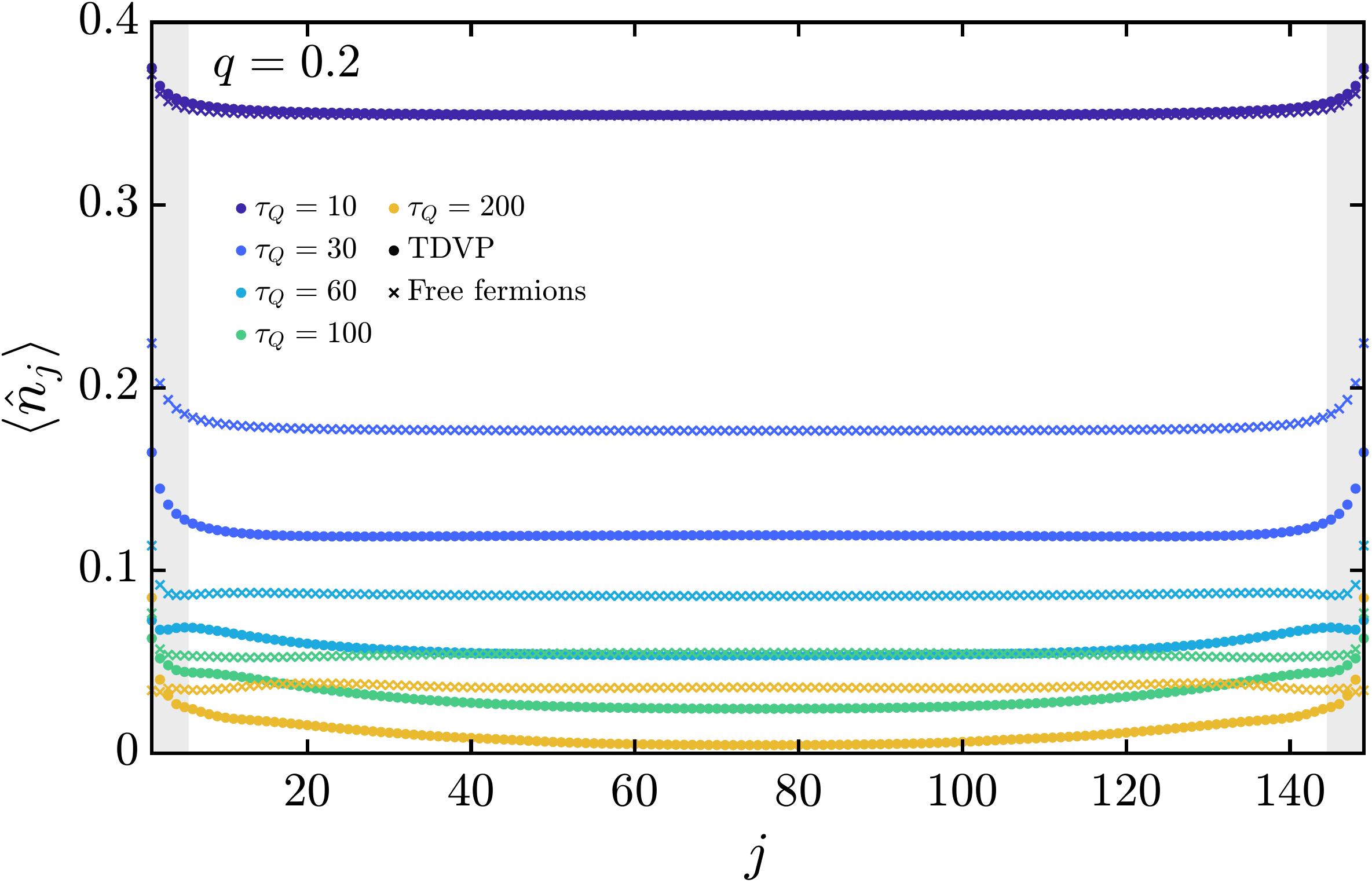} \\
    \includegraphics[width=0.45\textwidth]{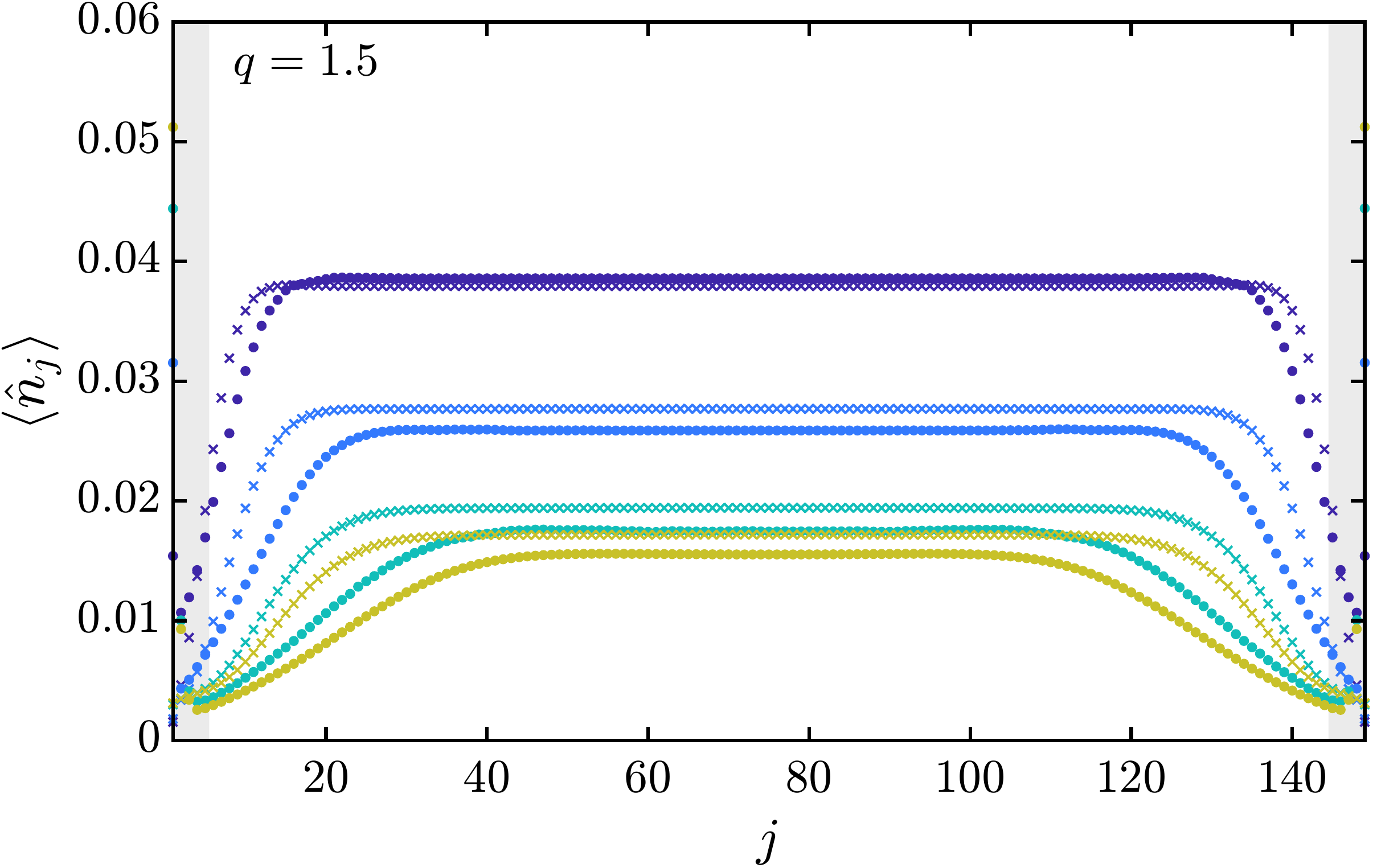} &
    \includegraphics[width=0.45\textwidth]{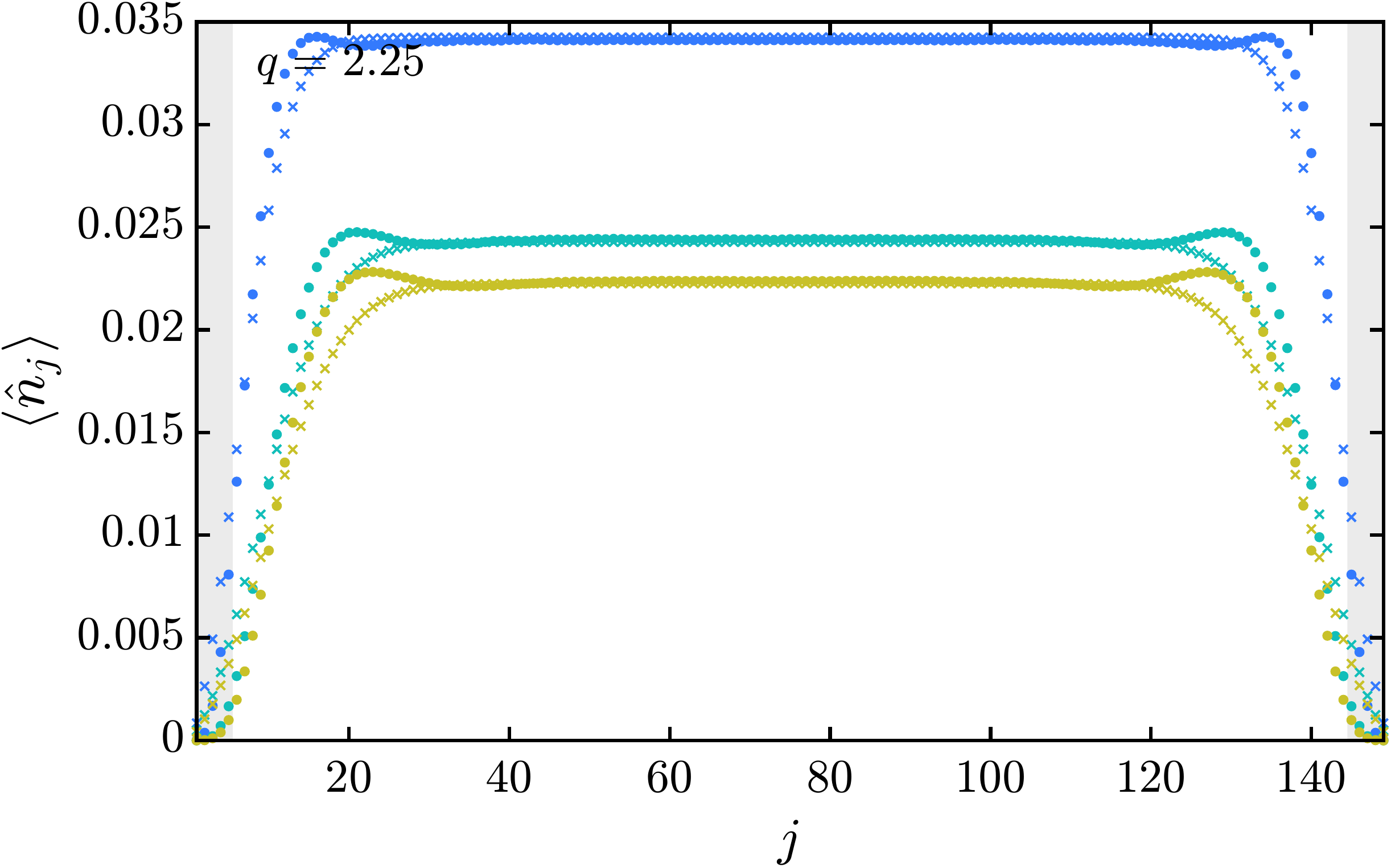} \\
    \end{tabular}
    \caption{Kink density $\langle\hat n_j\rangle$ at the end of the ramp versus the bond index $j$ for (a) $q=0.1$, (b) $q=0.2$, (c) $q=1.5$ and (d) $q=2.25$. Filled circles and crosses correspond to the TDVP simulations and to the free-fermion approximation on the same open chain, respectively, with one color per driving time as given in (b). The gray bands mark the five outermost bonds at each end.}
    \label{fig:kink_density}
\end{figure*}

In this section, we show the spatial distribution of the kinks at the end of the ramp, $\langle\hat n_j\rangle$, with the kink projectors $\hat n_j$ of Eq.~\eqref{eq:Ndef_app}, which sum to the mean kink number, $\kappa_1=\sum_j\langle\hat n_j\rangle$. Figure~\ref{fig:kink_density} compares the TDVP profiles with those of the free-fermion approximation on the same open chain for the four fractional orders of the main text.

For $q=1.5$ and $q=2.25$ the density forms a flat plateau in the bulk, which the TDVP and free-fermion data reproduce closely. Near both ends fewer kinks are created, in a layer of about $20$--$40$ bonds that widens as $\tau_Q$ grows. As this layer occupies a growing fraction of the chain at slow ramps, it steepens the decay of the mean at $L=150$, which accounts for about half of the difference between the TDVP and ring exponents at $q=1.5$ and for all of it at $q=2.25$, as discussed in Sec.~\ref{sec:cumulants_num}.

For $q=0.1$ and $q=0.2$ the TDVP profiles behave in the opposite way. The density grows towards both ends, and at slow ramps the profile becomes U-shaped, so that at $q=0.2$ and $\tau_Q=200$ it drops to about $0.005$ in the center of the chain and reaches about $0.04$ near the edges. The free-fermion profiles remain flat. This enhancement reflects the reduced coupling near the ends of the open chain, where a spin feels only $34\%$ of the coupling of the infinite chain at $q=0.2$, compared with $62\%$ at the center (Sec.~\ref{sec:Ising_model}), so that the ordering is weaker there.

The profile fixes the mean, while the variance in excess of the local contributions, $\kappa_2-\sum_j\langle\hat n_j\rangle(1-\langle\hat n_j\rangle)=\sum_{i\neq j}C_{ij}$, stems from the correlations between different bonds of Sec.~\ref{sec:kinkcorr}. The correlations of the middle bond in Sec.~\ref{sec:kinkcorr} show that the violation of the free-fermion bound at small $q$ originates in the bulk, while the enhanced density near the edges adds to the mean and variance of the whole chain.

\clearpage

\bibliography{biblio_new}

\end{document}